\documentclass[11pt]{article}

\usepackage{amsmath,amssymb,amsfonts,amsthm,mathtools}
\usepackage{booktabs,array,threeparttable,graphicx,float}
\usepackage{enumitem}
\usepackage{placeins}
\usepackage{geometry,setspace,natbib,hyperref,xcolor,microtype}
\hypersetup{
  colorlinks=true,citecolor=blue,linkcolor=blue,urlcolor=blue,
  pdftitle={Distributional Difference-in-Differences: Aggregation Before or After Quantile Inversion?},
  pdfauthor={Ulrich Hounyo},
  pdfsubject={Working paper on distributional difference-in-differences and nonlinear aggregation of quantile treatment effects},
  pdfkeywords={difference-in-differences; distributional treatment effects; quantile treatment effects; changes-in-changes; staggered adoption}
}

\newtheorem{corollary}{Corollary}
\newtheorem{proposition}{Proposition}

\newcommand{\E}{\mathbb{E}}
\newcommand{\Prb}{\mathbb{P}}

\newcommand{\DTT}{\mathrm{DTT}}
\newcommand{\QTT}{\mathrm{QTT}}
\newcommand{\Var}{\operatorname{Var}}
\newcommand{\Cov}{\operatorname{Cov}}

\title{\textbf{Distributional Difference-in-Differences:}\\
Aggregation Before or After Quantile Inversion?}
\author{
Ulrich Hounyo\thanks{Department of Economics, University at Albany - State University
of New York, Albany, NY 12222, USA. E-mail: \texttt{khounyo@albany.edu}.}
}
\date{August 22, 2026}

\begin{document}
\maketitle

\begin{abstract}
Staggered distributional difference-in-differences produces cohort-specific potential-outcome distributions, but applied work typically wants one overall quantile treatment effect. Two natural summaries---averaging cohort quantile treatment effects (QTTs) and mixing cohort distributions before inversion---use the same policy weights yet answer different target-population questions and can disagree even in sign. We derive the exact sharp interval for their gap conditional on cohort quantiles and weights, a globally sharp range-only envelope, and a locally sharp density-tilt representation. We develop joint smooth and mass-point-safe inference and show that neither estimator is uniformly more precise, even when the estimands coincide. In a same-object reconstruction of a public staggered-QTT application, holding data, identification, distributions, and weights fixed while changing only aggregation order reverses reported signs at several quantiles. Aggregation order is therefore part of the estimand and must be chosen before inversion.
\end{abstract}



\section{Introduction}
\label{sec:intro}

Staggered-adoption difference-in-differences (DiD) naturally produces cohort--time effects, while policy analysis often demands one overall summary. For distributional effects, that final aggregation step is itself an estimand choice. Fix an event time $e$ and prespecified positive policy weights $\omega_g^e$ on the contributing cohorts $\mathcal G_e$. Once the cohort-state potential-outcome distributions $F_{g,g+e}^{d}$, $d\in\{0,1\}$, have been identified, two natural overall quantile treatment effects are
\begin{align*}
\QTT_{\mathrm{avg}}^e(\tau)
&=\sum_{g\in\mathcal G_e}\omega_g^e
\big\{Q_{g,g+e}^{1}(\tau)-Q_{g,g+e}^{0}(\tau)\big\},\\
\QTT_{\mathrm{mix}}^e(\tau)
&=\left(\sum_{g\in\mathcal G_e}\omega_g^eF_{g,g+e}^{1}\right)^{-1}(\tau)
-\left(\sum_{g\in\mathcal G_e}\omega_g^eF_{g,g+e}^{0}\right)^{-1}(\tau).
\end{align*}
The first averages cohort-specific quantile effects under the policy weights. The second compares quantiles of the two pooled potential-outcome marginals generated by the same cohort lottery. They use the same cohort distributions and the same weights; only aggregation order differs. Because quantile inversion is nonlinear, they are generally different population objects.

A two-cohort calibration shows why the distinction matters. With equal cohort weights, the average cohort QTT can equal $+1$ at the 20th, 50th, and 80th percentiles while the pooled-population QTTs are approximately $0$, $+1$, and $+2$. The average-minus-mixture gap is then about $+1$, $0$, and $-1$. Nothing about the data, identifying assumptions, or target weights changes. Averaging QTTs answers the average-cohort quantile question; mixing first answers how a quantile of the pooled target population moves. Neither target dominates the other, and neither is automatically a distribution of individual treatment effects or a welfare ranking without additional economic structure.

The main theorem asks how far apart these targets can be even after the cohort quantiles and policy weights are fixed. Write $q_{g,d}(\tau)$ for the cohort $\tau$-quantile in state $d$, $\bar q_d=\sum_g\omega_g^e q_{g,d}$, and let $L_d$ and $R_d$ be the distances from $\bar q_d$ to the minimum and maximum active-support cohort quantiles. Proposition~\ref{prop:aggregationgap} proves the sharp interval
\begin{equation}
\boxed{
-\{L_0(\tau)+R_1(\tau)\}
\le \Delta_{\mathrm{agg}}^e(\tau)
\le R_0(\tau)+L_1(\tau),
}
\qquad
\Delta_{\mathrm{agg}}^e=\QTT_{\mathrm{avg}}^e-\QTT_{\mathrm{mix}}^e.
\label{eq:introsharp}
\end{equation}
Both endpoints are arbitrarily attainable by smooth Gaussian component distributions preserving every active-support cohort quantile and weight. Hence the exact worst-case magnitude conditional on those objects is
$B_{\mathrm{sharp}}^e=\max\{L_0+R_1,R_0+L_1\}$. Whenever cross-cohort quantile dispersion is nonzero, the same cohort quantiles and weights are compatible with either sign of the aggregation gap. A coarser bound remains globally sharp when only weights and state-specific quantile ranges are retained. This is a population statement about nonlinear aggregation, not an asymptotic approximation.

The theorem also explains the local mechanism. Mixture inversion replaces the prespecified policy weights, to first order, by positive state- and quantile-specific weights tilted toward cohorts with higher density at their own target quantiles. The resulting tilt--dispersion index has a sharp first-order envelope. Proposition~\ref{prop:aggregationinference} gives joint influence-function inference for the two QTTs and their gap with estimated cohort weights; Corollary~\ref{cor:masspointbands} supplies a CDF--weight projection route that remains valid at mass points; and Corollary~\ref{cor:aggregationprocess} gives process inference. Proposition~\ref{prop:efficiencyrisk} shows that neither estimator is uniformly more precise even when the two estimands coincide and quantifies the local quadratic-risk cost of reporting one aggregand when the other is the prespecified target.

The distinction is empirically operative. Current \texttt{qte} documentation defines staggered \texttt{ddid(..., gt\_type="qtt")} summaries by mixture-CDF aggregation \citep{callaway2026qte}. We independently reconstruct its group--time potential-outcome distributions in the public \texttt{mpdta} minimum-wage panel, attach every distribution and weight to its explicit $(g,t)$ label, and then change only the final aggregation operator. At $\tau=.10$, the average QTT is $-0.050$ and the mixture QTT is $+0.068$; signs also differ at other quantiles. This is not evidence that mixture aggregation is wrong. It shows that a reported distributional effect can change sign solely because the target functional changes; Section~\ref{sec:mpdta} reports the accompanying state-deletion sensitivity separately.

Our contribution begins after first-stage distributional identification. Recent work develops staggered distributional identification and uniform QTT/DTT inference \citep{djuazontsyawo2026}, doubly robust staggered QTT estimation and aggregation \citep{millergalvao2026}, and density-weighted quantile-regression/TWFE decompositions \citep{arias2026}; the broader DiD literature clarifies treatment timing, comparison groups, and heterogeneous average effects \citep{callawaysantanna2021,goodmanbacon2021,sunabraham2021,dechaisemartin2020,borusyak2024,roth2023whats,baker2026}. Arias's density weights arise from regression decompositions that can be non-convex across cohort--time effects. Here the starting point is prespecified positive policy weights on already identified cohort-state CDFs, and density tilting is induced by inversion of their positive mixture. We claim neither component aggregation operation nor generic inverse-CDF inference as new. The contribution is the sharp finite/global/local theory for the discrepancy between the two policy aggregation operators, inference on that discrepancy, and the statistical consequences of substituting one target for the other.

Distributional DiD still requires a credible route to each missing untreated CDF. Section~\ref{sec:framework} states the first-stage scope needed to make the aggregation problem well defined, including standardization before marginal quantile inversion. The Appendix retains the prespecified diagnostic workflow, generic max/stepdown machinery, simulations, and additional empirical validation. Those diagnostics can reject or qualify a first-stage route, but they never choose between the two aggregation targets. The resulting practitioner rule is short: recover the counterfactual distribution under a named route; choose the target population and aggregation order before inversion; when both summaries are substantively meaningful, report both and their gap; and use CDF projection rather than smooth quantile linearization when the outcome is nonregular.

\section{Setup and First-Stage Scope}
\label{sec:framework}

Observe periods $t=1,\ldots,T$ and let $G_i\in\mathcal G\cup\{\infty\}$ be unit $i$'s first treatment date, with absorbing treatment $D_{it}=1\{t\ge G_i\}$. Write $Y_{it}(g)$ for the potential outcome under first treatment at $g$ and $Y_{it}(\infty)$ for the untreated path. We maintain consistency/no interference and no anticipation over the baseline window. For treated cohort $g$ and post-treatment date $t\ge g$, define
\[
F_{g,t}^{1}(y)=\Prb\{Y_{it}(g)\le y\mid G_i=g\},\qquad
F_{g,t}^{0}(y)=\Prb\{Y_{it}(\infty)\le y\mid G_i=g\}.
\]
Consistency identifies $F_{g,t}^{1}$; the first-stage causal-design problem is to recover the missing $F_{g,t}^{0}$. The aggregation theory below is deliberately modular: it starts once the cohort--state distributions have been recovered under an independently defended distributional-DiD route.

The marginal distribution treatment effect on the treated (DTT) and quantile treatment effect on the treated (QTT) are
\begin{align}
\DTT_{g,t}(y)&=F_{g,t}^{1}(y)-F_{g,t}^{0}(y),
\label{eq:dtt}\\
\QTT_{g,t}(\tau)&=Q_{g,t}^{1}(\tau)-Q_{g,t}^{0}(\tau),\qquad
Q_{g,t}^{d}(\tau)=\inf\{y:F_{g,t}^{d}(y)\ge\tau\}.
\label{eq:qtt}
\end{align}
No rank-invariance or cross-state copula assumption is needed for these marginal objects once $(F_{g,t}^{1},F_{g,t}^{0})$ is identified. By contrast, the distribution of the individual effect $Y_{it}(g)-Y_{it}(\infty)$ requires additional dependence information. A conditional quantile contrast is also generally not a marginal QTT \citep{arias2026}: conditional identification must be followed by standardization of the conditional CDF and only then by inversion.

\paragraph{First-stage routes.}
Mean parallel trends alone does not recover $F_{g,t}^{0}$. One transparent route used in our robustness illustration is additive CDF parallel trends (CDF-PT). For baseline $s<g$ and an admissible comparison population $C$ untreated at both dates,
\begin{equation}
F_{g,t}^{0}(y)=F_{g,s}(y)+F_{C,t}(y)-F_{C,s}(y).
\label{eq:cdfid}
\end{equation}
The population right-hand side must itself be a proper CDF; finite-sample shape projection cannot repair a false identifying restriction. Changes-in-changes (CIC) instead uses a monotone nonseparable structure and, in its canonical two-period form, identifies
\begin{equation}
F_{g,t}^{0}(y)=F_{g,s}\!\left[F_{C,s}^{-1}\{F_{C,t}(y)\}\right],
\label{eq:cic}
\end{equation}
subject to the required support condition \citep{atheyimbens2006}. Other routes include panel change-distribution methods with dependence restrictions \citep{callawaylioka2018,callawayli2019}, distribution-regression DiD \citep{fernandezval2024}, common-CDF time-effect models \citep{kimwooldridge2025}, and partial identification \citep{fanyu2012}. These restrictions are generally non-nested and should not be treated as interchangeable estimators.

If a route identifies $F_{g,t}^{0}(y\mid x)$ conditionally, the target marginal counterfactual is
\begin{equation}
F_{g,t}^{0}(y)=\int F_{g,t}^{0}(y\mid x)\,dF_{X\mid G=g}(x),
\label{eq:standardize}
\end{equation}
and nonlinear marginal functionals are computed only after that integration. Thus the implementation order is
\[
\boxed{\text{recover }F_{g,t}^{0}\ \longrightarrow\ \text{standardize if needed}\ \longrightarrow\ \text{apply the nonlinear target}.}
\]
The Appendix gives the full identification map, comparison-set/support qualifications, influence functions, mass-point branch, and the prespecified STOP/PROCEED diagnostic theory. Those diagnostics assess the maintained route used to recover $F_{g,t}^{0}$; they never choose between the two aggregation targets studied next. Conversely, aggregation theory cannot rescue an unidentified first stage. This separation keeps the main paper focused on the estimand problem while retaining a fully auditable distributional-DiD workflow.

\section{Staggered Adoption, Nonlinear Aggregation, and Inference}
\label{sec:staggered}

Recent work studies QTT and distributional identification directly in staggered-adoption environments \citep{lilin2024,ciaccio2024,djuazontsyawo2026,millergalvao2026}. We condition on that first-stage identification problem being solved and ask a different question: which nonlinear population object is created when cohort information is aggregated? The answer depends on whether aggregation is applied before or after quantile inversion. This section contains the paper's central formal contribution: the estimand comparison, smooth and nonregular inference, and the connection back to disciplined post-gate reporting.

Modern DiD practice recommends building cohort-time causal effects before aggregation \citep{callawaysantanna2021}. Distributional DiD should inherit this design principle.

\subsection{Cohort-time building blocks}

For each cohort $g$ and post-treatment date $t\ge g$, estimate
\[
F_{g,t}^{1},\qquad F_{g,t}^{0},
\]
using comparison units valid for that specific contrast. The resulting $\DTT_{g,t}(y)$ and $\QTT_{g,t}(\tau)$ preserve heterogeneity by cohort and exposure length.

This organization also avoids importing the interpretation problems of conventional two-way fixed-effects regressions under heterogeneous treatment effects and staggered timing \citep{goodmanbacon2021,dechaisemartin2020,sunabraham2021,dechaisemartinsurvey2023,borusyak2024}. Related nonlinear DiD strategies for panel outcomes are discussed by \citet{wooldridge2023nonlinear}. A nonlinear or quantile two-way fixed-effects (TWFE) interaction has no automatic interpretation as a weighted average of cohort-time QTTs.

\subsection{Event-time DTTs}

Let $e=t-g$ and begin with prespecified nonnegative target-population weights among cohorts observed at event time $e$. Define the \emph{active target support}
\[
\mathcal G_e=\{g:g+e\le T,\ \omega_g^e>0\}.
\]
Zero-weight cohorts do not enter either aggregand and are omitted from all sums, minima, maxima, and ranges below. Thus, without loss of generality for the target under study, $\omega_g^e>0$ for every $g\in\mathcal G_e$ and $\sum_{g\in\mathcal G_e}\omega_g^e=1$. The weighting rule is prespecified, although its population shares may be estimated from the sample. A natural event-time DTT is
\[
\DTT^e(y)
=
\sum_{g\in\mathcal G_e}\omega_g^e\DTT_{g,g+e}(y).
\]
Because CDF differences are linear, this equals the difference between the correspondingly weighted potential-outcome CDFs when both target mixtures use the same cohort weights \(\omega_g^e\); using different treated and untreated weights defines a different estimand.

\subsection{Two different QTT aggregands}

Quantiles create a choice that does not arise for linear CDF contrasts. One estimand is the average cohort-specific QTT,
\begin{equation}
\QTT_{\mathrm{avg}}^e(\tau)
=
\sum_{g\in\mathcal G_e}\omega_g^e\QTT_{g,g+e}(\tau).
\label{eq:qavg}
\end{equation}
A different estimand first forms mixture distributions
\[
F_d^e(y)=\sum_{g\in\mathcal G_e}\omega_g^eF_{g,g+e}^{d}(y),\qquad d\in\{0,1\},
\]
and then defines
\begin{equation}
\QTT_{\mathrm{mix}}^e(\tau)
=
(F_1^e)^{-1}(\tau)-(F_0^e)^{-1}(\tau).
\label{eq:qmix}
\end{equation}
The distinction can be written as a target-population statement. If $G^\star$ is a cohort label drawn from the prespecified policy distribution $\Pr(G^\star=g)=\omega_g^e$, then
\[
\QTT_{\mathrm{avg}}^e(\tau)
=\E_{\omega^e}\!\left[Q_{G^\star,G^\star+e}^{1}(\tau)-Q_{G^\star,G^\star+e}^{0}(\tau)\right]
\]
is the policy-weighted average of cohort-specific quantile contrasts. By contrast, $\QTT_{\mathrm{mix}}^e(\tau)$ is the difference between the $\tau$-quantiles of the two pooled marginal distributions obtained after drawing the cohort label from the same policy distribution. This interpretation does not turn either quantity into a distribution of individual treatment effects, nor does it impose a welfare ordering without additional social preferences; it makes explicit which population functional the analyst chooses.

Generally,
\[
\QTT_{\mathrm{avg}}^e(\tau)\neq\QTT_{\mathrm{mix}}^e(\tau).
\]

Write \(q_{g,d}(\tau)=Q_{g,g+e}^d(\tau)\),
\(\bar q_d(\tau)=\sum_g\omega_g^e q_{g,d}(\tau)\), and
\(q_d^{\mathrm{mix}}(\tau)=(F_d^e)^{-1}(\tau)\). Define
\[
q_{d,\min}(\tau)=\min_{g\in\mathcal G_e}q_{g,d}(\tau),\qquad
q_{d,\max}(\tau)=\max_{g\in\mathcal G_e}q_{g,d}(\tau),
\]
\[
L_d(\tau)=\bar q_d(\tau)-q_{d,\min}(\tau),\qquad
R_d(\tau)=q_{d,\max}(\tau)-\bar q_d(\tau),\qquad
D_d(\tau)=\max\{L_d(\tau),R_d(\tau)\}.
\]

\paragraph{How to read the main proposition.}
Part (ii) is the headline finite-dispersion result: it gives the sharp interval and exact worst-case magnitude conditional on cohort quantiles and target weights. Parts (iii)--(vii) explain the local density-tilt mechanism, uniform extension, equality cases, policy-target error, and sharp first-order sensitivity. A reader interested primarily in the finite sharp bound can read parts (i)--(ii) and then proceed directly to the inference results.

\begin{proposition}[Sharp aggregation-gap bounds and local density tilt]
\label{prop:aggregationgap}
Fix $e$ and $\tau\in(0,1)$. Suppose the contributing cohort CDFs are continuous and strictly increasing on intervals containing their $\tau$-quantiles.
\begin{enumerate}[label=(\roman*)]
\item For $d\in\{0,1\}$,
\[
\min_{g\in\mathcal G_e} q_{g,d}(\tau)\le q_d^{\mathrm{mix}}(\tau)\le\max_{g\in\mathcal G_e} q_{g,d}(\tau).
\]
\item Let $H_d(\tau)=q_{d,\max}(\tau)-q_{d,\min}(\tau)=L_d(\tau)+R_d(\tau)$ and $\omega_{\min}^e=\min_{g\in\mathcal G_e}\omega_g^e>0$. Then the aggregation gap obeys the one-sided interval
\begin{equation}
-\{L_0(\tau)+R_1(\tau)\}
\le \Delta_{\mathrm{agg}}^e(\tau)
\le R_0(\tau)+L_1(\tau).
\label{eq:aggregationsharpinterval}
\end{equation}
Conditional on the fixed active-support weights and the fixed cohort quantile vectors $\{q_{g,0}(\tau),q_{g,1}(\tau)\}_g$, both endpoints are sharp over continuous strictly increasing cohort CDFs: whenever the relevant endpoint is nonzero, for every $\varepsilon>0$ there exist smooth Gaussian component distributions preserving those cohort quantiles for which the gap lies within $\varepsilon$ of that endpoint. Consequently, over the maintained class of continuous strictly increasing component CDFs that preserve the active-support cohort quantiles and weights,
\begin{equation}
\sup |\Delta_{\mathrm{agg}}^e(\tau)|
=B_{\mathrm{sharp}}^e(\tau)
=\max\{L_0(\tau)+R_1(\tau),\;R_0(\tau)+L_1(\tau)\}.
\label{eq:bsharp}
\end{equation}
It satisfies the nested bounds
\begin{align}
|\Delta_{\mathrm{agg}}^e(\tau)|
&\le B_{\mathrm{sharp}}^e(\tau)
\le D_0(\tau)+D_1(\tau) \nonumber\\
&\le (1-\omega_{\min}^e)\{H_0(\tau)+H_1(\tau)\}
\le H_0(\tau)+H_1(\tau).
\label{eq:aggregationbound}
\end{align}
When at least two cohorts contribute, the coefficient $1-\omega_{\min}^e$ in the range-only bound remains globally sharp if only the weights and state-specific ranges $H_0,H_1$ are fixed: for every $\varepsilon>0$ and $H_0,H_1>0$, smooth Gaussian cohort distributions can be chosen so that
\[
|\Delta_{\mathrm{agg}}^e(\tau)|
>(1-\omega_{\min}^e)(H_0+H_1)-\varepsilon.
\]
Thus \eqref{eq:aggregationsharpinterval} is the sharp fixed-quantile envelope, while the $(1-\omega_{\min}^e)$ inequality is the sharp coarser envelope when only weights and ranges are retained. In particular, if $H_0(\tau)+H_1(\tau)>0$, both endpoints in \eqref{eq:aggregationsharpinterval} are strictly nonzero and smooth component distributions can make the gap either positive or negative while preserving every cohort quantile and target weight. Cohort quantiles and weights alone therefore do not determine the sign of the aggregation discrepancy.
\item Suppose additionally that, for each $d\in\{0,1\}$, the contributing CDFs admit continuously differentiable densities $f_{g,d}$ on the interval from $\min_{g\in\mathcal G_e} q_{g,d}(\tau)$ to $\max_{g\in\mathcal G_e} q_{g,d}(\tau)$, with
\[
\inf_{g,y} f_{g,d}(y)\ge m_d>0,
\qquad
\sup_{g,y}|f'_{g,d}(y)|\le K_d<\infty.
\]
Let
\[
\bar f_d(\tau)=\sum_g\omega_g^e f_{g,d}\{q_{g,d}(\tau)\},
\qquad
\Delta f_d(\tau)=\max_g f_{g,d}\{q_{g,d}(\tau)\}-\min_g f_{g,d}\{q_{g,d}(\tau)\},
\]
and define
\[
A_d(\tau)=
\frac{\sum_g\omega_g^e f_{g,d}\{q_{g,d}(\tau)\}[q_{g,d}(\tau)-\bar q_d(\tau)]}{\bar f_d(\tau)},
\qquad
\widetilde\omega_{g,d}^e(\tau)=
\frac{\omega_g^e f_{g,d}\{q_{g,d}(\tau)\}}{\bar f_d(\tau)}.
\]
Equivalently, define the \emph{density-tilt operator} $\mathcal R_{d,\tau}$ on the cohort-weight simplex by
\[
\{\mathcal R_{d,\tau}(\omega^e)\}_g
=\frac{\omega_g^e f_{g,d}\{q_{g,d}(\tau)\}}
{\sum_h\omega_h^e f_{h,d}\{q_{h,d}(\tau)\}}
=\widetilde\omega_{g,d}^e(\tau).
\]
Then
\[
\begin{aligned}
q_d^{\mathrm{mix}}(\tau)-\bar q_d(\tau)
&=A_d(\tau)+r_d(\tau),\\
q_d^{\mathrm{mix}}(\tau)
&=\sum_g\widetilde\omega_{g,d}^e(\tau)q_{g,d}(\tau)+r_d(\tau),\\
|r_d(\tau)|
&\le
\frac{K_d}{2\bar f_d(\tau)}
\sum_g\omega_g^e\{q_d^{\mathrm{mix}}(\tau)-q_{g,d}(\tau)\}^2
\le \frac{K_d}{2m_d}H_d(\tau)^2.
\end{aligned}
\]
Thus the density tilt is the first-order term in quantile dispersion when cross-cohort density heterogeneity is not itself vanishing at the same rate. More precisely,
\[
|A_d(\tau)|
\le
\frac{\Delta f_d(\tau)H_d(\tau)}{4\bar f_d(\tau)}
\le
\frac{\Delta f_d(\tau)H_d(\tau)}{4m_d},
\]
and
\begin{equation}
\begin{aligned}
\Delta_{\mathrm{agg}}^e(\tau)
&\equiv\QTT_{\mathrm{avg}}^e(\tau)-\QTT_{\mathrm{mix}}^e(\tau)\\
&=A_0(\tau)-A_1(\tau)+\rho_e(\tau),\qquad
|\rho_e(\tau)|\le\sum_{d=0}^1\frac{K_dH_d(\tau)^2}{2m_d}.
\end{aligned}
\label{eq:aggregationlocal}
\end{equation}
Define the total-variation tilt and the \emph{tilt--dispersion index}
\[
V_d(\tau)=\frac12\sum_g|\widetilde\omega_{g,d}^e(\tau)-\omega_g^e|,
\qquad
\mathcal D_{\mathrm{tilt}}^e(\tau)=H_0(\tau)V_0(\tau)+H_1(\tau)V_1(\tau).
\]
Here $0\le V_d(\tau)\le1$; over cohorts with positive target weight, $V_d(\tau)=0$ if and only if the own-quantile density values $f_{g,d}\{q_{g,d}(\tau)\}$ are common across cohorts.
Because $A_d(\tau)=\sum_g\{\widetilde\omega_{g,d}^e(\tau)-\omega_g^e\}q_{g,d}(\tau)$ and both weight vectors sum to one,
\begin{equation}
|A_d(\tau)|\le H_d(\tau)V_d(\tau),
\qquad
|\Delta_{\mathrm{agg}}^e(\tau)|
\le \mathcal D_{\mathrm{tilt}}^e(\tau)
+\sum_{d=0}^1\frac{K_dH_d(\tau)^2}{2m_d}.
\label{eq:tiltdiagnosticbound}
\end{equation}
Hence $\widehat{\mathcal D}_{\mathrm{tilt}}^e(\tau)$, formed from estimated cohort quantiles, weights, and local densities, is a directly reportable first-order aggregation-divergence diagnostic. It is not a confidence bound unless estimation error and the curvature remainder are also accounted for.
\item Let $\mathcal T=[\underline\tau,\overline\tau]\subset(0,1)$ be compact. If the assumptions in part (iii) hold simultaneously on the union of the cohort-quantile ranges generated by $\tau\in\mathcal T$, with the same constants $m_d,K_d$, then the decomposition and bounds are uniform on $\mathcal T$. In particular,
\[
\sup_{\tau\in\mathcal T}|\rho_e(\tau)|
\le\sum_{d=0}^1\frac{K_d}{2m_d}\sup_{\tau\in\mathcal T}H_d(\tau)^2,
\]
and
\[
\sup_{\tau\in\mathcal T}|\Delta_{\mathrm{agg}}^e(\tau)|
\le
\sup_{\tau\in\mathcal T}\left\{\mathcal D_{\mathrm{tilt}}^e(\tau)
+\sum_{d=0}^1\frac{K_dH_d(\tau)^2}{2m_d}\right\}.
\]
\item The aggregation gap has no universal sign and can change sign across quantiles in the same design.
\item Define the state-specific aggregation distortion
\[
\kappa_d^e(\tau)=q_d^{\mathrm{mix}}(\tau)-\bar q_d(\tau).
\]
Then
\[
\Delta_{\mathrm{agg}}^e(\tau)=\kappa_0^e(\tau)-\kappa_1^e(\tau),
\]
so the two QTT aggregands coincide at $(e,\tau)$ if and only if $\kappa_0^e(\tau)=\kappa_1^e(\tau)$. This exact equality class includes, but is not limited to: a single contributing cohort; common cohort $\tau$-quantiles within each potential-outcome state; and a common treatment location shift across cohorts,
\[
F_{g,g+e}^1(y)=F_{g,g+e}^0(y-\delta_e),
\]
under which both aggregands equal $\delta_e$ for every $\tau$. If the policy target is $\QTT_{\mathrm{avg}}^e(\tau)$ but the mixture QTT is reported, the report-minus-target error is exactly $-\Delta_{\mathrm{agg}}^e(\tau)$; reversing the target reverses the sign.
\item The tilt--dispersion envelope is sharp for the first-order component. Fix $d$, the target weights $\omega^e$, and the induced tilted weights $\widetilde\omega_d^e(\tau)$, and retain the notation $V_d(\tau)$. Among all cohort-quantile vectors $q_d=(q_{g,d})_g$ satisfying $\max_{g\in\mathcal G_e} q_{g,d}-\min_{g\in\mathcal G_e} q_{g,d}\le H_d(\tau)$,
\[
\sup_{q_d}
\left|\sum_g\{\widetilde\omega_{g,d}^e(\tau)-\omega_g^e\}q_{g,d}\right|
=H_d(\tau)V_d(\tau).
\]
Consequently, allowing the two state-specific quantile vectors to vary separately,
\[
\sup_{q_0,q_1}|A_0(\tau)-A_1(\tau)|
=H_0(\tau)V_0(\tau)+H_1(\tau)V_1(\tau)
=\mathcal D_{\mathrm{tilt}}^e(\tau).
\]
When $V_d(\tau)>0$, equality is attained by assigning one endpoint of the admissible quantile range to cohorts for which $\widetilde\omega_{g,d}^e-\omega_g^e>0$ and the other endpoint to cohorts for which it is negative. Moreover, any strictly positive target and tilted weight vectors with common support can be embedded, at a fixed $\tau$, in smooth Gaussian cohort distributions having the prescribed cohort quantiles and own-quantile densities that generate the tilt. Hence along shrinking-dispersion sequences satisfying the smoothness conditions in part (iii), the exact aggregation gap can attain the first-order envelope up to the stated $O\{H_0(\tau)^2+H_1(\tau)^2\}$ curvature remainder. In this precise local sense, $\mathcal D_{\mathrm{tilt}}^e(\tau)$ is a sharp sensitivity envelope rather than merely a descriptive index.
\end{enumerate}
\end{proposition}

The proof, a shrinking-heterogeneity Gaussian calibration, and a two-cohort sign-reversal example are in the Appendix. Part (ii) is a finite-dispersion result, not a local approximation: after the active-support cohort quantiles and policy weights are fixed, $B_{\mathrm{sharp}}^e(\tau)$ equals the supremum of the gap magnitude over the maintained class of smooth component distributions. The wider $(1-\omega_{\min}^e)(H_0+H_1)$ envelope discards the locations of the weighted averages within the two cohort-quantile ranges and is sharp only for that coarser information set. Part (iii) then makes the local economic mechanism explicit: relative to the target-population weights $\omega_g^e$, mixture inversion tilts influence toward cohorts whose distributions are denser at their own $\tau$-quantile. Part (vii) upgrades $\mathcal D_{\mathrm{tilt}}^e(\tau)$ from a descriptive warning to a sharp first-order sensitivity envelope. The Taylor remainder clarifies scope: when density dispersion is itself $O\{H_d(\tau)\}$, the state-specific aggregation distortion is second order rather than first order. Part (iv) makes the deterministic local statement uniform over compact quantile ranges, while part (vi) gives an exact equality characterization and the exact policy-misreporting error.

The distinction is substantive. Equation \eqref{eq:qavg} asks for an average of cohort-specific quantile contrasts. Equation \eqref{eq:qmix} asks how treatment changes the quantile of an aggregate population formed by mixing cohorts. Neither dominates the other; the paper must state which policy question is being answered.

The distinction also clarifies the relation to recent staggered-QTT work. \citet{millergalvao2026} develop doubly robust cohort/time QTT estimators and aggregation schemes that summarize those QTTs, whereas \citet{djuazontsyawo2026} develop staggered DTT/QTT identification and uniform inference that remains valid for possibly non-continuous outcomes. \citet{arias2026} is especially close in vocabulary because density weighting also appears in his analysis of quantile-regression and recentered-influence-function TWFE implementations. The mechanisms are different. Arias's density weights arise in a regression decomposition and staggered TWFE can attach non-convex weights to cohort--time effects; here the starting weights are prespecified nonnegative policy weights, every cohort-state CDF is already identified, and density tilting is induced endogenously by quantile inversion of their positive mixture. Our object is the exact discrepancy between that mixture quantile and the target-weighted average of cohort quantiles. Thus the closest papers provide first-stage distributional identification, estimation, or weighting antecedents, while the present margin is operator noncommutativity, sharp discrepancy envelopes, inference on the gap, and the consequences of substituting one target for another.

\begin{table}[t]
\centering
\caption{Closest staggered distributional-DiD antecedents and the present margin}
\label{tab:daylight}
\small
\begin{tabular}{p{2.7cm}p{4.4cm}p{6.2cm}}
\toprule
Work & Primary object & Margin relative to this paper\\
\midrule
Miller--Galvao (2026) & Doubly robust cohort/time QTT identification, estimation, inference, and aggregation under staggered interventions & Supplies close QTT building blocks and aggregation summaries; does not make the average-QTT versus mixture-CDF noncommutativity and its gap the target of analysis.\\
Djuazon--Tsyawo (2026) & Staggered DTT/QTT identification and uniform inference, including non-continuous outcomes & Supplies close distributional targets and nonregular inference technology; the present paper uses generic CDF-band inversion as an input rather than claiming it as new.\\
Arias (2026) & Conditional/unconditional quantile DiD and density-weighted quantile-regression/recentered-influence-function TWFE decompositions & Density weighting arises from a regression decomposition and may be non-convex across cohort--time effects. Here density tilt is generated by inversion of a positive mixture whose prespecified policy weights and cohort-state CDFs are held fixed.\\
This paper & Given identified cohort-state CDFs, compare average-cohort QTT and mixture-CDF QTT & Sharp fixed-quantile and range-only discrepancy envelopes; sharp local tilt--dispersion envelope; exact equality and policy-misreporting identities; joint gap/process inference; no-universal-efficiency theorem and target-risk decomposition; projection of joint CDF--weight uncertainty through both aggregation maps.\\
\bottomrule
\end{tabular}
\begin{flushleft}\footnotesize
Notes: The table deliberately separates first-stage identification/estimation from the paper's aggregation-operator question. Neither component aggregation operation, generic CDF-band inversion, nor generic bootstrap/multiple-testing machinery is claimed as new.
\end{flushleft}
\end{table}

Proposition~\ref{prop:aggregationgap} studies the noncommutativity directly: averaging cohort quantile contrasts and inverting cohort-mixture CDFs need not answer the same policy question. Proposition~\ref{prop:aggregationinference} makes inference on that discrepancy operational, and Proposition~\ref{prop:efficiencyrisk} shows why an apparent precision advantage cannot be used to redefine the policy target.

\begin{proposition}[Influence-function inference for the average-versus-mixture aggregation gap]
\label{prop:aggregationinference}
Fix an event time \(e\), a quantile \(\tau\in(0,1)\), and a finite set of contributing cohorts \(\mathcal G_e\), with \(\min_{g\in\mathcal G_e}\omega_g^e>0\). Let \(\widehat\omega_g^e\) estimate \(\omega_g^e\), with \(\sum_g\widehat\omega_g^e=1\) and \(\widehat\omega_g^e\ge0\), and let \(\widehat F_{g,d}\) estimate \(F_{g,g+e}^d\), \(d\in\{0,1\}\). Suppose that, on fixed neighborhoods containing the cohort quantiles \(q_{g,d}\) and mixture quantiles \(q_d^{\mathrm{mix}}\),
\[
\sqrt L\{\widehat F_{g,d}(y)-F_{g,g+e}^d(y)\}
=L^{-1/2}\sum_{\ell=1}^L\psi_{\ell,g,d}(y)+o_p(1)
\]
uniformly in \(y\), jointly over the finite cohort--state collection. Assume that the normalized score processes are jointly locally stochastically equicontinuous at every cohort and mixture quantile, and
\[
\sqrt L(\widehat\omega_g^e-\omega_g^e)
=L^{-1/2}\sum_{\ell=1}^L\xi_{\ell,g}+o_p(1),
\qquad
\sum_g\xi_{\ell,g}=0.
\]
The statement permits independent, non-identically distributed clusters. For the natural sample-share estimator, let $A_{\ell g,L}^e$ be cluster $\ell$'s count or survey-weight contribution to cohort $g$, put $A_{\ell+,L}^e=\sum_hA_{\ell h,L}^e$, $\mu_{g,L}^e=L^{-1}\sum_\ell\E A_{\ell g,L}^e$, $\mu_{+,L}^e=\sum_g\mu_{g,L}^e>0$, and $\omega_{g,L}^e=\mu_{g,L}^e/\mu_{+,L}^e$. Under the corresponding denominator LLN and joint Lindeberg--Feller CLT,
\[
\xi_{\ell,g,L}^e=
\frac{A_{\ell g,L}^e-\E A_{\ell g,L}^e
-\omega_{g,L}^e\{A_{\ell+,L}^e-\E A_{\ell+,L}^e\}}
{\mu_{+,L}^e},
\qquad \sum_g\xi_{\ell,g,L}^e=0,
\]
gives the ratio expansion around the triangular-array target $\omega_{g,L}^e$. If a fixed limit $\omega_g^e$ is used in the display below, additionally require $\sqrt L(\omega_{g,L}^e-\omega_g^e)\to0$. The identically distributed formula is the special case recorded in the Appendix. No independence between weight and CDF scores is imposed. This heterogeneous-cluster version is the one relevant for applications with unequal state sizes or survey-weight exposure.
Assume the relevant cohort densities are positive and continuous at \(q_{g,d}\), the mixture densities
\[
f_d^e(y)=\sum_{g\in\mathcal G_e}\omega_g^e f_{g,d}(y)
\]
are positive and continuous at \(q_d^{\mathrm{mix}}\), and the finite vector of score evaluations and weight scores satisfies a joint mean-zero Gaussian CLT. Define
\[
\chi_{\ell,g,d}
=-\frac{\psi_{\ell,g,d}(q_{g,d})}{f_{g,d}(q_{g,d})}
\]
and
\[
\chi_{\ell,d}^{\mathrm{mix}}
=-\frac{
\sum_g\omega_g^e\psi_{\ell,g,d}(q_d^{\mathrm{mix}})
+\sum_g\xi_{\ell,g}F_{g,g+e}^d(q_d^{\mathrm{mix}})
}{f_d^e(q_d^{\mathrm{mix}})}.
\]
Define the aggregation discrepancy by \(\Delta_{\mathrm{agg}}^e=\QTT_{\mathrm{avg}}^e-\QTT_{\mathrm{mix}}^e\) and its plug-in estimator by \(\widehat\Delta_{\mathrm{agg}}^e=\widehat\QTT_{\mathrm{avg}}^e-\widehat\QTT_{\mathrm{mix}}^e\). Then, for the plug-in estimators obtained by replacing all CDFs, quantiles, and weights by their sample analogues,
\[
\sqrt L
\begin{pmatrix}
\widehat\QTT_{\mathrm{avg}}^e-\QTT_{\mathrm{avg}}^e\\
\widehat\QTT_{\mathrm{mix}}^e-\QTT_{\mathrm{mix}}^e\\
\widehat\Delta_{\mathrm{agg}}^e-\Delta_{\mathrm{agg}}^e
\end{pmatrix}
=
L^{-1/2}\sum_{\ell=1}^L
\begin{pmatrix}
\Gamma_{\ell}^{\mathrm{avg}}\\
\Gamma_{\ell}^{\mathrm{mix}}\\
\Gamma_{\ell}^{\mathrm{gap}}
\end{pmatrix}
+o_p(1),
\]
where
\begin{align*}
\Gamma_{\ell}^{\mathrm{avg}}
&=
\sum_g\omega_g^e(\chi_{\ell,g,1}-\chi_{\ell,g,0})
+\sum_g\xi_{\ell,g}\{q_{g,1}-q_{g,0}\},\\
\Gamma_{\ell}^{\mathrm{mix}}
&=\chi_{\ell,1}^{\mathrm{mix}}-\chi_{\ell,0}^{\mathrm{mix}},\\
\Gamma_{\ell}^{\mathrm{gap}}
&=\Gamma_{\ell}^{\mathrm{avg}}-\Gamma_{\ell}^{\mathrm{mix}}.
\end{align*}
Consequently the three estimators are jointly asymptotically normal with covariance given by the limiting covariance of the displayed influence vector. When all three are stacked, this covariance is necessarily singular because $\Gamma_{\ell}^{\mathrm{gap}}=\Gamma_{\ell}^{\mathrm{avg}}-\Gamma_{\ell}^{\mathrm{mix}}$ identically; fixed cohort weights may create additional degeneracy by setting the weight-score block to zero. No covariance inverse is required for the stated joint CLT or multiplier bootstrap. If the estimated cluster influence vectors are empirically $L_2$-consistent, the true influence array has convergent second moments and maximal contribution $o_p(\sqrt L)$, and the multipliers have mean zero, unit variance, and a finite $2+\kappa$ moment, the centered-score covariance is consistent and the cluster multiplier bootstrap is conditionally valid. The same conclusion holds after stacking any fixed finite grid of quantiles; Corollary~\ref{cor:aggregationprocess} gives the functional extension.
\end{proposition}

\begin{proposition}[Precision, target mismatch, and no universal efficiency ranking]
\label{prop:efficiencyrisk}
Under Proposition~\ref{prop:aggregationinference}, let $(Z_{\mathrm{avg}},Z_{\mathrm{mix}},Z_{\mathrm{gap}})'$ denote the Gaussian limit of the three scaled estimators, so $Z_{\mathrm{gap}}=Z_{\mathrm{avg}}-Z_{\mathrm{mix}}$. Write
\[
\sigma_a^2=\Var(Z_{\mathrm{avg}}),\qquad
\sigma_m^2=\Var(Z_{\mathrm{mix}}),\qquad
\sigma_\Delta^2=\Var(Z_{\mathrm{gap}}),\qquad
\sigma_{m\Delta}=\Cov(Z_{\mathrm{mix}},Z_{\mathrm{gap}}).
\]
Then:
\begin{enumerate}[label=(\roman*)]
\item The asymptotic-variance difference obeys the exact identity
\begin{equation}
\sigma_a^2-\sigma_m^2
=\sigma_\Delta^2+2\sigma_{m\Delta}.
\label{eq:varianceordering}
\end{equation}
Thus the mixture estimator is weakly more precise if and only if $\sigma_\Delta^2+2\sigma_{m\Delta}\ge0$. The sign is not determined by density tilting alone; it depends on the covariance of the mixture score with the score for the aggregation gap.
\item There is no universal efficiency ranking, even on the exact-equality class $\Delta_{\mathrm{agg}}^e(\tau)=0$. In independent repeated-cross-section designs with fixed cohort shares and smooth Gaussian cohort distributions, one can have either $\sigma_m^2<\sigma_a^2$ or $\sigma_a^2<\sigma_m^2$ while the two population QTT aggregands coincide exactly. In particular, at the median, common cohort medians with unequal cohort scales make mixture inversion strictly more efficient, whereas two equally weighted, well-separated location cohorts can make average-cohort inversion strictly more efficient. The Appendix gives closed-form two-cohort constructions.
\item Precision does not determine the economic target. Consider a triangular sequence and suppose the squared scaled estimation errors are uniformly integrable so that the asymptotic second moments equal the variances above. If the prespecified target is $\theta_{a,L}=\QTT_{\mathrm{avg},L}^e(\tau)$ and $\Delta_L=\theta_{a,L}-\theta_{m,L}$, then for a fixed nonzero target mismatch $\Delta_L\to\Delta\ne0$,
\[
\E(\widehat\theta_{m,L}-\theta_{a,L})^2\to\Delta^2,
\qquad
\E(\widehat\theta_{a,L}-\theta_{a,L})^2=O(L^{-1}).
\]
Under local mismatch $\sqrt L\,\Delta_L\to\delta$,
\begin{equation}
L\,\E(\widehat\theta_{a,L}-\theta_{a,L})^2\to\sigma_a^2,
\qquad
L\,\E(\widehat\theta_{m,L}-\theta_{a,L})^2\to\sigma_m^2+\delta^2.
\label{eq:localtargetrisk}
\end{equation}
Hence a local variance advantage of the mixture estimator compensates for using it to report an average-cohort target only when $\sigma_a^2-\sigma_m^2>\delta^2$. The symmetric statement holds when the mixture QTT is the prespecified target.
\end{enumerate}
\end{proposition}

Proposition~\ref{prop:efficiencyrisk} separates two decisions that should not be conflated. The policy or scientific question selects the estimand; precision describes the sampling cost of estimating that selected object. Equation~\eqref{eq:varianceordering} is directly estimable from the same cluster influence vectors already needed for Proposition~\ref{prop:aggregationinference}, while \eqref{eq:localtargetrisk} makes the cost of target substitution explicit. A shorter confidence interval for the other aggregand is not, by itself, a reason to change the target.

\begin{corollary}[Mass-point-safe post-gate QTT and aggregation bands by joint CDF--weight projection]
\label{cor:masspointbands}
Fix route $r$ and $\mathcal T\subset(0,1)$. Suppose simultaneous CDF bands $[L_{d,L}(\cdot;\delta),U_{d,L}(\cdot;\delta)]$, $d\in\{0,1\}$, have nondecreasing proper-CDF endpoints (after shape projection if needed) and jointly cover the identified post-treatment CDFs with asymptotic probability at least $1-\delta$. For any CDF $H$, write $Q_H(\tau)=\inf\{y:H(y)\ge\tau\}$. Then
\[
I_L^{\QTT}(\tau;\delta)=
[Q_{U_{1,L}}(\tau)-Q_{L_{0,L}}(\tau),\;
 Q_{L_{1,L}}(\tau)-Q_{U_{0,L}}(\tau)]
\]
has simultaneous unconditional noncoverage at most $\delta$ without continuity or density assumptions. If the region is reported only after a prespecified diagnostic gate, the Appendix's branch-bound calibration transfers the same joint CDF coverage to the selected reporting branch.

For staggered adoption, fix a finite cohort set $\mathcal G_e$ and let $\omega^e=(\omega_g^e)_{g\in\mathcal G_e}$ lie in the probability simplex. Suppose a random nonempty compact set $\mathcal W_L^e(\delta)$ and simultaneous cohort--state bands with proper-CDF endpoints satisfy
\[
\liminf_{L\to\infty}\Pr\!\left\{\omega^e\in\mathcal W_L^e(\delta),\;
L_{g,d,L}(y;\delta)\le F_{g,g+e}^{d}(y)\le U_{g,d,L}(y;\delta)\ \forall(g,d,y)\right\}\ge1-\delta.
\]
Then projection of this same joint region through weighted cohort-QTT averaging and through CDF mixing followed by generalized inversion yields simultaneous bands for $\QTT_{\mathrm{avg}}^e$, $\QTT_{\mathrm{mix}}^e$, and $\Delta_{\mathrm{agg}}^e$, with unconditional and post-gate noncoverage bounded as above. This result permits estimated cohort weights; fixed prespecified weights are the singleton case $\mathcal W_L^e=\{\omega^e\}$. Whenever the originating joint region has finite-sample coverage at least $1-\delta$, the same projection statements hold in finite samples. Exact projection endpoints and a Bonferroni construction from separate CDF and weight regions are given in the Appendix.
\end{corollary}

The inverse-CDF step follows the generic band transformation of \citet{chernozhukov2020discrete}. The paper-specific extension transfers coverage through the route-specific proceed event and jointly projects uncertainty in cohort distributions and cohort weights through the two noncommuting aggregation operations and their discrepancy. In plain terms, the smooth process result below is used only when the estimated CDFs are locally smooth over the relevant quantile range and their densities stay safely away from zero; otherwise the CDF-projection route is the primary inferential object rather than a density-based quantile linearization.

\begin{corollary}[Uniform inference for the aggregation-gap process]
\label{cor:aggregationprocess}
Fix $e$, a finite cohort set $\mathcal G_e$, and a compact quantile interval $\mathcal T=[\underline\tau,\overline\tau]\subset(0,1)$. Strengthen Proposition~\ref{prop:aggregationinference} as follows: the joint CDF-score representations hold uniformly on fixed outcome intervals containing $\{q_{g,d}(\tau),q_d^{\mathrm{mix}}(\tau):\tau\in\mathcal T\}$; the stacked normalized CDF-score processes and weight scores converge weakly to a tight mean-zero Gaussian element in the corresponding product sup-norm space whose CDF-score sample paths lie almost surely in the product space $\prod_{g,d}C(I_{g,d})$, where each compact interval $I_{g,d}$ contains the full cohort- and mixture-quantile ranges used for $\tau\in\mathcal T$; this is the tangential subspace on which the inverse-CDF maps are Hadamard differentiable in the sup norm; the cohort and mixture densities are uniformly continuous and bounded away from zero on those quantile ranges; and the cluster multiplier process consistently estimates the same Gaussian law conditionally on the data. Then
\[
\sqrt L\{\widehat\Delta_{\mathrm{agg}}^e-\Delta_{\mathrm{agg}}^e\}
\Rightarrow \mathbb G_{\Delta}^e
\quad\text{in }\ell^\infty(\mathcal T),
\]
where $\mathbb G_{\Delta}^e(\tau)$ is the Gaussian limit of $L^{-1/2}\sum_{\ell=1}^L\Gamma_{\ell}^{\mathrm{gap}}(\tau)$ with the pointwise influence function in Proposition~\ref{prop:aggregationinference}. Moreover, if $\widehat\Gamma_{\ell}^{\mathrm{gap}}(\tau)$ is uniformly empirically-$L_2$ consistent and the multiplier moment condition in Proposition~\ref{prop:aggregationinference} holds uniformly, then conditionally on the data
\[
L^{-1/2}\sum_{\ell=1}^L\zeta_\ell\widehat\Gamma_{\ell}^{\mathrm{gap}}(\cdot)
\Rightarrow_p\mathbb G_{\Delta}^e
\quad\text{in }\ell^\infty(\mathcal T).
\]
Therefore multiplier critical values for $\sup_{\tau\in\mathcal T}|\mathbb G_{\Delta}^e(\tau)/s_\Delta(\tau)|$ yield asymptotically valid simultaneous bands, provided $s_\Delta(\tau)$ is continuous and bounded away from zero on $\mathcal T$, $\sup_{\tau\in\mathcal T}|\widehat s_\Delta(\tau)-s_\Delta(\tau)|\to_p0$, and the CDF of the limiting supremum is continuous at its $(1-\gamma)$ quantile. The proof is an application of the functional and bootstrap delta methods for inverse-CDF and finite-mixture maps \citep{vanderVaartWellner1996}. This is a smooth-quantile process theorem; Corollary~\ref{cor:masspointbands} remains the appropriate route at atoms.
\end{corollary}

\paragraph{Operational smooth-versus-projection routing.}
The smooth and projection results are not meant to be selected by whichever gives a shorter interval. A conservative rule can be frozen before target-effect inspection. Let $b_L\downarrow0$ with $\sqrt L\,b_L\to\infty$ and flag an atom whenever the largest empirical point mass in any cohort--state distribution over the prespecified target-quantile neighborhoods exceeds $b_L$; known rounding or heaping on a measurement lattice is treated as an atom flag regardless of this numerical screen. Separately fix a density safety floor $m_*>0$ and a finite, prespecified bandwidth-multiplier set (for example $0.8,1,1.2$ times a baseline rule), and require every cohort and mixture density estimate over the relevant neighborhoods to exceed $m_*$ under every bandwidth. Use the smooth influence-function/process route only if both screens pass; otherwise use joint CDF--weight projection. Under an atomless law whose relevant densities are uniformly above $m_*+\varepsilon$, whose largest normalized observation weight is $o_p(b_L)$, and whose density estimates are uniformly consistent, this rule selects the smooth route with probability tending to one. Under any fixed atom of positive mass, the atom screen selects projection with probability tending to one. The boundary cases---vanishing atoms, densities close to zero or to $m_*$, and bandwidth-instability---are deliberately routed to projection; no uniform smooth-quantile claim is made over such local-to-nonregular sequences. The Appendix records the short selection-consistency argument.

\paragraph{Reporting after a prespecified diagnostic gate.}
If the aggregation gap is reported only on a frozen PROCEED branch, its unconditional smooth or projection-based confidence region can be carried into that branch using the branch-bound or fixed-dimensional Gaussian selective calibration developed in the Appendix. These corrections address only the prespecified diagnostic gate; ex-post choice of route, event time, quantile, or aggregation definition is a separate selection problem.

Proposition~\ref{prop:aggregationinference} is the pointwise fixed-cohort, smooth-quantile, many-cluster result; Corollary~\ref{cor:aggregationprocess} adds process-level inference, and Corollary~\ref{cor:masspointbands} supplies the complementary projection route at mass points with estimated cohort weights. Section~\ref{sec:mpdta} then turns to the public Callaway--Sant'Anna minimum-wage panel. The Appendix contains the expanded first-stage diagnostic checklist, simulations, and supporting validation exercises.

\section{Nonlinear Aggregation in a Staggered Minimum-Wage Panel}\label{sec:mpdta}

The central aggregation result is an estimand statement, so the empirical question is whether changing only aggregation order can visibly change a reported distributional effect in an actual staggered panel. We use the public \texttt{mpdta} panel distributed with the \texttt{did} package \citep{callawaysantanna2021,callawaysantanna2026did}: 500 counties observed from 2003--2007, with \texttt{lemp}---log county-level teen employment---as the outcome and state minimum-wage increases generating staggered treatment cohorts. The target distribution is the cross-county distribution of this county-level outcome, not an individual-level distribution of teenagers. Section~\ref{sec:mpdta} first holds the distributional first stage fixed and changes only the aggregation operator; the separate CDF-PT robustness exercise was added post-development and its placebo calculations are treated only as within-dataset compatibility diagnostics.

\subsection{Holding identification fixed: a same-object operator audit}

The practical relevance of aggregation order should not depend on changing the identifying model at the same time. Version 2.0.0 of \texttt{qte} documents \texttt{ddid(..., gt\_type="qtt")} as mixture-CDF aggregation under the Callaway--Li--Oka dependence-based distributional DiD construction \citep{callawaylioka2018,callaway2026qte}. We independently reconstruct the underlying \texttt{ddid} group--time counterfactual distributions from the public \texttt{mpdta} panel. As a validation, the reconstruction reproduces the documented overall and event-time-zero ATT values to four decimals; details are in the Appendix.

We then freeze the two placebo-supported event-time-zero cohorts, their recovered pairs $(\widehat F_{g,g}^{1},\widehat F_{g,g}^{0})$, and the supported-sample weights $(0.234,0.766)$. Quantiles are obtained by exact generalized inversion of explicitly label-aligned finite mixtures. Only the final aggregation operator changes. At $\tau=.10$, the average-cohort QTT is $-0.050$ while the mixture QTT is $+0.068$, giving an average-minus-mixture gap of $-0.118$. At $\tau=.60$ the estimates are $-0.074$ and $+0.005$, while at $\tau=.70$ they are $+0.012$ and $-0.009$. The median estimates are much closer, $-0.096$ and $-0.088$. The operator difference is therefore quantile specific rather than a uniform location shift.

\begin{figure}[h!]
\centering
\includegraphics[width=.76\textwidth]{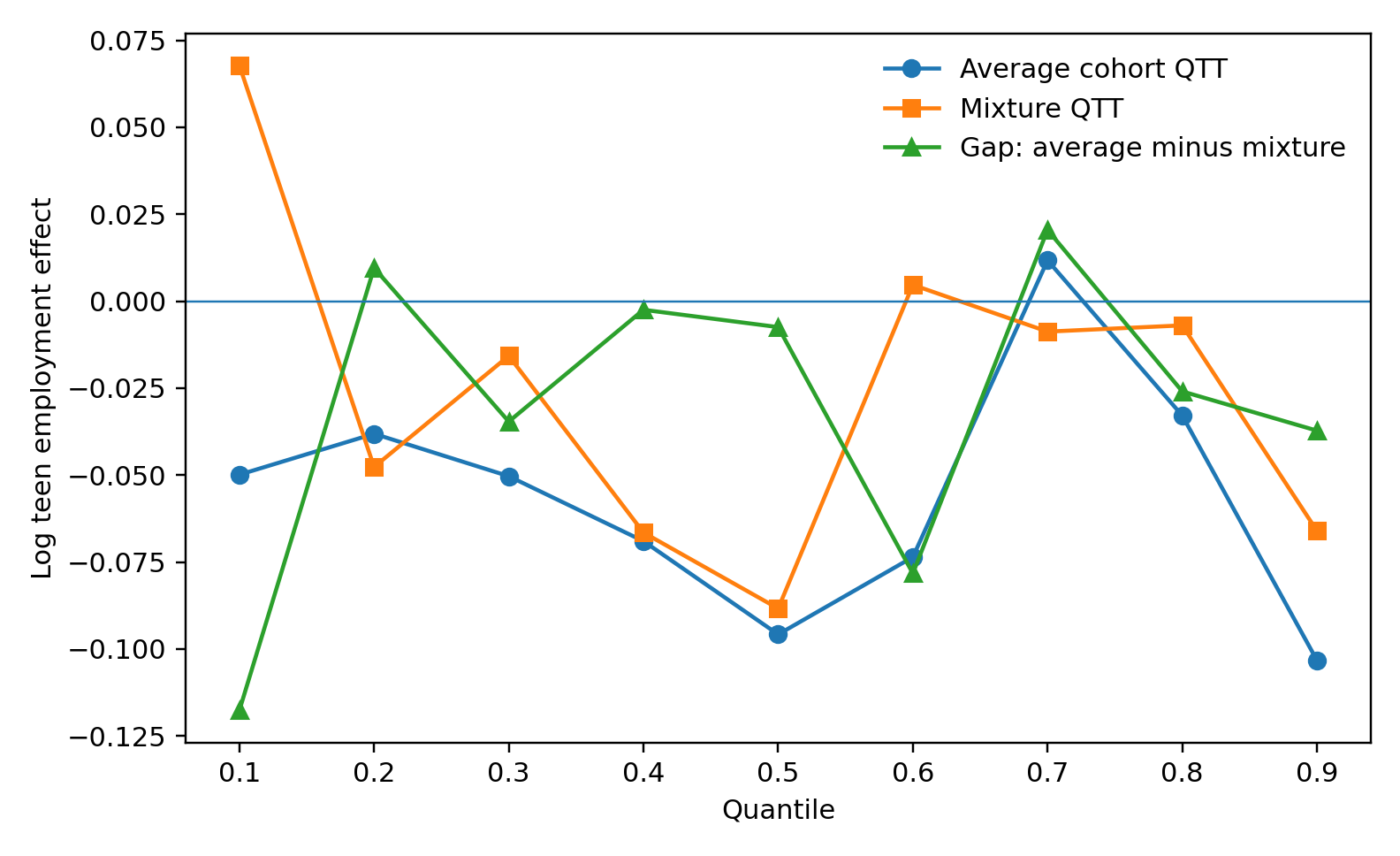}
\caption{Holding identification fixed and changing only aggregation order. The average-QTT and mixture-QTT curves use the same independently reconstructed event-time-zero potential-outcome distributions for the 2006 and 2007 cohorts and the same weights $(0.234,0.766)$. At $\tau=.10$, average-QTT $=-0.050$ and mixture-QTT $=+0.068$ (gap $=-0.118$). ``Gap'' is average minus mixture.}
\label{fig:qtesameobject}
\end{figure}

The lower-tail sign reversal is also the most stable of the displayed reversals. A state-delete-one sensitivity removes each of the 29 states in turn and recomputes the entire two-cohort first stage, supported-sample weights, and both operators. The $\tau=.10$ QTTs retain opposite signs in 28 of 29 deletions; the corresponding counts are 8 of 29 at $\tau=.60$ and 18 of 29 at $\tau=.70$. This is a sensitivity exercise, not formal few-cluster inference, and we do not describe sign disagreement as uniform across quantiles or leave-one-state samples.

The exercise is useful because it isolates the estimand issue: first-stage identification, data, recovered cohort distributions, cohort set, and weights are held fixed. The Appendix reports all nine quantiles, an all-post-period version, and the source-order audit motivating explicit $(g,t)$ alignment of every distribution and weight.

\subsection{Route-specific robustness and precision}

As a separate robustness exercise, we recover event-time-zero counterfactual CDFs under unconditional CDF-PT for the placebo-supported 2006 and 2007 cohorts, using the same supported-sample weights $(0.234,0.766)$. This changes the first-stage identifying route relative to the software audit and is therefore not used to attribute differences across the two exercises to aggregation. Within this CDF-PT construction, however, the two aggregation maps are again applied to exactly the same four cohort-state CDFs. The median average-cohort QTT is $-0.101$ and the mixture QTT is $-0.016$, a gap of $-0.085$ log points; at $\tau=.10$ the gap is $-0.113$, and at $\tau=.70$ and $.80$ the two point estimates have opposite signs. Never-treated controls produce similar point gaps. The associated within-dataset placebo calculations are far from rejection, but the dataset and specification were selected after theory development, so those diagnostics are compatibility checks rather than a prospectively size-controlled selection rule.

This robustness exercise is deliberately not advertised as a precise causal separation of the two aggregands. Policy timing is state-level, the panel contains only 29 states, and the 2006 supported cohort occupies only three treated states. No displayed CDF-PT gap is statistically distinguishable from zero under the reported state-level sensitivity calculations; at the median, deleting one state at a time moves the recomputed gap from $-0.127$ to $0.065$. The Appendix contains the complete threshold family, control-pool checks, state-jackknife curve, mass-point-safe simultaneous CDF projection, and a variance-equivalent cluster-count calibration. Their role is to document the empirical limitation rather than disguise it. The stronger empirical message in this section is therefore the same-object operator audit above: its sign reversal is a literal consequence of changing aggregation order while holding the recovered distributional objects fixed.

\FloatBarrier
\paragraph{Supporting validation material.}
The Appendix contains the material needed to audit the first stage without competing with the main aggregation argument: a prospectively specified Monte Carlo suite, the complete STOP/PROCEED and selective-inference theory, a frozen CPS minimum-wage design that stops because no prespecified cross-route arbitration rule licenses selecting an ex-post survivor, and a separately frozen NSW exercise that proceeds to a randomized benchmark. It also reports all mass-point, bootstrap-draw, comparison-support, and software-reconstruction details. These exercises test the distributional objects or the inferential machinery; none is used to redefine the aggregation target after seeing the data.

\section{Conclusion}
\label{sec:conclusion}

Staggered distributional DiD contains an economic choice that mean-effect aggregation does not: whether to average cohort-specific quantile effects under prespecified policy weights or to pool cohort potential-outcome distributions and then invert. These operations answer different questions. Their difference is not a numerical nuisance but a target-population discrepancy that can be large and can reverse sign.

The paper gives that discrepancy a sharp theory. The mixture quantile lies within the cohort-quantile range; conditional on the cohort quantiles and policy weights, the average-minus-mixture gap lies in a sharp one-sided interval, while a coarser weight--range envelope remains globally sharp when only ranges are retained; its leading local component is governed by density tilting; the tilt--dispersion envelope is sharp to first order; and exact equality of the two QTT aggregands does not imply equal precision. The joint influence representation yields pointwise and process inference, while joint CDF--weight projection remains valid when atoms or weak densities make smooth quantile inference unreliable. Precision and target choice are therefore separate decisions: a shorter interval for one aggregand is not a reason to report it when the policy question names the other.

The same-object minimum-wage reconstruction makes the distinction concrete. Holding the group--time potential-outcome distributions, supported cohort set, and weights fixed, changing only the aggregation operator reverses the sign at several quantiles; the lower-tail reversal survives 28 of 29 state deletions. A separate CDF-PT construction yields the same qualitative noncommutativity, although state-level precision is limited. The broader lesson is portable: recover the counterfactual distribution first, name the target population and aggregation order before inversion, report the aggregation gap when both summaries are meaningful, and use CDF projection rather than smooth quantile linearization when the data are nonregular.

The Appendix retains the supporting STOP/PROCEED theory, Monte Carlo evidence, CPS non-arbitration example, NSW randomized-benchmark validation, and the complete software audit. Keeping that material out of the main line is deliberate: the paper's primary contribution is the theory and empirical consequence of nonlinear aggregation in staggered distributional DiD.

\clearpage
\appendix
\section*{Supplementary Appendix}
\addcontentsline{toc}{section}{Supplementary Appendix}
\noindent This supplementary appendix proves the paper's nonlinear staggered-aggregation results---including exact equality conditions, density-tilt and compact-uniform bounds, pointwise and process-level influence-function inference, and mass-point-safe joint CDF--weight projection. It also houses the supporting route-specific diagnostic-gate, strong familywise-error-rate (FWER) stepdown, local-power, and post-gate conditional-inference results that are deliberately compressed in the main text. It records the CDF-PT and CIC influence functions, gives primitive scope conditions for the growing-family Gaussian-multiplier construction, and proves when fixed-dimensional Gaussian selective inversion is a connected interval. The appendix also documents the prospectively specified Monte Carlo suite, the 99-coordinate stress experiments, the frozen Current Population Survey (CPS) minimum-wage design, and the frozen National Supported Work (NSW) Demonstration validation. Additional material covers support, repeated-cross-section standardization, comparison-set construction, mass points, and implementation details.

\medskip

\paragraph{Notation and cross-reference convention.}
Proposition, Corollary, Remark, equation, and assumption numbers refer to the main text unless explicitly stated otherwise. Because the generic diagnostic results have been moved out of that main text, they are labeled Diagnostic Results D1--D4 and Diagnostic Specializations D5--D6 below. An identifying route is a maintained restriction used to recover the missing untreated distribution, and a placebo coordinate is a prespecified pre-treatment contrast that should be zero under that route. In the diagnostic proofs, \(r\) indexes a prespecified identifying route, \(L\) is the number of independent assignment/sampling clusters, \(\mathcal F\) is the fixed placebo family, \(t_L^r=(t_j^r)_{j\in\mathcal F}\) is the studentized placebo vector, and \(T_L^r=\|t_L^r\|_\infty\) is the max-\(|t|\) gate statistic. Conditions (S1)--(S3) denote, respectively, the joint Gaussian central limit theorem, standard-error consistency, and centered bootstrap validity stated in Diagnostic Result D1. We use PT for parallel trends, CLT for central limit theorem, DGP for data-generating process, LAN for local asymptotic normality, and ATT for average treatment effect on the treated. References cited by author and year in this appendix are listed in the bibliography of this working paper.

\section{Supporting diagnostic results moved from the main text}
\label{app:diagnostic-results}

\paragraph{Diagnostic Result D1 (finite-family global gate).}
Fix route \(r\) and a finite placebo family \(\mathcal F\). Suppose (S1) \(\sqrt L(\widehat\theta^r-\theta^r)\Rightarrow Z^r\) with a mean-zero Gaussian limit and positive marginal variances; (S2) every studentizing standard error is consistent and bounded away from zero in probability; and (S3) the centered studentized cluster bootstrap consistently estimates the joint law of the studentized Gaussian limit. Then the max-\(|t|\) bootstrap gate has asymptotic rejection probability \(\alpha\) under the complete placebo null and rejection probability tending to one under any fixed alternative with at least one nonzero coordinate. For finite \(J\), no separate anti-concentration assumption is required: the limiting maximum is atomless even if its Gaussian covariance matrix is singular.

\paragraph{Diagnostic Result D2 (Romano--Wolf localization).}
Let \(\mathcal H_0\subseteq\mathcal F\) be the true placebo nulls. If the centered studentized bootstrap consistently estimates the maximum over every nonempty subset of \(\mathcal H_0\) at the relevant critical values, uniformly over the intersection-null configurations considered, route-specific Romano--Wolf stepdown has asymptotic strong FWER at most \(\alpha\). Under the complete null its first step is the D1 global gate.

\paragraph{Diagnostic Result D3 (local power).}
Under (S1)--(S3), consider \(\theta_L^r=h^r/\sqrt L\) under laws mutually contiguous with the null and satisfying \(\sqrt L(\widehat\theta^r-\theta_L^r)\Rightarrow Z^r\) with the same centered Gaussian limit. Writing \(D_r=\operatorname{diag}(s_1^r,\ldots,s_J^r)\), \(W^r=D_r^{-1}Z^r\), and \(c_{1-\alpha}(R_r)\) for the null max-\(|t|\) critical value,
\[
\Pr(T_L^r>\widehat c_{1-\alpha}^r)\to
\Pr\!\left\{\max_j\left|W_j^r+h_j^r/s_j^r\right|>c_{1-\alpha}(R_r)\right\}.
\]
If \(R_r\) is positive definite and \(h^r\ne0\), the limit exceeds \(\alpha\); if the largest standardized drift diverges, power tends to one.

\paragraph{Diagnostic Result D4 (post-gate inference).}
Let \(G_L^r\) denote the proceed event for a route that identifies scalar target \(\beta_0^r\). (i) If \(p_r^{\mathrm{pass}}=\liminf_L\Pr(G_L^r)>0\) and an unconditional confidence set has asymptotic noncoverage at most \(\delta\), then conditional noncoverage given \(G_L^r\) is at most \(\delta/p_r^{\mathrm{pass}}\). The same conclusion applies to the random-level calibration described in the proof below under nesting and local uniform validity. (ii) Under the complete placebo null, if the scalar target statistic and finite diagnostic vector converge jointly to a centered nonsingular Gaussian law and the covariance matrix and gate critical value are consistently estimated, the conditional Gaussian pivot \(H_r\) defined below is asymptotically Uniform\((0,1)\) given \(G_L^r\). Inverting it yields nominal conditional coverage. Under the same centered symmetric Gaussian law, conventional two-sided symmetric Gaussian intervals are weakly conservative after passing the gate.

\paragraph{Diagnostic Specialization D5 (CDF-PT placebo QTTs).}
For a fixed finite collection of CDF-PT placebo QTTs, a joint cluster-level empirical-CDF asymptotic-linear representation with a Lindeberg moment condition, local stochastic equicontinuity at the finitely many target arguments, stable comparison sets and weights, proper and locally differentiable CDF-PT counterfactuals, positive target densities, nondegenerate standard errors, and bootstrap consistency for the same local CDF-score process imply (S1)--(S3), and hence D1.

\paragraph{Diagnostic Specialization D6 (CIC placebo QTTs).}
For a fixed finite collection of CIC placebo QTTs, the analogous cluster-level CDF-score CLT/bootstrap conditions, stable comparison sets and weights, interior support for every probability index used by the transport map, and positive locally continuous densities at every transport and quantile argument imply Hadamard differentiability of the CIC and inverse-CDF maps, hence (S1)--(S3) and D1.

These diagnostic statements are supporting infrastructure rather than the main text's novelty. Their proofs and numerical calibrations are retained here so that the shortened main text does not weaken the scope qualifications attached to the empirical workflow.

\section{Proofs of the Formal Results}

\subsection{Familywise validity and consistency}

Under the global null, conditions (S1)--(S2) and Slutsky's theorem imply
\[
T_L^r\Rightarrow
M^r\equiv\max_{j\in\mathcal F}|Z_j^r/s_j^r|.
\]
Assumption (S3) implies that, conditionally on the data, the bootstrap statistic \(T_L^{r,*}\) converges weakly in probability to the same \(M^r\). The quantile step remains valid even if the Gaussian covariance matrix is singular. To see this explicitly, let \(W^r=(Z_j^r/s_j^r)_j\) and write \(W^r=AX\), where \(X\sim N(0,I_q)\), \(q=\operatorname{rank}\{\operatorname{Var}(W^r)\}\ge1\), and \(A\) has full column rank. Then \(x\mapsto\|Ax\|_\infty\) is a norm on \(\mathbb R^q\). For every \(c>0\), the sphere \(\{x:\|Ax\|_\infty=c\}\) is the boundary of the convex norm ball \(\{x:\|Ax\|_\infty\le c\}\) and has \(q\)-dimensional Lebesgue measure zero. Because \(X\) has an everywhere-positive density, \(\Pr\{\|AX\|_\infty=c\}=0\), so the law is atomless. Moreover, for every \(0\le a<b\), the annulus \(\{x:a<\|Ax\|_\infty<b\}\) is a nonempty open set and has strictly positive probability. Thus the CDF of \(M^r\) is continuous and strictly increasing on \((0,\infty)\); its \(1-\alpha\) quantile is unique. Conditional weak convergence therefore yields
\[
\widehat c_{1-\alpha}^r\overset{p}\longrightarrow c_{1-\alpha}^r.
\]
Therefore
\[
\Pr(T_L^r>\widehat c_{1-\alpha}^r)
\to
\Pr(M^r>c_{1-\alpha}^r)=\alpha.
\]
This is global-null familywise size control: under the complete placebo null, rejection is exactly the event that at least one coordinate of the frozen family crosses the common max-\(t\) critical value. Strong FWER localization under partial null configurations is supplied separately by the stepdown proposition.

Under a fixed alternative with \(\theta_{j_0}^r\neq0\),
\[
\left|\frac{\sqrt L\,\widehat\theta_{j_0}^r}{\widehat s_{j_0}^r}\right|
=
\frac{\sqrt L|\theta_{j_0}^r|}{s_{j_0}^r}+O_p(1)\to\infty,
\]
whereas \(\widehat c_{1-\alpha}^r=O_p(1)\). Hence rejection probability converges to one. \(\square\)

\subsection{Strong-FWER stepdown localization}

\paragraph{Proof of the stepdown proposition in the main text.}
Let \(\mathcal H_0\subseteq\mathcal F\) be the set of true coordinate nulls. At the first step at which a true null could be rejected, every previously removed hypothesis is false, so the active set \(I\) still contains \(\mathcal H_0\). For every bootstrap draw,
\[
\max_{j\in I}|t_j^{*}|
\ge
\max_{j\in\mathcal H_0}|t_j^{*}|,
\]
hence the active-set bootstrap critical value is no smaller than the critical value for the true-null subset. By the assumed subsetwise bootstrap validity, the latter controls
\(\max_{j\in\mathcal H_0}|t_j|\) at level \(\alpha+o(1)\). Thus the probability of the first false rejection is at most \(\alpha+o(1)\). The monotone adjusted-\(p\) construction preserves this bound at later steps. This is the standard Romano--Wolf argument applied to the route-specific studentized placebo vector. \(\square\)

\subsection{Post-gate conditional inference}

\paragraph{Proof of Diagnostic Result D4(i).}
Write \(E_L(\delta)=\{\beta_0^r\notin C_L^r(\delta)\}\). For every \(L\) with \(\Pr(G_L^r)>0\),
\[
\Pr\{E_L(\delta)\mid G_L^r\}
=
\frac{\Pr\{E_L(\delta)\cap G_L^r\}}{\Pr(G_L^r)}
\le
\frac{\Pr\{E_L(\delta)\}}{\Pr(G_L^r)}.
\]
Taking upper limits and writing
\[
p_r^{\mathrm{pass}}\equiv\liminf_{L\to\infty}\Pr(G_L^r)>0,
\]
with \(\limsup\Pr\{E_L(\delta)\}\le\delta\), gives
\[
\limsup_{L\to\infty}\Pr\{E_L(\delta)\mid G_L^r\}
\le \delta/p_r^{\mathrm{pass}}.
\]
If \(0<\underline p_r\le p_r^{\mathrm{pass}}\), setting \(\delta=\gamma\underline p_r\) gives the first calibrated statement.

For the data-dependent version, let \(\widehat p_{r,L}\to_p p_r^\star\in(0,p_r^{\mathrm{pass}}]\), put \(\delta_r^\star=\gamma p_r^\star\), and assume the confidence-set family is nested in \(\delta\) and asymptotically level-valid in a neighborhood of \(\delta_r^\star\). Fix \(\varepsilon>0\) small enough that \(\delta_r^\star+\varepsilon\) remains in that neighborhood. Since \(\delta_L=\gamma\widehat p_{r,L}\to_p\delta_r^\star\),
\[
\Pr\{\delta_L>\delta_r^\star+\varepsilon\}=o(1).
\]
On the complementary event, nesting implies
\[
\{\beta_0^r\notin C_L^r(\delta_L)\}
\subseteq
\{\beta_0^r\notin C_L^r(\delta_r^\star+\varepsilon)\}.
\]
Therefore
\[
\limsup_{L\to\infty}
\Pr\{\beta_0^r\notin C_L^r(\delta_L)\}
\le \delta_r^\star+\varepsilon.
\]
Letting \(\varepsilon\downarrow0\) gives unconditional noncoverage at most \(\gamma p_r^\star\), and the branch bound then gives conditional noncoverage at most \(\gamma p_r^\star/p_r^{\mathrm{pass}}\le\gamma\). The nesting condition is what prevents the random level from selecting data realizations on which a merely pointwise-valid confidence procedure fails.

Under the complete placebo null, Diagnostic Result D1 gives \(\Pr(G_L^r)\to1-\alpha\), so \(p_r^{\mathrm{pass}}=1-\alpha\) and \(\delta=\gamma(1-\alpha)\) is valid. Off the null, the proceed probability can be smaller; therefore \(1-\alpha\) is not a generally valid lower calibration. A centered null-bootstrap proceed frequency should not automatically be interpreted as a consistent estimator of the off-null proceed probability: consistency or asymptotic conservativeness of any \(\widehat p_{r,L}\) used in the adaptive rule must be established for the calibration scheme at hand. The same argument applies to an unconditional simultaneous confidence band. \(\square\)

\paragraph{Proof of Diagnostic Result D4(ii).}
Work under the complete placebo null stated in part (ii) and suppress route notation. Let
\[
G_L=\{\|t_L\|_\infty\le\widehat c\},
\qquad
G=\{\|T\|_\infty\le c\}.
\]
Positive definiteness of \(\Omega\) implies that \(T\) has a continuous density, hence
\(\Pr(\|T\|_\infty=c)=0\). Joint weak convergence and
\(\widehat c\overset p\to c\) therefore imply
\[
\Pr(G_L\triangle\{\|t_L\|_\infty\le c\})\to0,
\qquad
\Pr(G_L)\to\Pr(G)=p^\circ>0.
\]
One way to see the first statement is to bound the symmetric difference, for any
\(\varepsilon>0\), by
\[
\Pr(|\widehat c-c|>\varepsilon)
+
\Pr\!\left(
\big|\|t_L\|_\infty-c\big|\le\varepsilon
\right),
\]
and then let \(L\to\infty\) followed by \(\varepsilon\downarrow0\).

Define
\[
H(u;\Omega,c)
=
\Pr_\Omega(U\le u\mid \|T\|_\infty\le c).
\]
Because \(\Omega\) is positive definite and \(p^\circ>0\), this conditional distribution is continuous. Continuity of multivariate Gaussian rectangle probabilities in \((u,\Omega,c)\), together with
\(\widehat\Omega\overset p\to\Omega\) and
\(\widehat c\overset p\to c\), gives the plug-in convergence needed below. For \(x\in[0,1]\),
\[
\Pr\!\left[
H\{U_L;\widehat\Omega,\widehat c\}\le x,\ G_L
\right]
\to
\Pr\!\left[
H(U;\Omega,c)\le x,\ G
\right].
\]
By the probability integral transform applied to the conditional law of \(U\mid G\),
the right-hand side equals \(x\,p^\circ\). Dividing by
\(\Pr(G_L)\to p^\circ\) yields
\[
\Pr\!\left[
H\{U_L;\widehat\Omega,\widehat c\}\le x
\mid G_L
\right]\to x.
\]
This proves conditional asymptotic uniformity and hence coverage of the inverted confidence set. Connected-interval shape additionally requires monotonicity of the fully recomputed candidate-parameter pivot. In particular, \(\widehat s_\beta(\beta)\) may depend on the hypothesized value, so the interval claim requires that this dependence, the covariance estimate, and the selection correction be first-order invariant over the inversion neighborhood or otherwise preserve monotonicity.
For a fixed number of targets, the union bound applied conditionally on the same
gate gives the Bonferroni statement. \(\square\)

\paragraph{Proof of the connected-selective-interval remark.}
Fix a positive-definite Gaussian covariance matrix $\Omega$ and a finite gate critical value $c$ with $\Pr_\Omega(\|T\|_\infty\le c)>0$. Because $\Omega$ is positive definite, the conditional law $T\mid U=u$ is Gaussian with a positive-definite Schur-complement covariance matrix for every finite $u$. Hence the conditional density of $T\mid U=u$ is everywhere positive on $\mathbb R^J$, and therefore
\[
0<\Pr_\Omega(\|T\|_\infty\le c\mid U=u)<1
\qquad\text{for every }u\in\mathbb R.
\]
The selected scalar density is consequently
\[
f_{U\mid G}(u)
=
\frac{\phi(u)\Pr_\Omega(\|T\|_\infty\le c\mid U=u)}
{\Pr_\Omega(\|T\|_\infty\le c)},
\]
where $\phi$ is the standard-normal density. It is strictly positive for every finite $u$, so
$u\mapsto H(u;\Omega,c)$ is continuous and strictly increasing from zero to one.

Now suppose that over the candidate parameter region the estimated standard error is constant, $\widehat s_\beta(\beta)=\widehat s_\beta>0$, and that $\widehat\Omega$ and $\widehat c$ are held fixed across candidate values. Then
\[
U_L(\beta)=\frac{\sqrt L(\widehat\beta-\beta)}{\widehat s_\beta}
\]
is strictly decreasing in $\beta$, while $H\{U_L(\beta);\widehat\Omega,\widehat c\}$ is strictly decreasing as a composition of a strictly increasing function with $U_L(\beta)$. The equal-tail acceptance condition
\[
\gamma/2\le H\{U_L(\beta);\widehat\Omega,\widehat c\}\le1-\gamma/2
\]
therefore has a connected preimage (possibly empty or unbounded only if the candidate parameter space itself is truncated). On the unrestricted real line it is a nonempty finite closed interval because the pivot tends to one and zero at the two parameter extremes. More generally, any candidate-wise recomputation that preserves monotonicity of the full pivot gives the same connectedness conclusion. If standard errors, covariance estimates, or the selection correction vary with $\beta$ in a way that destroys monotonicity, Diagnostic Result D4(ii) still guarantees a confidence set after inversion but does not imply that the set is connected. $\square$

\subsection{Appendix-only growing-family Gaussian-multiplier template}
\label{sec:hdselectivetemplate}

The following result is retained as a sufficient-condition template rather than as a numbered proposition in the main text. It is not invoked for the CPS application.

\paragraph{Growing-family selective template.}

Let $J=J_L$ and $p_L=J_L+1$, possibly with $J_L>L$. Under the complete placebo null, suppose the joint standardized target--diagnostic vector $Y_L^r=(U_L^r,t_L^{r\prime})'$ and a Gaussian-multiplier analogue
\[
Y_L^{r,*}=L^{-1/2}\sum_{\ell=1}^L\zeta_\ell\widehat X_{\ell,L}^r
=(U_L^{r,*},T_L^{r,*\prime})',\qquad \zeta_\ell\stackrel{iid}{\sim}N(0,1),
\]
admit approximation by the same centered Gaussian proxy $Z_L^r$ with unit marginal variances; its covariance may be singular. Writing $\mathcal A_{p_L}^{\rm re}$ for axis-aligned hyperrectangles, assume
\[
\sup_{A\in\mathcal A_{p_L}^{\rm re}}|\Pr(Y_L^r\in A)-\Pr(Z_L^r\in A)|\to0,
\qquad
\sup_{A\in\mathcal A_{p_L}^{\rm re}}|\Pr^*(Y_L^{r,*}\in A)-\Pr(Z_L^r\in A)|\to_p0.
\]
Let $c_L$ be the $(1-\alpha)$ quantile of $\|Z_{T,L}^r\|_\infty$, $a_L=1+\mathbb E\|Z_{T,L}^r\|_\infty$, and let $\widehat c_{1-\alpha,L}^{r,*}$ be the multiplier quantile. If
\[
a_L|\widehat c_{1-\alpha,L}^{r,*}-c_L|\to_p0,
\]
define
\[
\widehat H_{r,L}^*(u)=
\frac{\Pr^*\{U_L^{r,*}\le u,\ \|T_L^{r,*}\|_\infty\le\widehat c_{1-\alpha,L}^{r,*}\}}
{\Pr^*\{\|T_L^{r,*}\|_\infty\le\widehat c_{1-\alpha,L}^{r,*}\}}.
\]
Then, on $G_{L,\mathrm{HD}}^r=\{\|t_L^r\|_\infty\le\widehat c_{1-\alpha,L}^{r,*}\}$,
\[
\widehat H_{r,L}^*\{U_L^r(\beta_0^r)\}\mid G_{L,\mathrm{HD}}^r
\Rightarrow\operatorname{Uniform}(0,1),\qquad
\Pr(G_{L,\mathrm{HD}}^r)\to1-\alpha.
\]
Thus fully recomputed multiplier-pivot inversion gives conditional coverage $1-\gamma$ without covariance inversion.

The statement is intentionally high level. The next paragraphs give a primitive score route under which its rectangle approximations and threshold stability follow. No claim is made that these primitive conditions hold for the empirical CPS cluster-score array.

\paragraph{Proof of the appendix growing-family selective template.}
Suppress route notation. Let
\[
M_L=\|t_L\|_\infty,
\qquad
M_L^Z=\|Z_{T,L}\|_\infty,
\qquad
M_L^*=\|T_L^*\|_\infty,
\]
and let \(c_L\) and \(\widehat c_L\) denote the oracle Gaussian and multiplier critical values in the template. Because all Gaussian marginal variances are one, \(M_L^Z\) can be written as the maximum of the \(2J_L\) signed Gaussian coordinates \((Z_{T,L}',-Z_{T,L}')'\). The Gaussian anti-concentration result in Theorem~3 and Comment~4 of Chernozhukov, Chetverikov, and Kato (2015), applied to the signed vector $(Z_{T,L}',-Z_{T,L}')'$, therefore gives the explicit equal-variance bound
\begin{equation}
\sup_x
\Pr\{|M_L^Z-x|\le \varepsilon\}
\le 4\varepsilon a_L,
\qquad
a_L=1+\mathbb{E}M_L^Z.
\label{eq:hdanticoncproof}
\end{equation}
No nonsingularity of \(\Omega_L\) is required for this bound.

First replace the random gate threshold by the oracle threshold. For every fixed \(\eta>0\), on the event
\(a_L|\widehat c_L-c_L|\le\eta\),
\[
\{M_L\le\widehat c_L\}\triangle\{M_L\le c_L\}
\subseteq
\{|M_L-c_L|\le\eta/a_L\}.
\]
The event on the right is the difference of two hyperrectangles. Hence the assumed Gaussian hyperrectangle approximation together with \eqref{eq:hdanticoncproof} implies
\[
\limsup_{L\to\infty}
\Pr\bigl(
\{M_L\le\widehat c_L\}\triangle\{M_L\le c_L\}
\bigr)
\le 4\eta.
\]
Critical-value stability lets \(L\to\infty\) and then \(\eta\downarrow0\), giving
\begin{equation}
\Pr\bigl(G_{L,\rm HD}\triangle G_L^0\bigr)\to0,
\qquad
G_L^0=\{M_L\le c_L\}.
\label{eq:hdgatesymdiff}
\end{equation}
Since the maximum of finitely many nondegenerate Gaussian coordinates has a continuous distribution for each \(L\), \(\Pr(M_L^Z\le c_L)=1-\alpha\). The rectangle approximation therefore also gives
\[
\Pr(G_L^0)\to1-\alpha,
\qquad
\Pr(G_{L,\rm HD})\to1-\alpha.
\]

Define the oracle conditional Gaussian CDF
\[
H_L^0(u)
=
\frac{\Pr\{Z_{U,L}\le u,\ M_L^Z\le c_L\}}{1-\alpha}.
\]
The numerator event is an axis-aligned hyperrectangle in \(\mathbb R^{1+J_L}\). Consequently the assumed Gaussian hyperrectangle approximation, uniformly in \(u\), gives
\begin{equation}
\sup_u
\left|
\Pr(U_L\le u\mid G_L^0)-H_L^0(u)
\right|
\to0.
\label{eq:hdactualcond}
\end{equation}
The CDF \(H_L^0\) is continuous: for every \(u\),
\(
\Pr\{Z_{U,L}=u,\ M_L^Z\le c_L\}
\le \Pr(Z_{U,L}=u)=0
\),
regardless of whether the joint Gaussian covariance is singular.

Next compare the multiplier selective CDF with the same oracle CDF. Conditional on the data, the assumed multiplier hyperrectangle approximation applies uniformly to the random rectangle
\(
\{U_L^*\le u,\ M_L^*\le\widehat c_L\}
\),
because the supremum is over all hyperrectangles. Thus the multiplier numerator differs uniformly in \(u\) from
\(
\Pr\{Z_{U,L}\le u,\ M_L^Z\le\widehat c_L\}
\)
by at most \(\rho_L^*\). On the event
\(a_L|\widehat c_L-c_L|\le\eta\), changing \(\widehat c_L\) to \(c_L\) changes any such Gaussian rectangle probability by at most the shell probability in \eqref{eq:hdanticoncproof}, hence by at most \(4\eta\). The same argument applies to the denominator. Moreover, by definition of the generalized multiplier quantile,
\(
\Pr^*(M_L^*\le\widehat c_L)\ge1-\alpha
\)
for every sample. Hence
\begin{equation}
\sup_u
|\widehat H_L^*(u)-H_L^0(u)|
\overset p\longrightarrow0.
\label{eq:hdbootcond}
\end{equation}

If \(V_L\) had exactly the conditional CDF \(H_L^0\), continuity would imply \(H_L^0(V_L)\sim\operatorname{Uniform}(0,1)\). Equation \eqref{eq:hdactualcond} therefore yields
\[
H_L^0(U_L)\mid G_L^0
\Rightarrow \operatorname{Uniform}(0,1).
\]
The symmetric-difference result \eqref{eq:hdgatesymdiff} transfers this conclusion from \(G_L^0\) to \(G_{L,\rm HD}\), because both proceed probabilities converge to \(1-\alpha>0\). Finally, \eqref{eq:hdbootcond} and the same positive-probability argument imply that replacing \(H_L^0\) by \(\widehat H_L^*\) changes the pivot by \(o_p(1)\), also conditionally on proceeding. This proves the conditional uniform limit and the inversion result.

For primitive conditions, the hyperrectangle central limit theorem and Gaussian multiplier bootstrap of Chernozhukov, Chetverikov, and Kato (2017) apply to independent cluster-score arrays under coordinate variance lower bounds and suitable moment/tail control even when \(p_L\gg L\). Under their exponential-tail condition, a familiar sufficient growth restriction is
\[
B_L^2\log^7(p_LL)=o(L),
\]
where \(B_L\) is the coordinate tail envelope. Estimated-score, asymptotic-linearization, and studentization errors must additionally be negligible uniformly in the coordinates on the Gaussian anti-concentration scale. The separate critical-stability assumption can be verified by adding a local quantile-separation condition. Writing \(F_L^Z(x)=\Pr(M_L^Z\le x)\), a concrete sufficient formulation is that, for every fixed \(\varepsilon>0\), there exists \(\eta_\varepsilon>0\) such that, for all sufficiently large \(L\),
\[
F_L^Z(c_L-\varepsilon/a_L)\le 1-\alpha-\eta_\varepsilon,
\qquad
F_L^Z(c_L+\varepsilon/a_L)\ge 1-\alpha+\eta_\varepsilon.
\]
If the conditional multiplier max-CDF converges uniformly to \(F_L^Z\), its generalized \(1-\alpha\) quantile must then lie in \([c_L-\varepsilon/a_L,c_L+\varepsilon/a_L]\) with probability approaching one. Since \(\varepsilon\) is arbitrary, this yields \(a_L|\widehat c_L-c_L|\to_p0\). This lower-separation requirement is logically distinct from Gaussian anti-concentration: anti-concentration controls the probability cost of a small threshold perturbation, whereas quantile separation ensures that bootstrap CDF error cannot move the critical value by more than the required \(1/a_L\) scale. The theorem does not claim that these asymptotic rate conditions are numerically sharp at \(L=51\). \(\square\)

\paragraph{A primitive score route for a growing fixed-threshold CDF-PT family.}
The high-dimensional assumption can be connected directly to the mass-point-robust CDF diagnostics used in the paper. Let
\(
X_{\ell,L}=(X_{\ell0,L},X_{\ell1,L},\ldots,X_{\ell J_L,L})'
\)
denote the oracle standardized cluster influence vector for one prespecified scalar post-treatment target (coordinate zero) and the \(J_L\) fixed-threshold CDF-PT placebo contrasts. Suppose clusters are independent, the coordinates are centered, and for constants \(b>0\) and \(B_L\ge1\), uniformly in \(j=0,\ldots,J_L\),
\begin{align*}
L^{-1}\sum_{\ell=1}^L\mathbb{E} X_{\ell j,L}^2 &\ge b,\\
L^{-1}\sum_{\ell=1}^L\mathbb{E}|X_{\ell j,L}|^{2+k} &\le B_L^k,\qquad k=1,2,\\
\mathbb{E}\exp\{|X_{\ell j,L}|/B_L\} &\le2,\qquad \ell=1,\ldots,L.
\end{align*}
These are the variance, moment, and exponential-tail conditions entering the hyperrectangle Gaussian approximation in Theorem~2.1 and Corollary~2.1 and the Gaussian-multiplier bootstrap in Theorem~4.1, Remark~4.1, and Corollary~4.1 of Chernozhukov, Chetverikov, and Kato (2017). For rectangles the simple-convex-set conditions in Corollary~4.1 hold with coordinate-sparse normals, while Remark~4.1 records the rectangle-specific covariance-comparison bound. If
\[
B_L^2\log^7(p_LL)=o(L),
\qquad p_L=J_L+1,
\]
then their results give the oracle Gaussian and multiplier rectangle approximations required in the appendix growing-family selective template. To transfer those approximations from oracle influence sums to the feasible studentized statistics, it is sufficient that the combined asymptotic-linearization, nuisance-estimation, and studentization remainder satisfy
\[
\sqrt{\log p_L}\,\|Y_L-S_L\|_\infty=o_p(1),
\qquad
\sqrt{\log p_L}\,\|Y_L^*-S_L^*\|_\infty=o_{p^*}(1)
\quad\text{in probability},
\]
where \(S_L\) and \(S_L^*\) are the corresponding oracle score sum and Gaussian-multiplier score sum. The Gaussian boundary bound for hyperrectangles is of order \(\delta\sqrt{\log p_L}\) for a coordinatewise boundary perturbation \(\delta\), so these rates transfer the oracle rectangle approximations to the feasible statistics. They are deliberately stronger than merely requiring an \(o_p(a_L^{-1})\) perturbation of the max statistic: the theorem assumes approximation uniformly over \emph{all} hyperrectangles, not only over the single gate boundary. Since \(a_L\lesssim\sqrt{\log p_L}\) for unit-variance Gaussian coordinates, the displayed rates are also sufficient for the gate-threshold perturbations used above. Together with the local quantile-separation condition, these conditions imply the appendix growing-family selective template.

For fixed-threshold CDF-PT coordinates, no inverse-CDF derivative or positive outcome density is needed: each placebo coordinate is a linear contrast of CDF values. This is precisely why the fixed-dollar-threshold family is useful with heaped wages. Bounded individual indicators, however, do \emph{not} by themselves make \(B_L\) bounded when cluster sizes, survey weights, or cluster leverage can grow. The required envelope is a condition on the \emph{cluster-level standardized influence scores}; it must be checked or justified for the sampling design rather than inferred mechanically from \(1\{Y\le y\}\in[0,1]\). This distinction prevents the high-dimensional theorem from being misread as an automatic finite-51-jurisdiction guarantee. In particular, the active CPS replication materials do not contain a verified bound for the standardized cluster-score envelope $B_L$ or the feasible-score remainder on the $\sqrt{\log p_L}$ scale. The appendix growing-family selective template is therefore not invoked as an empirical validity theorem for the CPS $J=99,L=51$ array; the paper reports separate design-matched stress experiments and wild-cluster sensitivity calculations instead.

\paragraph{Proof of the mass-point-safe CDF-inversion corollary in the main text.}
For each state \(d\), let
\[
E_{d,L}(\delta)
=
\{L_{d,L}(y;\delta)\le F_d^r(y)\le U_{d,L}(y;\delta)
\ \text{for all }y\in\mathcal Y\}.
\]
On \(E_{d,L}(\delta)\), monotonicity of the left-inverse operation reverses the CDF order:
\[
U_{d,L}^{\leftarrow}(\tau;\delta)
\le
(F_d^r)^{\leftarrow}(\tau)
\le
L_{d,L}^{\leftarrow}(\tau;\delta)
\qquad\text{for every }\tau\in[0,1].
\]
This is the generic distribution-band inversion result of Chernozhukov, Fern\'andez-Val, Melly, and W\"uthrich (2020); it uses only monotonicity and therefore permits jumps and flat regions. Hence, on \(E_{0,L}(\delta)\cap E_{1,L}(\delta)\), both potential-outcome quantile functions are covered simultaneously, and interval arithmetic gives
\[
Q_1^r(\tau)-Q_0^r(\tau)
\in
[U_{1,L}^{\leftarrow}(\tau)-L_{0,L}^{\leftarrow}(\tau),
 L_{1,L}^{\leftarrow}(\tau)-U_{0,L}^{\leftarrow}(\tau)]
\]
for every \(\tau\). Therefore the event that the QTT band fails anywhere is contained in the event that at least one CDF band fails anywhere. The unconditional probability bound follows immediately from the joint CDF-band condition in Corollary~1 of the main text. For any proceed event with positive probability,
\[
\Pr(E_{\QTT,L}^c\mid G_L^r)
\le
\frac{\Pr(E_{\QTT,L}^c)}{\Pr(G_L^r)}
\le
\frac{\Pr((E_{0,L}\cap E_{1,L})^c)}{\Pr(G_L^r)},
\]
and taking limsups gives the displayed post-gate bound associated with Corollary~1 of the main text. Deterministic and random-level calibrations then follow exactly as in Diagnostic Result D4(i).

We now prove the staggered-adoption statement with possibly estimated weights. Let \(\mathcal G_e\) be the finite contributing cohort set and let \(\Delta_{\mathcal G_e}\) denote its probability simplex. Let \(\mathcal W_L^e\subseteq\Delta_{\mathcal G_e}\) be the random nonempty compact weight region in the corollary and define the joint event
\[
E_L=\left\{\omega^e\in\mathcal W_L^e,\quad
L_{g,d,L}(y)\le F_{g,d}(y)\le U_{g,d,L}(y)
\ \text{for every }g,d,y\right\}.
\]
All statements below are deterministic on \(E_L\).

For each cohort define the inverted QTT endpoints
\begin{align*}
\ell_{g,L}(\tau)
&=U_{g,1,L}^{\leftarrow}(\tau)-L_{g,0,L}^{\leftarrow}(\tau),\\
 u_{g,L}(\tau)
&=L_{g,1,L}^{\leftarrow}(\tau)-U_{g,0,L}^{\leftarrow}(\tau).
\end{align*}
Then \(\QTT_{g,g+e}(\tau)\in[\ell_{g,L}(\tau),u_{g,L}(\tau)]\) simultaneously. Because every admissible weight vector is nonnegative and sums to one, the average-cohort QTT is covered by
\begin{equation}
I_{\mathrm{avg},L}^e(\tau)
=
\left[
\inf_{w\in\mathcal W_L^e}\sum_gw_g\ell_{g,L}(\tau),\quad
\sup_{w\in\mathcal W_L^e}\sum_gw_gu_{g,L}(\tau)
\right].
\label{eq:weightrobustavgband}
\end{equation}
Indeed the true \(\omega^e\) is one feasible vector on \(E_L\), and weighted interval addition is monotone in every cohort endpoint.

For mixture aggregation define the weight-robust CDF envelopes
\begin{align}
\underline F_{d,L}^e(y)
&=\inf_{w\in\mathcal W_L^e}\sum_gw_gL_{g,d,L}(y),\\
\overline F_{d,L}^e(y)
&=\sup_{w\in\mathcal W_L^e}\sum_gw_gU_{g,d,L}(y).
\label{eq:weightrobustcdfenvelopes}
\end{align}
On \(E_L\),
\[
\underline F_{d,L}^e(y)
\le\sum_g\omega_g^eL_{g,d,L}(y)
\le F_d^e(y)
\le\sum_g\omega_g^eU_{g,d,L}(y)
\le\overline F_{d,L}^e(y).
\]
The envelopes are nondecreasing. If the cohort band endpoints have the usual CDF boundary limits and are right-continuous, the envelopes inherit these properties: for \(y_n\downarrow y\), compactness of \(\mathcal W_L^e\) and finiteness of \(\mathcal G_e\) give
\[
\sup_{w\in\mathcal W_L^e}\left|\sum_gw_gH_g(y_n)-\sum_gw_gH_g(y)\right|
\le\max_g|H_g(y_n)-H_g(y)|\to0
\]
for either collection of band endpoints \(H_g\). Thus generalized inversion applies directly; otherwise the same conclusion follows after the standard monotone/range/right-continuous shape projection of the envelope bounds. Hence
\begin{equation}
I_{\mathrm{mix},L}^e(\tau)
=
\left[
(\overline F_{1,L}^e)^{\leftarrow}(\tau)-(\underline F_{0,L}^e)^{\leftarrow}(\tau),\quad
(\underline F_{1,L}^e)^{\leftarrow}(\tau)-(\overline F_{0,L}^e)^{\leftarrow}(\tau)
\right]
\label{eq:weightrobustmixband}
\end{equation}
covers \(\QTT_{\mathrm{mix}}^e(\tau)\) simultaneously over \(\tau\in\mathcal T\). Finally, writing the preceding intervals as \([a_\ell,a_u]\) and \([m_\ell,m_u]\),
\begin{equation}
I_{\Delta,L}^e(\tau)
=
[a_\ell-m_u,\ a_u-m_\ell]
=
I_{\mathrm{avg},L}^e(\tau)\ominus I_{\mathrm{mix},L}^e(\tau)
\label{eq:weightrobustgapband}
\end{equation}
covers \(\Delta_{\mathrm{agg}}^e(\tau)\). The common-weight dependence between the two aggregands can make this Minkowski band conservative, but cannot invalidate coverage.

All three containments hold on the same joint event \(E_L\); therefore their simultaneous unconditional failure probability is asymptotically no larger than \(\delta\), and the post-gate branch bound follows exactly as above. If the CDF bands and weight region are constructed separately with asymptotic failure probabilities at most \(\delta_F\) and \(\delta_W\), respectively, the union bound gives the joint result with \(\delta=\delta_F+\delta_W\), without any independence requirement. If instead the originating joint event has finite-sample probability at least \(1-\delta\), every deterministic containment above holds on that same event and therefore gives the corresponding finite-sample projection guarantee. When \(\mathcal W_L^e=\{\omega^e\}\), equations \eqref{eq:weightrobustavgband}--\eqref{eq:weightrobustgapband} reduce to the fixed-weight formulas from the previous stage. Intersecting any inverted quantile interval with known marginal support preserves coverage. \(\square\)

\paragraph{Implementation with estimated cohort shares.}
The corollary deliberately separates the deterministic projection argument from construction of the weight region. In a cluster bootstrap or multiplier implementation, one direct route is to recompute the cohort shares together with every cohort-state CDF and construct a joint confidence region for the stacked CDF-and-weight object. A simpler conservative route is to combine a simultaneous cohort-CDF band with a confidence region for \(\omega^e\) using \(\delta_F+\delta_W=\delta\). No independence between the two regions is required. For polyhedral or convex weight regions, the average-QTT endpoints in \eqref{eq:weightrobustavgband} and the pointwise mixture-CDF envelopes in \eqref{eq:weightrobustcdfenvelopes} are linear programs; if the region is represented by finitely many extreme points, the extrema are attained at those points. The validity statement concerns the population weight vector covered by the region, not the plug-in sample weights treated as fixed.

\paragraph{Derivation for the centered-symmetric-gate remark in the main text.}
Suppress route notation and let
\[
A=\{(u,t): |u|\le z_{1-\gamma/2}\},
\qquad
B=\{(u,t): \|t\|_\infty\le c\}.
\]
Both \(A\) and \(B\) are convex and centrally symmetric subsets of \(\mathbb R^{1+J}\). Under Diagnostic Result D4(ii), \(Y=(U,T')'\) is a centered Gaussian vector with unit marginal variance for \(U\). The Gaussian correlation inequality therefore gives
\[
\Pr(Y\in A\cap B)\ge \Pr(Y\in A)\Pr(Y\in B).
\]
Because \(\Pr(Y\in B)=p^\circ>0\) and \(\Pr(Y\in A)=\Pr\{|U|\le z_{1-\gamma/2}\}=1-\gamma\), division by \(p^\circ\) yields
\[
\Pr\{|U|\le z_{1-\gamma/2}\mid \|T\|_\infty\le c\}\ge 1-\gamma.
\]
This is the one-coordinate-versus-rectangle normal correlation inequality covered by Khatri (1967), and it also follows immediately from Royen's (2014) full Gaussian correlation theorem for symmetric convex sets. The nonsingularity imposed in Diagnostic Result D4(ii) is more than sufficient for the form used here. The sample conditional-coverage statement follows from the same joint weak convergence, critical-value convergence, and boundary-probability argument used in Diagnostic Result D4(ii). This conclusion is specific to a \emph{centered} Gaussian law and a \emph{symmetric} gate. Under a shifted local diagnostic law, the acceptance event remains symmetric geometrically but the probability measure is no longer centered, so this argument does not determine the sign of the selection distortion. \(\square\)

\paragraph{Equivalent polyhedral representation.}
For computation or comparison with polyhedral selective inference, let
\(\eta=(1,0,\ldots,0)'\),
\(a=\Omega\eta/(\eta'\Omega\eta)\),
\(z=(I-a\eta')Y\), and \(V=\eta'Y\).
Then \(V\perp z\) and \(Y=z+aV\). Define
\[
A_{\mathrm{sel}}
=
\begin{pmatrix}
0 & I_J\\
0 & -I_J
\end{pmatrix},
\qquad
b_{\mathrm{sel}}=c\mathbf 1_{2J}.
\]
The rectangular event \(\|T\|_\infty\le c\) is exactly \(A_{\mathrm{sel}}Y\le b_{\mathrm{sel}}\). Substituting \(Y=z+aV\) gives \(2J\) scalar linear inequalities in \(V\). Conditional on \(z\) and selection, \(V\) is therefore Gaussian truncated to the interval obtained from those inequalities; rows with zero coefficient \((A_{\mathrm{sel}}a)_k=0\) restrict only \(z\), while empty lower or upper index sets use the conventions \(\max\varnothing=-\infty\) and \(\min\varnothing=+\infty\). This gives the familiar truncated-Gaussian representation (Lee et al., 2016). The marginal conditional CDF in the main theorem avoids requiring nonzero target--diagnostic correlations and is convenient for plug-in Gaussian simulation.

\begin{table}[htbp]
\centering
\caption{When post-gate selection correction matters in the Gaussian limit}
\label{tab:selectivegate}
\small
\begin{tabular}{rrrrrr}
\toprule
Gate level & Corr$(U,T_j)$ & Proceed prob. & Naive coverage & Selective coverage & Selective 97.5\% cutoff\\
\midrule
0.05 & 0.55 & 0.950 & 0.960 & 0.950 & 1.877 \\
0.05 & 0.75 & 0.950 & 0.969 & 0.950 & 1.791 \\
0.05 & 0.90 & 0.950 & 0.983 & 0.950 & 1.692 \\
0.10 & 0.55 & 0.900 & 0.965 & 0.950 & 1.831 \\
0.10 & 0.75 & 0.900 & 0.978 & 0.950 & 1.699 \\
0.10 & 0.90 & 0.900 & 0.994 & 0.950 & 1.538 \\
0.20 & 0.55 & 0.800 & 0.970 & 0.950 & 1.767 \\
0.20 & 0.75 & 0.800 & 0.987 & 0.950 & 1.572 \\
0.20 & 0.90 & 0.800 & 0.999 & 0.948 & 1.326 \\
\bottomrule
\end{tabular}
\begin{flushleft}\footnotesize
Notes: The diagnostic family has $J=99$ coordinates generated by a common-factor Gaussian model with pairwise diagnostic correlation $0.90$. The target statistic has the displayed correlation with each diagnostic coordinate. For each row the max-$|t|$ critical value gives the stated gate level under the complete null. Naive coverage uses the ordinary $|U|\le1.96$ rule conditional on passing. Selective coverage uses the exact conditional Gaussian 2.5\% and 97.5\% quantiles. Reported coverages are verified with 500,000 Monte Carlo draws; quadrature determines the critical values and selective cutoffs. At $(\alpha,\rho)=(.20,.90)$, naive conditional coverage is essentially one because passing strongly truncates the component of the target statistic correlated with the diagnostic family, whereas the selective rule restores 95\% coverage.
\end{flushleft}
\end{table}

Table~\ref{tab:selectivegate} shows both mild and consequential regimes. With a 5\% gate, the ordinary conditional coverage in the calibration ranges from 0.959 to 0.983 as target--diagnostic correlation moves from 0.55 to 0.90, while the selective pivot returns coverage near 0.950; the realistic payoff is therefore roughly 1--3 percentage points of tightening toward nominal coverage, not validity rescue. With strong common dependence in the 99-coordinate family, or a looser gate, passing substantially truncates the component of the target statistic correlated with the diagnostics: the ordinary 95\% interval can then overcover by several percentage points and, in the strongest calibration shown, almost never miss. The selective Gaussian cutoff adapts to that truncation and restores approximately 95\% conditional coverage. Under the centered symmetric gate maintained in Diagnostic Result D4(ii), this is the relevant first-order distortion: passing selects a correlated target statistic toward the center. We therefore do not manufacture an anti-conservative local-drift example that would change the maintained selection problem and simultaneously move away from the placebo-null regime.

\begin{table}[H]
\centering
\caption{Finite-cluster stress test for the 99-coordinate Gaussian post-gate pivot}
\label{tab:selectivefinite}
\small
\begin{tabular}{rrrrr}
\toprule
Clusters & Replications & Proceed rate & Naive cond. coverage & Selective coverage\\
\midrule
20 & 100,000 & 0.903 & 0.966 & 0.947 \\
30 & 100,000 & 0.920 & 0.966 & 0.948 \\
50 & 100,000 & 0.933 & 0.969 & 0.951 \\
\bottomrule
\end{tabular}
\begin{flushleft}
\footnotesize Notes: Each replication contains $L$ i.i.d. Gaussian cluster-score vectors with $J=99$ diagnostic coordinates. Diagnostic pairwise correlation is 0.90 and each target--diagnostic correlation is 0.75, generated by a one-common-factor representation. Target and diagnostics are studentized by sample cluster standard deviations. The gate uses the asymptotic 5\% max-$|t|$ critical value $c=2.671$. The selective pivot uses the known limiting common-factor covariance, so the exercise isolates finite-$L$ Gaussian/studentization error rather than covariance-estimation error. There are 100,000 replications per row; the Monte Carlo s.e. of selective coverage is below 0.001. This is not a fixed-$L$ validity theorem.
\end{flushleft}
\end{table}

Table~\ref{tab:selectivefinite} stress-tests the fixed-dimensional limiting selective pivot at the cluster counts most relevant for the applications using the same \(J=99\) diagnostic dimension as the CPS gate. The diagnostic scores follow the common-factor Gaussian calibration used above, with pairwise diagnostic correlation 0.90 and target--diagnostic correlation 0.75. Each finite sample is studentized by its cluster-level sample standard deviations. To isolate the approximation error from finite \(L\) and studentization, the selective calculation uses the \emph{known limiting structured covariance}; it therefore does not study covariance-estimation error. Selective conditional coverage is 0.947, 0.948, and 0.951 for \(L=20,30,50\), respectively.

\begin{table}[htbp]
\centering
\caption{Finite-cluster stress test for the growing-family Gaussian multiplier selective pivot ($J=99$).}
\label{tab:hdmultiplierselective}
\begin{tabular}{rrrrrr}
\toprule
$L$ & Proceed & Naive cond. & Multiplier-selective & MC s.e. & Mean $c^*$ \\
 & rate & coverage & coverage &  &  \\
\midrule
20 & 0.894 & 0.973 & 0.947 & 0.003 & 2.583 \\
30 & 0.912 & 0.968 & 0.943 & 0.003 & 2.615 \\
50 & 0.925 & 0.970 & 0.947 & 0.003 & 2.636 \\
\bottomrule
\end{tabular}
\par\vspace{2pt}\begin{minipage}{0.96\textwidth}\footnotesize Notes: $5{,}000$ outer replications and $1{,}999$ Gaussian multiplier draws per replication; seed 202610120. The diagnostic dimension is $J=99$. Diagnostic scores have pairwise correlation $0.90$, and each diagnostic has correlation $0.75$ with the scalar target score. Both observed and multiplier scores are studentized using the sample cluster standard deviations. The multiplier law is generated directly from the centered empirical cluster-score matrix and is therefore generally singular when $J+1>L$; no covariance inversion or factor structure is imposed. Coverage is conditional on the multiplier-calibrated gate proceeding. The experiment is a finite-sample stress test, not a fixed-$L$ theorem.
\end{minipage}
\end{table}

Table~\ref{tab:hdmultiplierselective} instead implements the appendix growing-family selective template directly. It uses the centered empirical cluster-score matrix and fresh Gaussian multipliers to generate the joint target--diagnostic law. When \(J+1>L\), that conditional Gaussian law is singular by construction, but no inverse is needed. At \(J=99\), multiplier-selective conditional coverage is 0.947, 0.943, and 0.947 for \(L=20,30,50\), respectively, while ordinary symmetric Gaussian intervals conditionally overcover. The small undercoverage at \(L=30\) is within roughly two Monte Carlo standard errors of 0.95. This second experiment therefore addresses the covariance-estimation/singularity issue that the oracle experiment intentionally held fixed, while remaining a finite-sample stress test rather than a fixed-\(L\) theorem.

\subsection{Local alternatives and power}

\paragraph{Proof of Diagnostic Result D3.}
Under the local sequence \(\theta_L^r=h^r/\sqrt L\), Diagnostic Result D3 assumes mutual contiguity of the fixed-drift local law and the null law, together with the same centered Gaussian limit for \(\sqrt L(\widehat\theta^r-\theta_L^r)\). A local asymptotic normality (LAN) log-likelihood-ratio expansion with a tight Gaussian central sequence is one sufficient route to contiguity, but the theorem imposes contiguity directly so that no parametric likelihood is required. If \(\mathcal R_L\) denotes any studentization or centered-bootstrap approximation remainder satisfying \(\mathcal R_L=o_{P_{L,0}}(1)\), mutual contiguity---equivalently here the standard Le Cam first-lemma implication for null-negligible events along the fixed local sequence---implies \(\mathcal R_L=o_{P_{L,h}}(1)\): for every \(\varepsilon>0\), the events \(\{|\mathcal R_L|>\varepsilon\}\) have null probability tending to zero and therefore local-law probability tending to zero. Hence (S2)--(S3), established under the null, continue to hold to first order along each fixed local sequence. Then
\[
\frac{\sqrt L\,\widehat\theta_j^r}{\widehat s_j^r}
=
\frac{\sqrt L(\widehat\theta_j^r-\theta_{L,j}^r)}{\widehat s_j^r}
+
\frac{h_j^r}{\widehat s_j^r}.
\]
By (S1)--(S2), the vector of observed studentized statistics converges to
\[
W^r+D_r^{-1}h^r.
\]
The bootstrap remains centered at the estimated law, so (S3) and continuity of the finite Gaussian maximum imply
\[
\widehat c_{1-\alpha}^r
\overset{p}\longrightarrow
c_{1-\alpha}(R_r),
\]
the null max-\(|t|\) critical value. The continuous mapping theorem and continuity at the critical value therefore give the local-power display in Diagnostic Result D3.

For \(h^r=0\), the limit equals \(\alpha\). If \(R_r\) is positive definite, the acceptance region
\[
\mathcal A_c=[-c,c]^J
\]
has nonempty interior and is convex and symmetric. Because the acceptance rectangle is bounded with nonempty interior and the Gaussian law is nondegenerate under positive-definite \(R_r\), Anderson's (1955) strict Gaussian shift inequality for a symmetric convex set implies that
\[
\Pr(W^r+D_r^{-1}h^r\in\mathcal A_c)
<
\Pr(W^r\in\mathcal A_c)
\]
for any nonzero displacement, giving \(\pi_r(h^r)>\alpha\).

For any coordinate \(j\), the rejection event contains
\[
\left\{
|W_j^r+h_j^r/s_j^r|>c_{1-\alpha}(R_r)
\right\}.
\]
Because \(W_j^r\) is standard normal, the probability of this marginal event is exactly the coordinatewise expression displayed below. Taking the largest such probability yields the conservative familywise lower bound. Finally, along any sequence of drift vectors whose largest standardized component diverges, the corresponding marginal rejection probability converges to one. \(\square\)

\paragraph{Coordinatewise lower bound.}
The local-power function also satisfies
\[
\pi_r(h^r)
\ge
\max_{j\in\mathcal F}
\left[
1-\Phi\!\left\{c_{1-\alpha}(R_r)-|h_j^r|/s_j^r\right\}
+
\Phi\!\left\{-c_{1-\alpha}(R_r)-|h_j^r|/s_j^r\right\}
\right].
\]

because rejection of any one coordinate implies rejection of the max-\(|t|\) statistic. This bound does not require independence across coordinates.

\paragraph{Interpretation.}
The local-power result does not rank correlation matrices by a scalar ``effective number of tests.'' Correlation changes both the max-\(|t|\) critical value and the probability of the shifted Gaussian vector leaving the acceptance rectangle. The relevant power calculation is therefore route- and design-specific.

\subsection{Route-specific sufficient conditions for the stop rule}

\paragraph{Explicit CDF-PT influence function.}
Let \(q_1(\tau)\) be the observed treated placebo quantile and \(q_0(\tau)\) the additive-CDF-PT counterfactual quantile. If the four CDF estimators entering the contrast admit cluster influence functions \(\psi_{T,t}(y)\), \(\psi_{T,s}(y)\), \(\psi_{C,t}(y)\), and \(\psi_{C,s}(y)\), then the first-order influence function of the placebo QTT is
\[
\phi_{\tau}^{\mathrm{CDFPT}}
=
-\frac{\psi_{T,t}\{q_1(\tau)\}}{f_{T,t}\{q_1(\tau)\}}
+
\frac{\psi_{T,s}\{q_0(\tau)\}
+\psi_{C,t}\{q_0(\tau)\}
-\psi_{C,s}\{q_0(\tau)\}}
{f_0\{q_0(\tau)\}}.
\]
Thus simultaneous inference uses the covariance of the stacked cluster contributions across the frozen family.

\paragraph{Proof of Diagnostic Specialization D5.}
For each fixed coordinate, stack the empirical treated, treated-baseline, comparison-post, and comparison-baseline CDF evaluations entering the additive CDF-PT construction. By the assumed product-sup-norm cluster expansion, the Lyapunov condition yields the required finite-dimensional Gaussian limits, while local stochastic equicontinuity controls the score process in shrinking neighborhoods of the target quantiles. The counterfactual CDF map
\[
(F_{T,s},F_{C,t},F_{C,s})\mapsto F_{T,s}+F_{C,t}-F_{C,s}
\]
is affine and therefore Hadamard differentiable. At a quantile \(q(\tau)\) with density \(f\{q(\tau)\}>0\), the inverse-CDF map is Hadamard differentiable tangentially to continuous perturbations, with derivative
\[
h\mapsto-\frac{h\{q(\tau)\}}{f\{q(\tau)\}}.
\]
The local stochastic-equicontinuity condition permits replacement of the empirical score process evaluated at an estimated quantile by its value at the corresponding population quantile. Applying the functional delta method to the observed and counterfactual quantiles therefore gives, for one coordinate,
\[
\sqrt L\{\widehat q_1(\tau)-q_1(\tau)\}
=
-\frac{1}{\sqrt L}\sum_{\ell=1}^L
\frac{\psi_{T,t,\ell}\{q_1(\tau)\}}{f_{T,t}\{q_1(\tau)\}}
+o_p(1),
\]
and
\[
\sqrt L\{\widehat q_0(\tau)-q_0(\tau)\}
=
-\frac{1}{\sqrt L}\sum_{\ell=1}^L
\frac{
\psi_{T,s,\ell}\{q_0(\tau)\}
+\psi_{C,t,\ell}\{q_0(\tau)\}
-\psi_{C,s,\ell}\{q_0(\tau)\}
}{
f_0\{q_0(\tau)\}
}
+o_p(1).
\]
Therefore
\[
\sqrt L\{\widehat{\QTT}(\tau)-\QTT(\tau)\}
=
L^{-1/2}\sum_{\ell=1}^L\phi_{\tau,\ell}^{\mathrm{CDFPT}}+o_p(1),
\]
with \(\phi_{\tau,\ell}^{\mathrm{CDFPT}}\) as in the explicit CDF-PT influence-function display above. Stacking the finitely many coordinates yields the joint asymptotic linear representation used by the max-\(t\) statistic. Assumptions on the standard errors give (S2). Bootstrap consistency for the underlying CDF-score vector, together with the bootstrap delta method through the same affine and inverse-CDF maps, gives (S3). Positive marginal variances and Gaussianity imply that the maximum of the absolute studentized finite Gaussian vector has no point masses and hence has a continuous CDF. Diagnostic Result D1 then applies. \(\square\)

\paragraph{Proof of Diagnostic Specialization D6.}
For one coordinate, write the CIC counterfactual CDF as
\[
F_{0}(y)=F_{T,s}\!\left[
F_{C,s}^{-1}\{F_{C,t}(y)\}
\right].
\]
The support-interiority condition ensures that the inverse \(F_{C,s}^{-1}\) is evaluated away from the boundary of its probability range. Positivity and continuity of the relevant densities make both the inverse-CDF map and the outer composition locally Hadamard differentiable. For a perturbation \((h_T,h_s,h_t)\), writing \(x_y=F_{C,s}^{-1}\{F_{C,t}(y)\}\), the first-order perturbation of the CIC counterfactual CDF is
\[
h_T(x_y)
+
f_{T,s}(x_y)
\frac{h_t(y)-h_s(x_y)}{f_{C,s}(x_y)}.
\]
Because the diagnostic family is finite, local smoothness and support interiority need only hold in neighborhoods of the finitely many transport and target-quantile arguments used by the family; no global uniform-in-\(y\) condition is required for the stated corollary. Consequently, the map from the stacked empirical CDFs \((F_{T,s},F_{C,s},F_{C,t})\) to \(F_0\), and then from \(F_0\) to its target quantile, is Hadamard differentiable. The observed treated evaluation-date quantile is handled by the same inverse-CDF derivative, using the positive density of \(F_{T,t}\) at that quantile. Combining the assumed joint cluster CLT with the functional delta method therefore yields a joint Gaussian linear representation for the finite CIC placebo-QTT vector. Consistent nondegenerate standard errors give (S2), and bootstrap consistency for the stacked empirical-CDF scores propagates through the differentiable CIC and quantile maps by the bootstrap delta method, giving (S3). The same finite-Gaussian continuity argument completes the max-statistic continuity step. Diagnostic Result D1 completes the proof. \(\square\)

\subsection{Proof of Proposition 1: nonlinear aggregation}

Fix an event time $e$. Throughout this proof $\mathcal G_e$ denotes the active target support defined in the main text. The weights $\omega_g^e$, $g\in\mathcal G_e$, are fixed, finite in number, strictly positive, and sum to one; zero-weight cohorts are excluded because they enter neither aggregand.

\paragraph{Part (i): bracketing.}
Fix $d$ and $\tau$, suppress $(d,e,\tau)$ where harmless, and write $q_{\min}=\min_gq_g$ and $q_{\max}=\max_gq_g$. If $y<q_{\min}$, then $y<q_g$ for every $g$, so the generalized-quantile definition implies $F_g(y)<\tau$ for every component. Hence $F^e_d(y)<\tau$ and $q_d^{\mathrm{mix}}\ge q_{\min}$. By continuity at each component $\tau$-quantile, $F_g(q_g)=\tau$; monotonicity gives $F_g(q_{\max})\ge\tau$ for every $g$, hence $F_d^e(q_{\max})\ge\tau$ and $q_d^{\mathrm{mix}}\le q_{\max}$.

\paragraph{Part (ii): sharp fixed-quantile interval and coarser range-only sharpness.}
For state $d$, write
\[
q_{d,\min}=\min_{g\in\mathcal G_e}q_{g,d},\qquad q_{d,\max}=\max_{g\in\mathcal G_e}q_{g,d},\qquad
L_d=\bar q_d-q_{d,\min},\qquad R_d=q_{d,\max}-\bar q_d.
\]
Part (i) implies
\[
-L_d\le \kappa_d^e\equiv q_d^{\mathrm{mix}}-\bar q_d\le R_d.
\]
Because
\[
\Delta_{\mathrm{agg}}^e=\kappa_0^e-\kappa_1^e,
\]
subtracting the two state-specific intervals gives immediately
\[
-(L_0+R_1)\le \Delta_{\mathrm{agg}}^e\le R_0+L_1.
\]
Hence
\[
|\Delta_{\mathrm{agg}}^e|
\le B_{\mathrm{sharp}}^e
\equiv\max\{L_0+R_1,R_0+L_1\}.
\]
Since $D_d=\max\{L_d,R_d\}$,
\[
B_{\mathrm{sharp}}^e\le D_0+D_1.
\]
Now put $H_d=L_d+R_d$ and $\omega_{\min}=\min_g\omega_g^e$. At least one cohort attaining $q_{d,\min}$ has weight at least $\omega_{\min}$, so
\[
L_d=\sum_g\omega_g^e(q_{g,d}-q_{d,\min})\le(1-\omega_{\min})H_d.
\]
Likewise, at least one cohort attaining $q_{d,\max}$ has weight at least $\omega_{\min}$, and therefore
\[
R_d=\sum_g\omega_g^e(q_{d,\max}-q_{g,d})\le(1-\omega_{\min})H_d.
\]
Thus $D_d\le(1-\omega_{\min})H_d$, which yields the nested inequalities stated in Proposition~1(ii).

We next prove that the first interval is sharp while holding the cohort quantile vector itself fixed. Fix one state and suppress $d$. If all $q_g$ are equal, then $L=R=0$ and the claim is trivial. Otherwise let $q_{\min}<q_{\max}$ and preserve every prescribed $q_g$. For any $\eta>0$, assign to each cohort a Gaussian CDF with $\tau$-quantile exactly $q_g$. To make the mixture quantile approach $q_{\min}$, choose the scale of every cohort attaining $q_{\min}$ sufficiently small that its CDF at $q_{\min}+\eta$ is arbitrarily close to one, and choose the scales of all remaining cohorts sufficiently large that their CDF values at the same point are arbitrarily close to $\tau$. At $q_{\min}$, the minimum-quantile components equal $\tau$ and every component with a larger quantile is strictly below $\tau$, so the mixture CDF is strictly below $\tau$. At $q_{\min}+\eta$, the preceding scale choices make the mixture CDF exceed $\tau$. Continuity therefore places $q^{\mathrm{mix}}$ within $\eta$ of $q_{\min}$. Hence $\kappa=q^{\mathrm{mix}}-\bar q$ can approach $-L$ arbitrarily closely.

The upper endpoint is analogous. Make the scales of cohorts attaining $q_{\max}$ sufficiently small that their CDFs at $q_{\max}-\eta$ are arbitrarily close to zero, while making every other scale sufficiently large that its CDF there is arbitrarily close to $\tau$. The mixture CDF at $q_{\max}-\eta$ can then be made strictly below $\tau$, whereas at $q_{\max}$ it is strictly above $\tau$ whenever some cohort has a lower quantile. Thus $q^{\mathrm{mix}}$ can approach $q_{\max}$ and $\kappa$ can approach $R$. Applying the upper construction independently in state $0$ and the lower construction in state $1$ makes $\Delta_{\mathrm{agg}}^e$ approach $R_0+L_1$; reversing the constructions makes it approach $-(L_0+R_1)$. The Gaussian locations are chosen as $\mu_g=q_g-\sigma_g\Phi^{-1}(\tau)$, so every prescribed cohort quantile is preserved exactly and all component CDFs remain smooth and strictly increasing. This proves sharpness of the one-sided interval in Proposition~1(ii) and therefore $\sup|\Delta_{\mathrm{agg}}^e|=B_{\mathrm{sharp}}^e$ conditional on the fixed active-support quantile vectors and weights. Because $\mathcal G_e$ is the active support, every contributing weight is positive; hence $H_d>0$ implies both $L_d>0$ and $R_d>0$. Hence whenever $H_0+H_1>0$, the sharp interval has a strictly negative lower endpoint and a strictly positive upper endpoint: fixed cohort quantiles and weights alone do not determine the sign of the aggregation gap.

For completeness, the coarser coefficient $1-\omega_{\min}$ is itself sharp when only the weights and state-specific ranges are retained. Suppose at least two cohorts contribute and choose $g_\star$ with $\omega_{g_\star}^e=\omega_{\min}$. Fix $H>0$. To approach the lower state-specific distortion, set $q_{g_\star}=0$ and $q_g=H$ for $g\ne g_\star$, so $\bar q=(1-\omega_{\min})H$. For any $\eta\in(0,H)$, make the scale of $g_\star$ sufficiently small that $F_{g_\star}(\eta)$ is arbitrarily close to one and every other scale sufficiently large that $F_g(\eta)$ is arbitrarily close to $\tau$. The mixture quantile can then be placed in $(0,\eta)$, so its distortion approaches $-(1-\omega_{\min})H$. To approach the upper distortion, set $q_{g_\star}=H$ and every other $q_g=0$, then make the $g_\star$ scale small and the remaining scales large; the mixture quantile approaches $H$, while $\bar q=\omega_{\min}H$, so the distortion approaches $(1-\omega_{\min})H$.

Apply the upper construction to state $0$ and the lower construction to state $1$, with ranges $H_0,H_1$. For every $\varepsilon>0$ the scales can be chosen so that
\[
|\Delta_{\mathrm{agg}}^e|>(1-\omega_{\min})(H_0+H_1)-\varepsilon.
\]
Thus no smaller universal coefficient can multiply $H_0+H_1$ when only the fixed weights and the two ranges are retained. This completes part (ii).

\paragraph{Part (iii): density tilt, remainder, and tilt--dispersion diagnostic.}
Fix $d$ and again suppress indices. To distinguish the density-curvature bound $K_d$ from the cluster count used elsewhere, write $K_f=K_d$. Let $q_g$ denote component quantiles, $q_m$ the mixture quantile, $\bar q=\sum_g\omega_gq_g$, and $f_g$ the component densities. By part (i), all points between $q_g$ and $q_m$ remain inside the assumed smoothness interval. Taylor's theorem applied to each CDF around its own quantile gives
\[
F_g(q_m)-\tau
=f_g(q_g)(q_m-q_g)+\frac12f'_g(\xi_g)(q_m-q_g)^2
\]
for some $\xi_g$ between $q_g$ and $q_m$. Weighted summation and $\sum_g\omega_gF_g(q_m)=\tau$ imply
\[
0=\bar f(q_m-\bar q)-\sum_g\omega_gf_g(q_g)(q_g-\bar q)+\mathcal R_f,
\qquad
\bar f=\sum_g\omega_gf_g(q_g),
\]
where
\[
|\mathcal R_f|
\le\frac{K_f}{2}\sum_g\omega_g(q_m-q_g)^2
\le\frac{K_f}{2}H^2.
\]
Because $\bar f\ge m>0$, with $r=-\mathcal R_f/\bar f$,
\[
q_m-\bar q=A+r,
\qquad
A=\frac{\sum_g\omega_gf_g(q_g)(q_g-\bar q)}{\bar f},
\qquad
|r|\le\frac{K_f}{2\bar f}\sum_g\omega_g(q_m-q_g)^2\le\frac{K_fH^2}{2m}.
\]
Define $\widetilde\omega_g=\omega_gf_g(q_g)/\bar f$. These weights are nonnegative and sum to one, and
\[
A=\sum_g(\widetilde\omega_g-\omega_g)q_g,
\qquad
q_m=\sum_g\widetilde\omega_gq_g+r.
\]
Thus the density-tilt operator $\mathcal R_{d,\tau}$ in the main text maps target weights to $\widetilde\omega$. On the support of positive target weights, $\widetilde\omega=\omega$ if and only if $f_g(q_g)=\bar f$ for every such cohort, equivalently the own-quantile densities are common across those cohorts.
This establishes the density-tilted representation. The phrase ``first order'' refers to the component in $H$ when density heterogeneity does not shrink at the same rate; the displayed remainder is uniformly quadratic in the component-quantile range under the stated curvature and density lower bounds.

For the covariance bound, set $X_g=f_g(q_g)$ and $Y_g=q_g$. The numerator of $A$ is $\operatorname{Cov}_\omega(X_g,Y_g)$. Cauchy--Schwarz together with the weighted Popoviciu inequality yields
\[
|\operatorname{Cov}_\omega(X_g,Y_g)|
\le \sqrt{\operatorname{Var}_\omega(X_g)\operatorname{Var}_\omega(Y_g)}
\le \frac14(\max_gX_g-\min_gX_g)(\max_gY_g-\min_gY_g),
\]
and hence
\[
|A|\le\frac{\Delta f\,H}{4\bar f}\le\frac{\Delta f\,H}{4m}.
\]

For the total-variation form, put $a_g=\widetilde\omega_g-\omega_g$. Then $\sum_ga_g=0$. For the midrange $c=(q_{\max}+q_{\min})/2$,
\[
|A|=\left|\sum_ga_g(q_g-c)\right|
\le \frac{H}{2}\sum_g|a_g|=HV,
\qquad
V=\frac12\sum_g|\widetilde\omega_g-\omega_g|.
\]
Applying the state-specific expansion for $d=0,1$ and using
\[
\Delta_{\mathrm{agg}}^e
=(q_0^{\mathrm{mix}}-\bar q_0)-(q_1^{\mathrm{mix}}-\bar q_1)
\]
gives the local aggregation-gap display in Proposition~1(iii) and, by the preceding total-variation inequality plus the remainder bounds, its tilt--dispersion bound. The plug-in $\widehat{\mathcal D}_{\mathrm{tilt}}$ is therefore a descriptive first-order divergence diagnostic. No coverage statement follows from plugging in estimated quantiles or densities unless their estimation error and the curvature remainder are incorporated.

Along any sequence with $H_d\to0$, $\Delta f_d=O(H_d)$, $m_d$ uniformly bounded away from zero, and $K_d$ uniformly bounded, both $A_d$ and $r_d$ are $O(H_d^2)$; this proves the qualification in part (iii).

\paragraph{Part (iv): compact-uniform extension.}
Suppose the smoothness assumptions hold on the union of the component-quantile ranges generated by the compact $\mathcal T$, with common lower-density and curvature constants. Every step above is then valid for every $\tau\in\mathcal T$ with the same deterministic constants. Taking suprema of the pointwise remainder inequality gives
\[
\sup_{\tau\in\mathcal T}|\rho_e(\tau)|
\le\sum_{d=0}^1\frac{K_d}{2m_d}\sup_{\tau\in\mathcal T}H_d(\tau)^2.
\]
Taking suprema in the pointwise tilt--dispersion inequality gives the second display in part (iv). No stochastic equicontinuity is needed for this deterministic uniform statement.

\paragraph{Part (v): sign.}
There is no universal sign. With two equally weighted unit-variance Gaussian cohorts, untreated means $-2$ and $2$, and treatment location shifts $0$ and $2$, respectively, the average-cohort QTT equals $1$ at every quantile, while the mixture QTT is approximately $0$, $1$, and $2$ at $\tau=.2,.5,.8$. Thus $\Delta_{\mathrm{agg}}^e$ changes from positive to zero to negative within the same design.

\paragraph{Part (vi): exact equality class.}
By definition
\[
\kappa_d^e(\tau)=q_d^{\mathrm{mix}}(\tau)-\bar q_d(\tau),
\]
so direct subtraction gives
\[
\Delta_{\mathrm{agg}}^e(\tau)=\kappa_0^e(\tau)-\kappa_1^e(\tau).
\]
Therefore the two QTT aggregands coincide if and only if $\kappa_0^e(\tau)=\kappa_1^e(\tau)$. This is an exact necessary-and-sufficient characterization, not an approximation.

The stated subclasses follow immediately. With one contributing cohort, each $\kappa_d$ is zero. If all cohort $\tau$-quantiles are identical within state $d$, bracketing in part (i) forces the mixture quantile to equal that common value, so again $\kappa_d=0$. Finally, if $F_{g,1}(y)=F_{g,0}(y-\delta_e)$ for every cohort and the same weights form both mixtures, then
\[
F_1^e(y)=\sum_g\omega_gF_{g,0}(y-\delta_e)=F_0^e(y-\delta_e).
\]
Hence $q_1^{\mathrm{mix}}(\tau)=q_0^{\mathrm{mix}}(\tau)+\delta_e$, while every cohort QTT equals $\delta_e$; both aggregate QTTs therefore equal $\delta_e$ for every $\tau$.

\paragraph{Part (vii): sharpness of the tilt--dispersion envelope.}
Fix a state $d$ and suppress $(d,e,\tau)$. Put $a_g=\widetilde\omega_g-\omega_g$. Then $\sum_ga_g=0$ and $V=\frac12\sum_g|a_g|=\sum_{a_g>0}a_g=-\sum_{a_g<0}a_g$. Translation of every $q_g$ by the same constant does not change $\sum_ga_gq_g$, so normalize the admissible interval to $[0,H]$. For every $q_g\in[0,H]$,
\[
\sum_ga_gq_g
\le H\sum_{a_g>0}a_g=HV,
\]
and similarly $\sum_ga_gq_g\ge-HV$. If $V>0$, the upper bound is attained by setting $q_g=H$ when $a_g>0$ and $q_g=0$ when $a_g<0$; the lower bound reverses the endpoint assignments. If $V=0$, both sides are zero. Therefore
\[
\sup_{\max q-\min q\le H}\left|\sum_ga_gq_g\right|=HV.
\]
Applying this separately to $d=0$ and $d=1$ and choosing the endpoint assignments so that $A_0$ and $A_1$ have opposite signs yields
\[
\sup|A_0-A_1|=H_0V_0+H_1V_1=\mathcal D_{\mathrm{tilt}}^e.
\]

It remains to verify attainability inside the smooth density class rather than only in the algebra of weights. Fix strictly positive target weights $\omega_g$ and strictly positive desired tilted weights $\widetilde\omega_g$ on the same finite support. Choose any $C>0$ and set the desired own-quantile density value to $c_g=C\widetilde\omega_g/\omega_g$. Let $z_\tau=\Phi^{-1}(\tau)$ and define a Gaussian cohort distribution with
\[
\sigma_g=\frac{\phi(z_\tau)}{c_g},\qquad
\mu_g=q_g-\sigma_g z_\tau.
\]
Its $\tau$-quantile is exactly $q_g$ and its density at that quantile is exactly $c_g$. Hence the induced density-tilt weights satisfy
\[
\frac{\omega_gc_g}{\sum_h\omega_hc_h}=\widetilde\omega_g.
\]
The Gaussian densities are smooth and strictly positive; because the cohort set is finite, on every compact interval joining the prescribed quantiles they admit finite common curvature bounds and a positive common density lower bound. For a shrinking family of endpoint-separated quantiles with the $c_g$ fixed, those constants can be taken uniformly for sufficiently small ranges. Part (iii) therefore gives
\[
\Delta_{\mathrm{agg}}^e=A_0-A_1+O(H_0^2+H_1^2),
\]
and the endpoint construction makes $|A_0-A_1|=\mathcal D_{\mathrm{tilt}}^e$. Thus the exact gap approaches the sharp first-order envelope up to the quadratic curvature remainder. This proves part (vii) and completes Proposition~1. $\square$

\paragraph{Population calibration of the local expansion.}
To isolate the local representation from the deliberately nonlocal sign-reversal example, consider two equally weighted Gaussian components at $\tau=.25$. Fix their standard deviations at $0.7$ and $1.4$, and choose means so that their component $\tau$-quantiles are $-\lambda/2$ and $+\lambda/2$. Then $\bar q(\tau)=0$, $H(\tau)=\lambda$, and the own-quantile densities give the exact leading term $A(\tau)=-\lambda/6$. Table~\ref{tab:aggregationlocalcal} reports the exact mixture quantile. The remainder divided by $\lambda^2$ stabilizes as $\lambda\downarrow0$, while the leading term shrinks linearly, matching Proposition~1(iii).

\begin{table}[htbp]
\centering
\caption{Local calibration of the density-tilted mixture-quantile expansion}
\label{tab:aggregationlocalcal}
\begin{tabular}{rrrrr}
\toprule
$\lambda$ & $A(\tau)$ & $q^{\mathrm{mix}}(\tau)-\bar q(\tau)$ & remainder & remainder/$\lambda^2$\\
\midrule
1.000 & -0.1667 & -0.2333 & -0.0666 & -0.0666\\
0.500 & -0.0833 & -0.1004 & -0.0171 & -0.0684\\
0.250 & -0.0417 & -0.0460 & -0.0044 & -0.0697\\
0.125 & -0.0208 & -0.0219 & -0.0011 & -0.0705\\
\bottomrule
\end{tabular}
\begin{minipage}{0.91\linewidth}\small
\emph{Notes:} Equal cohort weights, $\tau=.25$, Gaussian component standard deviations $0.7$ and $1.4$. The component $\tau$-quantiles are fixed at $-\lambda/2$ and $+\lambda/2$, so $\bar q(\tau)=0$ and $H(\tau)=\lambda$; the component means are chosen accordingly. Because the own-quantile densities are proportional to the inverse standard deviations, the density-tilt term is exactly $A(\tau)=-\lambda/6$. The final column stabilizes as $\lambda\downarrow0$, illustrating the quadratic remainder in Proposition~5(iii). This is a deterministic population calculation, not a Monte Carlo experiment.
\end{minipage}
\end{table}

\subsection{Primitive influence function for estimated cohort shares}

For the sample-share estimator stated in Proposition~2, first suppose the cluster contribution vectors are identically distributed across clusters. Let $A_{\ell g}^e\ge0$ be cluster $\ell$'s contribution to cohort $g$ at event time $e$, $A_{\ell+}^e=\sum_hA_{\ell h}^e$, $\mu_g^e=\mathbb E(A_{\ell g}^e)$, and $\mu_+^e=\sum_h\mu_h^e>0$. Then $\omega_g^e=\mu_g^e/\mu_+^e$ and
\[
\widehat\omega_g^e=\frac{\bar A_g^e}{\bar A_+^e},\qquad
\bar A_g^e=L^{-1}\sum_{\ell=1}^LA_{\ell g}^e.
\]
Assume a joint cluster-level CLT for $(A_{\ell g}^e)_{g\in\mathcal G_e}$ and a law of large numbers for $A_{\ell+}^e$. A first-order expansion of the ratio map $r_g(a)=a_g/(\sum_h a_h)$ around $\mu=(\mu_h)_h$ gives
\[
\sqrt L(\widehat\omega_g^e-\omega_g^e)
=L^{-1/2}\sum_{\ell=1}^L\xi_{\ell,g}^e+o_p(1),
\qquad
\xi_{\ell,g}^e=\frac{A_{\ell g}^e-\omega_g^eA_{\ell+}^e}{\mu_+^e}.
\]
Indeed, $\partial r_g/\partial a_h=(1\{g=h\}-\omega_g^e)/\mu_+^e$, so the displayed influence function follows by the multivariate delta method. Moreover,
\[
\sum_g\xi_{\ell,g}^e
=\frac{A_{\ell+}^e-A_{\ell+}^e\sum_g\omega_g^e}{\mu_+^e}=0.
\]

For independent but non-identically distributed cluster contributions, the same argument has a triangular-array form. Write
\[
\mu_{g,L}^e=L^{-1}\sum_{\ell=1}^L\mathbb E A_{\ell g,L}^e,
\qquad
\mu_{+,L}^e=\sum_g\mu_{g,L}^e,
\qquad
\omega_{g,L}^e=\mu_{g,L}^e/\mu_{+,L}^e,
\]
and suppose $\inf_L\mu_{+,L}^e>0$. If the denominator obeys the required LLN and the centered cluster array obeys the joint Lindeberg--Feller CLT, then
\[
\sqrt L(\widehat\omega_g^e-\omega_{g,L}^e)
=L^{-1/2}\sum_{\ell=1}^L\xi_{\ell,g,L}^e+o_p(1),
\]
with
\[
\xi_{\ell,g,L}^e=
\frac{
A_{\ell g,L}^e-\mathbb E A_{\ell g,L}^e
-\omega_{g,L}^e\{A_{\ell+,L}^e-\mathbb E A_{\ell+,L}^e\}
}{\mu_{+,L}^e}.
\]
The adding-up identity remains exact cluster by cluster:
\[
\sum_g\xi_{\ell,g,L}^e=0.
\]
If the proposition is stated around a fixed limit $\omega_g^e$, additionally require $\sqrt L(\omega_{g,L}^e-\omega_g^e)\to0$; otherwise $\omega_{g,L}^e$ is naturally interpreted as the triangular-array target. Under identical cluster distributions this centered array reduces to the preceding score because $\mathbb E A_{\ell g}^e=\omega_g^e\mu_+^e$.

No independence between the weight scores and the cohort-state CDF scores is assumed: both are functions of the same cluster history, and their covariance is part of the joint CLT required in Proposition~2. For unequal survey weights, $A_{\ell g}^e$ is the corresponding cluster-level weighted exposure contribution, so the same ratio expansion applies provided the required second moments and Lindeberg condition hold. $\square$

\subsection{Influence-function inference for the average-versus-mixture aggregation gap}

We prove Proposition~2. The component influence expansions are standard inverse-map ingredients; the object of interest here is their joint covariance and difference, which delivers inference on the average-versus-mixture aggregation gap. Fix \(e\) and \(\tau\), suppress the event-time superscript on the weights, and write \(F_{g,d}\) for \(F_{g,g+e}^d\). The assumed uniform asymptotic-linear representation, local stochastic equicontinuity, and positivity of \(f_{g,d}(q_{g,d})\) imply the standard inverse-map expansion
\[
\sqrt L(\widehat q_{g,d}-q_{g,d})
=-\frac{L^{-1/2}\sum_{\ell=1}^L\psi_{\ell,g,d}(q_{g,d})}{f_{g,d}(q_{g,d})}+o_p(1)
=L^{-1/2}\sum_{\ell=1}^L\chi_{\ell,g,d}+o_p(1).
\]
Because the number of contributing cohorts is fixed, products of the \(O_p(L^{-1/2})\) weight and quantile estimation errors are \(o_p(L^{-1/2})\). Hence
\begin{align*}
\sqrt L(\widehat\QTT_{\mathrm{avg}}^e-\QTT_{\mathrm{avg}}^e)
={}&
\sum_g\omega_g\sqrt L\{(\widehat q_{g,1}-q_{g,1})-(\widehat q_{g,0}-q_{g,0})\}\\
&+\sum_g\sqrt L(\widehat\omega_g-\omega_g)(q_{g,1}-q_{g,0})+o_p(1)\\
={}&L^{-1/2}\sum_{\ell=1}^L\Gamma_{\ell}^{\mathrm{avg}}+o_p(1).
\end{align*}

For the mixture object, define \(F_d^e(y)=\sum_g\omega_gF_{g,d}(y)\) and \(\widehat F_d^e(y)=\sum_g\widehat\omega_g\widehat F_{g,d}(y)\). Uniformly on a neighborhood of \(q_d^{\mathrm{mix}}\),
\begin{align*}
\sqrt L\{\widehat F_d^e(y)-F_d^e(y)\}
={}&
\sum_g\omega_g\sqrt L\{\widehat F_{g,d}(y)-F_{g,d}(y)\}\\
&+\sum_g\sqrt L(\widehat\omega_g-\omega_g)F_{g,d}(y)+o_p(1),
\end{align*}
where the cross-products are again \(o_p(1)\) because \(|\mathcal G_e|\) is fixed. Local stochastic equicontinuity permits evaluation of the leading empirical process at \(q_d^{\mathrm{mix}}\) after replacing the estimated inverse argument by its population value. Applying the inverse-map expansion to the mixture CDF then gives
\[
\sqrt L(\widehat q_d^{\mathrm{mix}}-q_d^{\mathrm{mix}})
=L^{-1/2}\sum_{\ell=1}^L\chi_{\ell,d}^{\mathrm{mix}}+o_p(1),
\]
because
\[
\{F_d^e\}'(q_d^{\mathrm{mix}})
=\sum_g\omega_g f_{g,d}(q_d^{\mathrm{mix}})
=f_d^e(q_d^{\mathrm{mix}})>0.
\]
Subtracting the \(d=0\) expansion from the \(d=1\) expansion yields
\[
\sqrt L(\widehat\QTT_{\mathrm{mix}}^e-\QTT_{\mathrm{mix}}^e)
=L^{-1/2}\sum_{\ell=1}^L\Gamma_{\ell}^{\mathrm{mix}}+o_p(1).
\]
The aggregation-gap expansion follows by subtraction.

The assumed joint CLT for the finite vector of primitive score evaluations and weight scores, followed by a linear map, gives joint asymptotic normality of the three displayed estimators. The resulting three-dimensional covariance is necessarily singular because the gap influence function is identically the average influence function minus the mixture influence function; this linear dependence is harmless because neither the CLT nor the multiplier approximation requires covariance inversion. For the multiplier statement, write \(\Gamma_\ell=(\Gamma_\ell^{\mathrm{avg}},\Gamma_\ell^{\mathrm{mix}},\Gamma_\ell^{\mathrm{gap}})'\) and impose the additional conditions stated in Proposition~2:
\[
L^{-1}\sum_{\ell=1}^L\Gamma_\ell\Gamma_\ell'\to_p V,\qquad
\max_{\ell\le L}\|\Gamma_\ell\|/\sqrt L\to_p0,\qquad
L^{-1}\sum_{\ell=1}^L\|\widehat\Gamma_{\ell}-\Gamma_{\ell}\|^2=o_p(1).
\]
Let \(\widetilde\Gamma_\ell=\widehat\Gamma_\ell-L^{-1}\sum_{k=1}^L\widehat\Gamma_k\). The empirical-\(L_2\) condition and Cauchy--Schwarz imply
\[
L^{-1}\sum_{\ell=1}^L
\widetilde\Gamma_\ell\widetilde\Gamma_\ell'
\to_p V
\quad\text{and}\quad
\max_{\ell\le L}\|\widetilde\Gamma_\ell\|/\sqrt L\to_p0.
\]
Hence, conditional on the data, the finite-dimensional multiplier central limit theorem applies to
\[
L^{-1/2}\sum_{\ell=1}^L\zeta_\ell\widetilde\Gamma_\ell,
\]
for independent multipliers with mean zero, variance one, and a finite \(2+\kappa\) moment for some \(\kappa>0\). Its conditional law converges to \(N(0,V)\), the same limit as the original influence sum. Stacking a fixed finite number of quantile-indexed influence vectors simply enlarges the finite vector and gives the stated simultaneous extension. \(\square\)

\paragraph{Proof of Proposition 3: precision and target risk.}
For part (i), the influence identity in Proposition~2 gives $Z_{\mathrm{avg}}=Z_{\mathrm{mix}}+Z_{\mathrm{gap}}$. Taking variances yields
\[
\sigma_a^2
=\sigma_m^2+\sigma_\Delta^2+2\sigma_{m\Delta},
\]
which is the variance identity stated in Proposition 3. No sign restriction on $\sigma_{m\Delta}$ is imposed by Proposition~2, so the identity alone cannot produce a universal ranking.

Part (ii) gives constructive smooth examples showing that this indeterminacy is substantive rather than formal. Consider independent repeated cross sections, a fixed event time, two equally weighted cohorts, and total sample size $n$ in each potential-outcome state, split equally across cohorts. The two states are sampled independently, so the QTT variance is the sum of the corresponding state-quantile variances. It is enough to compare one state.

First suppose the two cohort distributions share the same median $q$ but have positive densities $f_1(q)\ne f_2(q)$ there. The average of the two cohort sample medians and the median of the equal-weight mixture estimate the same population median. Standard quantile linearization gives
\[
\operatorname{AVar}\{\sqrt n(\widehat q_{\mathrm{avg}}-q)\}
=\frac18\left\{f_1(q)^{-2}+f_2(q)^{-2}\right\},
\]
whereas stratified-mixture inversion gives
\[
\operatorname{AVar}\{\sqrt n(\widehat q_{\mathrm{mix}}-q)\}
=\frac{1}{\{f_1(q)+f_2(q)\}^2}.
\]
Convexity of $x\mapsto x^{-2}$, or direct Cauchy--Schwarz, implies that the latter is weakly smaller, with strict inequality when the densities differ. Taking, for example, centered Gaussian cohorts with unequal scales gives a smooth strict example. Applying a common location shift in the treated state makes both QTT aggregands equal the same treatment shift while preserving the strict mixture-precision advantage.

For the reverse ordering, let the two equally weighted cohort distributions in state $d$ be $N(\mu_d-a,1)$ and $N(\mu_d+a,1)$ and take the median. Symmetry implies both the average cohort median and the mixture median equal $\mu_d$. The average-median variance is
\[
\operatorname{AVar}\{\sqrt n(\widehat q_{\mathrm{avg}}-\mu_d)\}
=\frac{1}{4\phi(0)^2}.
\]
At the mixture median $\mu_d$, put $p=\Phi(a)$. The stratified empirical-mixture CDF has scaled variance $p(1-p)$ and mixture density $\phi(a)$, so
\[
\operatorname{AVar}\{\sqrt n(\widehat q_{\mathrm{mix}}-\mu_d)\}
=\frac{p(1-p)}{\phi(a)^2}.
\]
At $a=2$, these values are approximately $1.57$ and $7.63$, respectively, so average-cohort inversion is strictly more precise. Again, applying the same location shift between untreated and treated states makes the two QTT aggregands coincide while the strict ordering is inherited by the QTT estimators. This proves that both directions occur on the exact-equality class.

For part (iii), write $\theta_{a,L}$ and $\theta_{m,L}$ for the two population aggregands and $\Delta_L=\theta_{a,L}-\theta_{m,L}$. Consistency and $\Delta_L\to\Delta\ne0$ give
\[
\widehat\theta_{m,L}-\theta_{a,L}
=(\widehat\theta_{m,L}-\theta_{m,L})-\Delta_L\overset{p}\longrightarrow-\Delta,
\]
while $\widehat\theta_{a,L}-\theta_{a,L}=O_p(L^{-1/2})$. Under the stated uniform-integrability condition this convergence transfers to second moments, giving the fixed-mismatch claim.

Under $\sqrt L\Delta_L\to\delta$,
\[
\sqrt L(\widehat\theta_{m,L}-\theta_{a,L})
=\sqrt L(\widehat\theta_{m,L}-\theta_{m,L})-\sqrt L\Delta_L
\Rightarrow Z_{\mathrm{mix}}-\delta.
\]
The asymptotic linear representation is centered and the uniform-integrability assumption transfers weak convergence to second moments, so
\[
L\mathbb{E}(\widehat\theta_{m,L}-\theta_{a,L})^2\to\mathbb{E}(Z_{\mathrm{mix}}-\delta)^2=\sigma_m^2+\delta^2.
\]
The correctly targeted estimator similarly has limiting scaled risk $\sigma_a^2$. Comparing the two limits gives the stated criterion. Interchanging $a$ and $m$ proves the symmetric case. $\square$

\paragraph{Proof of the uniform aggregation-gap process corollary.}
Let $\mathcal T=[\underline\tau,\overline\tau]$ and collect the finite family of cohort-state CDFs in $F=(F_{g,d})_{g,d}$ and the cohort weights in $\omega$. For state $d$, the finite-mixture map is
\[
\mathcal M_d(F,\omega)(y)=\sum_g\omega_gF_{g,d}(y).
\]
As a map on the product sup-norm space, $\mathcal M_d$ is continuously Hadamard differentiable. In direction $(h,\eta)$ satisfying $\sum_g\eta_g=0$, its derivative is
\[
\dot{\mathcal M}_d[h,\eta](y)
=\sum_g\omega_gh_{g,d}(y)+\sum_g\eta_gF_{g,d}(y).
\]
Under the corollary's uniform positivity and continuity assumptions, the inverse-CDF map restricted to the compact interior probability interval $\mathcal T$ is Hadamard differentiable tangentially to $C(I)$, where $I$ is a compact outcome interval containing $\{Q_G(\tau):\tau\in\mathcal T\}$ in its interior and the domain carries the sup norm. In particular, the limiting perturbation must have a continuous version on $I$, not merely pointwise continuity at a finite set of quantiles. Its derivative at a CDF $G$ with density $g$ is
\[
\dot Q_G[h](\tau)=-\frac{h\{Q_G(\tau)\}}{g\{Q_G(\tau)\}},
\qquad \tau\in\mathcal T,
\]
uniformly in $\tau$. The same formula applies cohort by cohort and to the mixture CDF $F_d^e=\mathcal M_d(F,\omega)$.

Composing these maps, the derivative of the average-cohort QTT process is
\[
\sum_g\omega_g\left[-\frac{h_{g,1}(q_{g,1})}{f_{g,1}(q_{g,1})}
+\frac{h_{g,0}(q_{g,0})}{f_{g,0}(q_{g,0})}\right]
+\sum_g\eta_g(q_{g,1}-q_{g,0}),
\]
where all quantile arguments depend on $\tau$. The derivative of the mixture-QTT process is the difference across $d=1,0$ of
\[
-\frac{\sum_g\omega_gh_{g,d}(q_d^{\mathrm{mix}})
+\sum_g\eta_gF_{g,d}(q_d^{\mathrm{mix}})}
{f_d^e(q_d^{\mathrm{mix}})}.
\]
Subtracting the two derivatives gives exactly the quantile-indexed influence function $\Gamma_\ell^{\mathrm{gap}}(\tau)$ in Proposition~2.

By assumption, the primitive CDF-score/weight process converges weakly in the product sup-norm space to a tight Gaussian element whose sample paths lie in the continuous tangent space needed by the inverse map. The functional delta method therefore yields
\[
\sqrt L(\widehat\Delta_{\mathrm{agg}}^e-\Delta_{\mathrm{agg}}^e)
\Rightarrow\mathbb G_\Delta^e
\quad\text{in }\ell^\infty(\mathcal T).
\]
The assumed conditional multiplier-process convergence, uniform empirical-$L_2$ consistency of the estimated influence functions, and the bootstrap delta method give the conditional analogue. If $s_\Delta(\tau)$ is continuous and bounded away from zero and its estimator is uniformly consistent, studentization is a continuous map on $\ell^\infty(\mathcal T)$. At continuity points of the CDF of the limiting supremum---in particular at the $(1-\gamma)$ quantile imposed in the corollary---conditional weak convergence implies consistency of the multiplier critical value, and the continuous mapping theorem yields the desired simultaneous coverage. This proof requires smooth positive densities over the compact quantile range; it does not cover atoms, for which the CDF-projection corollary is deliberately used instead. $\square$

\paragraph{Post-gate aggregation-discrepancy inference (supporting result).}
Proposition~2 gives
\[
\sqrt L\{\widehat\Delta_{\mathrm{agg}}^e(\tau)-\Delta_{\mathrm{agg}}^e(\tau)\}
=L^{-1/2}\sum_{\ell=1}^L\Gamma_{\ell}^{\mathrm{gap}}(\tau)+o_p(1).
\]
For the assumption-lean route, apply Diagnostic Result D4(i) with $\beta^r=\Delta_{\mathrm{agg}}^e(\tau)$ to any valid unconditional interval (or, on a finite quantile grid, to a valid unconditional simultaneous band). This is only the event-probability inequality and requires no additional Gaussian approximation.

For fixed-dimensional Gaussian selective inference, stack the asymptotically linear placebo vector with the aggregation-gap score. Under the complete placebo null and the joint covariance conditions in Diagnostic Result D4(ii), the resulting target--diagnostic vector has exactly the joint Gaussian limit required there, so inversion of the conditional pivot gives the stated conditional coverage. For a growing diagnostic family, include the aggregation-gap score as the scalar target coordinate in the appendix growing-family selective template; the result then follows only if its joint hyperrectangle, multiplier, feasible-score, and critical-value conditions all hold. As emphasized in the main text, those primitive growing-family conditions are not claimed to have been verified for the CPS cluster array.

For several prespecified quantiles, either begin from an unconditional joint band and use the branch bound or apply a prespecified conditional multiplicity correction to scalar selective confidence sets. Nothing in this composition adjusts for choosing the identifying route, event time, quantile, aggregation definition, or diagnostic family after observing the data. This proves the supporting post-gate statement and its scope qualification. $\square$

\paragraph{Numerical derivative audit for Proposition~2.}
As a source-level check on the estimated-weight terms, we independently perturb a four-cohort Gaussian system in cohort means, scales, and normalized cohort weights and compare central finite-difference derivatives of \(\QTT_{\mathrm{avg}}\), \(\QTT_{\mathrm{mix}}\), and their gap with the analytical influence formulas above. Across 2,000 random perturbation configurations, the largest absolute discrepancies are below \(2\times10^{-9}\) for all three derivatives. This is an algebra audit, not a Monte Carlo coverage claim; the checker and its output are included in the replication package.

\section{Supplementary Design and Interpretation Details}

\paragraph{Treatment paths and anticipation.}
With staggered adoption, potential outcomes are indexed by first treatment date rather than by a single binary state. Consistency/no interference maps observed outcomes to the realized treatment path. If anticipation is plausible, the baseline window used by the diagnostic must precede the anticipation horizon; the solution is to redefine the design window rather than reinterpret an affected lead as untreated.

\paragraph{Comparison populations and repeated cross-sections.}
Never-treated and not-yet-treated units are both legitimate comparison candidates only when untreated at every date entering the relevant contrast. Under repeated cross-sections, period-specific samples must represent a stable population or be standardized to an explicitly chosen target covariate distribution. Treatment-induced employment, migration, survey participation, or other selection can change the target population and therefore requires an explicit selection argument.

\paragraph{Marginal versus joint treatment-effect objects.}
Identification of \((F_{g,t}^{1},F_{g,t}^{0})\) is sufficient for marginal distributional functionals such as threshold probabilities, QTTs, inequality indices, or expected shortfall. The distribution of individual treatment effects requires the joint distribution of the two potential outcomes and hence additional dependence restrictions.

\paragraph{Mass points, tails, and transformations.}
Economically meaningful mass points should not be smoothed away merely to invoke density-based quantile asymptotics. The mass-point-safe corollary in the main text instead transfers simultaneous CDF-band coverage through generalized inversion and Minkowski differences, following the generic discrete-outcome construction of Chernozhukov, Fern\'andez-Val, Melly, and W\"uthrich (2020). For staggered adoption it also projects a jointly valid cohort-weight region, so average-QTT, mixture-QTT, and aggregation-gap bands remain valid when population cohort shares are estimated. Thin-tail QTTs should not be default outputs when support is weak. Additive CDF-PT is generally scale dependent, whereas CIC is invariant to strictly increasing transformations (Athey and Imbens, 2006); outcome-scale sensitivity is therefore identification sensitivity.

\paragraph{Operational routing at atoms and weak-density quantiles.}
The main text proposes an ex ante conservative routing screen rather than a post hoc choice between two interval procedures. Let $\widehat p_{\max,L}$ be the largest empirical point mass over the finitely many cohort--state distributions and prespecified outcome neighborhoods needed for the target range, and choose deterministic $b_L\downarrow0$ with $\sqrt L b_L\to\infty$. Under an atomless continuously distributed observed outcome, exact point masses are zero in population and, absent measurement rounding, $\widehat p_{\max,L}=o_p(b_L)$ provided the largest normalized observation weight is $o_p(b_L)$; this holds, for example, with uniformly bounded observation weights and total sample size proportional to $L$. If instead some support point has fixed population mass $p_0>0$, consistency of its empirical mass gives $\widehat p_{\max,L}\to_p p_0$, hence $\Pr(\widehat p_{\max,L}>b_L)\to1$. Known rounding or heaping is conservatively treated as discreteness without relying on the exact-mass asymptotics.

For the density screen, fix $m_*>0$ and a finite prespecified collection of bandwidth rules. If every relevant population density is bounded below by $m_*+\varepsilon$ and each corresponding density estimator is uniformly consistent on the target neighborhoods, the minimum estimated density across cohorts, states, neighborhoods, and the finite bandwidth collection exceeds $m_*$ with probability tending to one. Conversely, if a relevant smooth density is separated below $m_*$, the screen fails with probability tending to one. Combining the two screens therefore makes the routing decision asymptotically deterministic under separated regular and fixed-atom regimes. This argument is pointwise in those separated regimes. It does not provide uniform smooth-quantile validity for vanishing atoms, shrinking density floors, measurement processes whose rounding vanishes with sample size, or other local-to-nonregular sequences. Those cases are routed to the CDF-projection construction by design.

\paragraph{Simulation and bootstrap draw-count map.}
The numerical counts differ because the exercises have different purposes; they are not tuned to obtain favorable results. The following table makes the mapping explicit.
\begin{center}
\small
\begin{tabular}{p{6.8cm}r p{5.0cm}}
\toprule
Exercise & Draws & Purpose\\
\midrule
Design 8 wild-cluster gate, per outer replication & 499 & Finite-cluster size/power stress\\
\texttt{mpdta} pre-treatment max-$|t|$ compatibility calculation & 49,999 & Stable global reference $p$-value\\
\texttt{mpdta} joint CDF projection band & 29,999 & Simultaneous post-treatment CDF band\\
CPS baseline state-history pairs bootstrap & 499 & Frozen primary familywise gate\\
CPS wild-cluster-$t$ sensitivity & 9,999 & Familywise sensitivity to resampling/studentization\\
CPS Romano--Wolf localization & 19,999 & Lower Monte Carlo noise in adjusted coordinate $p$-values\\
NSW placebo label permutation & 49,999 & Randomization-reference gate\\
NSW placebo Rademacher multiplier & 49,999 & Non-exchangeability-robust asymptotic reference\\
NSW threshold-scale validation-gap multiplier & 49,999 & Simultaneous CDF-scale comparison\\
NSW stratified pairs bootstrap & 4,999 & Descriptive QTT sampling-variability ranges\\
\bottomrule
\end{tabular}
\end{center}

\paragraph{Compact reporting checklist.}
The main text compresses the practitioner workflow to save space. For implementation, the following checklist records the minimum information needed to make a distributional DiD analysis auditable.
\begin{center}
\small
\begin{tabular}{p{3.2cm}p{10.0cm}}
\toprule
Item & Minimum information to report\\
\midrule
Target & Estimand, target population, cohort/event time, and exact definition of the missing \(F_{g,t}^{0}\)\\
Identification & Named nonlinear route, comparison design, anticipation restriction, overlap/support conditions, and covariates\\
Diagnostics & Frozen placebo family, multiplicity rule, comparison support, composition, and mass points\\
Estimator/inference & Construction and any standardization of \(\widehat F_{g,t}^{0}\), sampling unit, resampling level, and post-gate adjustment\\
Aggregation & Contributing cohorts, event-time weights, average-QTT versus mixture-QTT target, exact plug-in gap, and (when smooth density estimation is credible) the tilt--dispersion diagnostic\\
Sensitivity/limits & Independently defensible alternative routes, scale dependence, and causal objects not identified\\
\bottomrule
\end{tabular}
\end{center}

\section{Identification Map and First-Stage Scope}
\label{app:identificationmap}
The main text deliberately compresses first-stage identification so that the aggregation theorem appears early. For reference, the maintained routes used or discussed in the empirical workflow are summarized here.
\begin{center}
\small
\begin{tabular}{p{2.8cm}p{4.3cm}p{6.0cm}}
\toprule
Route & Identifying restriction & Main implication for practice\\
\midrule
CDF-PT & Additive parallel changes of untreated CDFs & Threshold-scale restriction; the implied population counterfactual must itself be a proper CDF; standardize conditional CDFs before inversion.\\
CIC & Stable latent ranks under monotone time maps & Requires transport support; invariant to strictly increasing transformations of the outcome.\\
Changes + dependence & Untreated-change law plus a dependence restriction & Panel dependence carries identifying information and is part of the maintained causal model.\\
Distribution regression & No interaction on a chosen probability-link scale & Identification depends on the link specification; marginal targets require CDF standardization.\\
Partial identification & Set of admissible counterfactual CDFs & Report bounds or identified sets rather than unsupported point QTTs.\\
\bottomrule
\end{tabular}
\end{center}
The routes are generally non-nested. Their common role in this paper is only to recover the cohort--state marginal potential-outcome distributions that feed the aggregation operators. Comparison-set construction, overlap, anticipation, composition, repeated-cross-section standardization, mass points, and any route-specific dependence assumptions remain first-stage requirements and are not altered by Proposition~1.

\section{Local-Power Calibration}

To make Diagnostic Result D3 quantitative, we use the same 99-coordinate correlation structure as the stop-rule Monte Carlo:
\[
R
=
R_e\otimes R_\tau,
\qquad
(R_e)_{kk'}=0.65^{|k-k'|},
\qquad
(R_\tau)_{\ell\ell'}=0.75^{|\ell-\ell'|}.
\]
Using 500,000 null Gaussian draws gives a 5\% max-\(|t|\) critical value of 3.383. Power calculations use 150,000 draws per drift pattern. A localized drift shifts only \((e,\tau)=(-8,0.8)\); a diffuse drift shifts the upper three quantiles at \(e=-10,\ldots,-6\), for 15 coordinates total.

\begin{table}[htbp]\centering
\caption{Asymptotic local-power calibration for the 99-coordinate max-$|t|$ rule}\label{tab:localpower}
\small\begin{tabular}{rrr}\toprule
Noncentrality & Localized & Diffuse\\\midrule
0 & 0.050 & 0.051\\
1 & 0.057 & 0.112\\
2 & 0.126 & 0.453\\
3 & 0.381 & 0.881\\
4 & 0.747 & 0.995\\
\bottomrule\end{tabular}\end{table}

\begin{figure}[htbp]
\centering
\includegraphics[width=.72\textwidth]{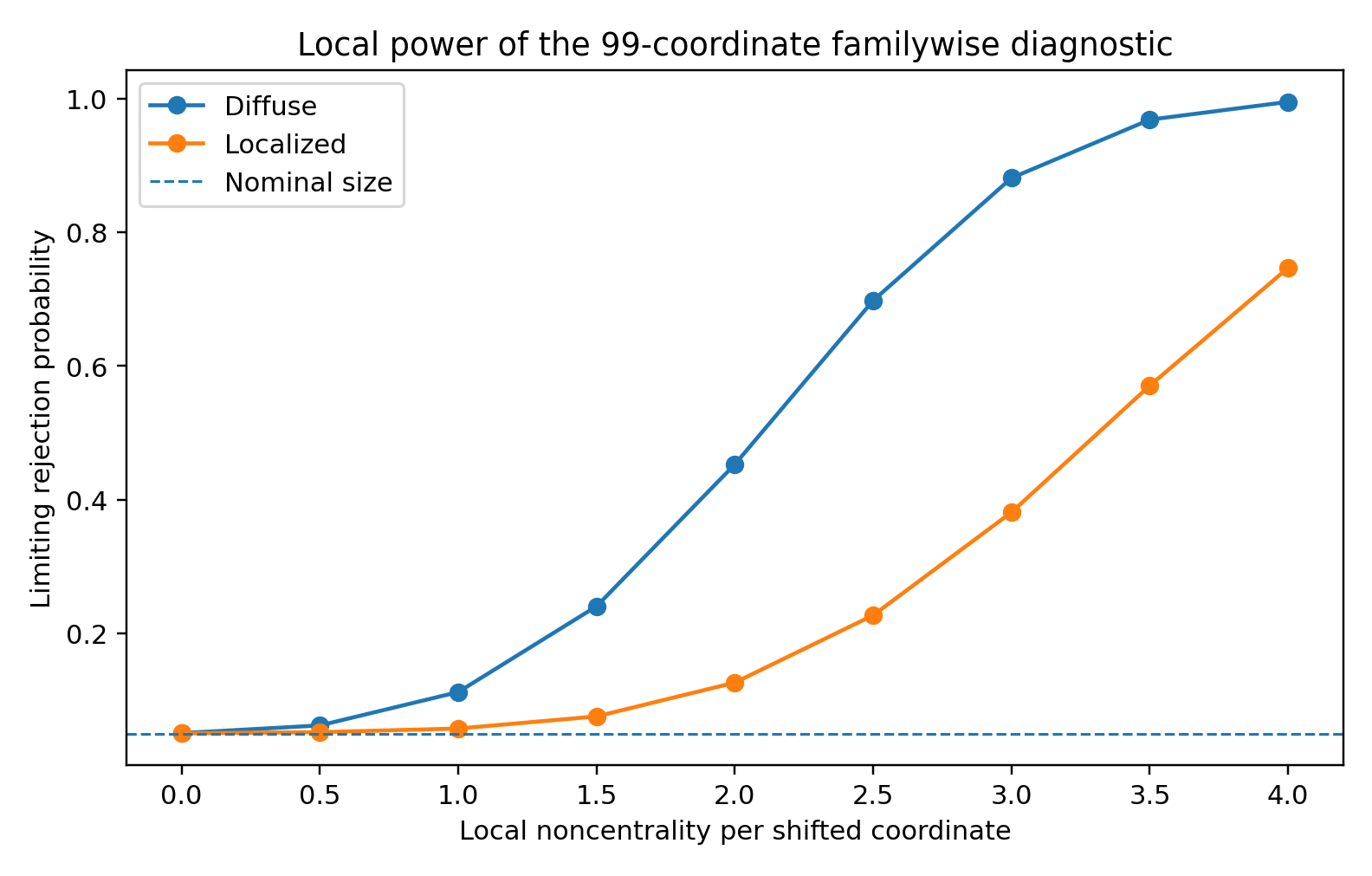}
\caption{Limiting power under \(L^{-1/2}\)-local violations. The same standardized drift applied to several neighboring coordinates is more detectable in this calibration because several correlated components of the max statistic move simultaneously. This is a calibration result, not a universal ordering across covariance structures.}
\label{fig:localpower}
\end{figure}

The finite-\(L\) Design-8 results are directionally consistent with the Gaussian limit but remain below it at \(L=30\), particularly for larger localized shifts. For example, per-cluster score means \(0.5,0.7,0.9\) imply Gaussian noncentralities \(\sqrt{30}\delta=2.74,3.83,4.93\); the observed finite-sample rejection rates are \(0.223,0.552,0.854\), compared with limiting powers \(0.293,0.692,0.945\). This gap is one reason the paper reports direct finite-cluster simulation rather than relying on the limiting power function alone.

\section{Comparison with a Quadratic Omnibus Diagnostic}

The max-\(|t|\) statistic is not the only way to aggregate the frozen placebo family. As a transparent benchmark, define
\[
Q=\sum_{j=1}^{99} t_j^2.
\]
We calibrate its 5\% critical value from the same correlated Gaussian null used for Figure~\ref{fig:localpower}; hence this is not a chi-square approximation that ignores dependence. The null simulation uses 400,000 Gaussian draws and each power calculation uses 120,000 draws. The resulting 5\% critical values are 3.384 for max-\(|t|\) and 166.878 for \(Q\).

\begin{table}[H]
\centering
\caption{Local-power comparison of max-\(|t|\) and quadratic diagnostics}
\label{tab:omnibuscompare}
\small
\begin{tabular}{lrrr}
\toprule
Pattern & Noncentrality & Max-\(|t|\) & Quadratic\\
\midrule
Localized & 0 & 0.050 & 0.050\\
Localized & 1 & 0.057 & 0.051\\
Localized & 2 & 0.127 & 0.057\\
Localized & 3 & 0.383 & 0.070\\
Localized & 4 & 0.746 & 0.087\\
Diffuse & 0 & 0.051 & 0.050\\
Diffuse & 1 & 0.113 & 0.102\\
Diffuse & 2 & 0.454 & 0.372\\
Diffuse & 3 & 0.880 & 0.843\\
Diffuse & 4 & 0.994 & 0.995\\
\bottomrule
\end{tabular}
\end{table}

In this calibration, max-\(|t|\) is substantially more sensitive to a single-coordinate violation. It also has higher power for the reported moderate diffuse alternatives, although the difference disappears for the largest diffuse drift. The table should not be read as a general optimality result. Quadratic procedures can be preferable for other covariance matrices and other dense local directions. The comparison instead clarifies why max-\(|t|\) is well matched to the paper's operational question: whether any prespecified identifying implication is sufficiently contradicted to trigger the stop rule.

\section{Prospectively Specified Monte Carlo Suite}

The simulation suite is prospectively specified. The complete machine-readable specification and master executable program accompany the replication files. All DGP parameters, sample sizes, numerical grids, replication counts, bootstrap counts, and random seeds are fixed and reported below.

\begin{table}[htbp]
\centering
\caption{Prospectively specified Monte Carlo evidence}
\label{tab:prospectivemc}
\scriptsize
\begin{tabular}{c p{2.5cm} p{6.1cm} p{3.1cm}}
\toprule
Design & Maintained setting & Representative prospective result & Lesson\\
\midrule
2 & Exact additive CDF PT & At $\tau=.50$, truth 0.350; mean estimate 0.347; bias -0.003. & CDF-DiD works under its own restriction.\\
3 & CIC correct, CDF PT false & At $\tau=.90$, CIC mean 0.249 for truth .250; CDF-PT mean 0.848. & Routes are not interchangeable.\\
4 & Dependence matters & Wrong-independence pseudo-truth is -0.285/0.885 at $\tau=.10/.90$; means -0.287/0.880. & Marginals do not determine change-based counterfactuals.\\
5 & Conditional identification & At $\tau=.50$, standardized mean 0.774 for truth 0.776; unadjusted mean 1.074. & Standardize conditional CDFs before inversion.\\
6 & Staggered aggregation & At $e=0$, average-minus-mixture truth is 0.464 at $\tau=.10$ and -0.608 at $\tau=.90$. & Aggregation gap has no universal sign.\\
7 & State-history inference & With 1,000 replications, family coverage is 0.946, 0.954, 0.949 for 30, 40, 50 clusters. & Resample at the assignment level.\\
8 & 99-coordinate diagnostic gate & Null rejection is 0.052, 0.051, 0.044 for 20, 30, 50 clusters. & The frozen max-$|t|$ gate controls size in the stress design.\\
\bottomrule
\end{tabular}
\begin{flushleft}\footnotesize
Notes: Designs 2--6 use 1,000 outer replications. Design 7 uses a 1,000-replication precision extension and 199 whole-history bootstrap draws per replication. Design 8 uses the frozen fully studentized wild-cluster implementation for the 99-coordinate family. DGP parameters, sample sizes, numerical grids, and random seeds are recorded with the replication files.
\end{flushleft}
\end{table}

\paragraph{Design 2: exact additive CDF parallel trends.}
Let \(F_A=N(0,1)\) and \(F_B=N(1.5,0.8^2)\). Control mixture weights are \((.65,.35)\) at baseline and \((.50,.50)\) post; treated untreated weights are \((.40,.60)\) and \((.25,.75)\). Hence both groups transfer exactly .15 probability mass from \(A\) to \(B\), so
\[
F_{T,1}^0-F_{T,0}=F_{C,1}-F_{C,0}.
\]
Treatment adds 0.35 to the treated post outcome. Each group-period cell contains 800 observations, \(R=1000\), seed 20269502, and CDF inversion uses a frozen 1,601-point grid on \([-4,6]\).

\begin{tabular}{lrrrrrrr}
\toprule
estimand & tau & truth & mean\_estimate & bias & mc\_sd & rmse & mc\_se\_mean \\
\midrule
CDFPT\_QTT & 0.100 & 0.350 & 0.349 & -0.001 & 0.219 & 0.219 & 0.007 \\
CDFPT\_QTT & 0.250 & 0.350 & 0.351 & 0.001 & 0.127 & 0.127 & 0.004 \\
CDFPT\_QTT & 0.500 & 0.350 & 0.347 & -0.003 & 0.084 & 0.084 & 0.003 \\
CDFPT\_QTT & 0.750 & 0.350 & 0.348 & -0.002 & 0.078 & 0.078 & 0.002 \\
CDFPT\_QTT & 0.900 & 0.350 & 0.347 & -0.003 & 0.093 & 0.093 & 0.003 \\
\bottomrule
\end{tabular}

\paragraph{Design 3: CIC correct, additive CDF-PT false.}
For treated units \(U_T\sim N(.4,1)\); for comparison units \(U_C\sim N(-.3,.8^2)\). Untreated outcomes obey
\[
Y_0=h_0(U)=U,\qquad
Y_1(0)=h_1(U)=\exp(.18+.45U),
\]
and treatment adds 0.25. Stable within-group latent ranks make CIC correct, while the nonlinear transformation and different latent distributions invalidate additive CDF-PT. Each group-period cell contains 1,000 observations, \(R=1000\), seed 20269503.

\begin{tabular}{lrrrrrrr}
\toprule
estimand & tau & truth & mean\_estimate & bias & mc\_sd & rmse & mc\_se\_mean \\
\midrule
CIC\_QTT & 0.100 & 0.250 & 0.251 & 0.001 & 0.033 & 0.033 & 0.001 \\
CIC\_QTT & 0.250 & 0.250 & 0.251 & 0.001 & 0.037 & 0.037 & 0.001 \\
CIC\_QTT & 0.500 & 0.250 & 0.251 & 0.001 & 0.051 & 0.051 & 0.002 \\
CIC\_QTT & 0.750 & 0.250 & 0.252 & 0.002 & 0.088 & 0.088 & 0.003 \\
CIC\_QTT & 0.900 & 0.250 & 0.249 & -0.001 & 0.201 & 0.201 & 0.006 \\
CDFPT\_QTT & 0.100 & 0.250 & 0.149 & -0.101 & 0.026 & 0.104 & 0.001 \\
CDFPT\_QTT & 0.250 & 0.250 & 0.288 & 0.038 & 0.027 & 0.047 & 0.001 \\
CDFPT\_QTT & 0.500 & 0.250 & 0.449 & 0.199 & 0.033 & 0.201 & 0.001 \\
CDFPT\_QTT & 0.750 & 0.250 & 0.635 & 0.385 & 0.048 & 0.388 & 0.002 \\
CDFPT\_QTT & 0.900 & 0.250 & 0.848 & 0.598 & 0.074 & 0.602 & 0.002 \\
\bottomrule
\end{tabular}

\paragraph{Design 4: dependence is identifying information.}
The DGP is
\[
Y_0(0)\sim N(0,1),\qquad
\Delta Y(0)\sim N(.2,1),
\]
with baseline/change correlations .75 for treated units and \(-.50\) for comparison units. Treatment adds .30. The correct treated untreated post distribution therefore has variance 3.5, whereas an estimator that combines the treated-baseline and comparison-change marginals as independent uses variance 2. Its population pseudo-truth is
\[
QTT_{\rm wrong}(\tau)=.30+\{\sqrt{3.5}-\sqrt2\}\Phi^{-1}(\tau).
\]
The finite-sample experiment uses 1,000 treated and 1,000 comparison units, \(R=1000\), seed 20269504.

\noindent\resizebox{\linewidth}{!}{\begin{tabular}{lrrrrrrr}
\toprule
estimand & tau & truth & mean\_estimate & bias & mc\_sd & rmse & mc\_se\_mean \\
\midrule
Oracle\_QTT & 0.100 & 0.300 & 0.300 & -0.000 & 0.000 & 0.000 & 0.000 \\
Oracle\_QTT & 0.250 & 0.300 & 0.300 & -0.000 & 0.000 & 0.000 & 0.000 \\
Oracle\_QTT & 0.500 & 0.300 & 0.300 & -0.000 & 0.000 & 0.000 & 0.000 \\
Oracle\_QTT & 0.750 & 0.300 & 0.300 & -0.000 & 0.000 & 0.000 & 0.000 \\
Oracle\_QTT & 0.900 & 0.300 & 0.300 & -0.000 & 0.000 & 0.000 & 0.000 \\
WrongIndependence\_QTT & 0.100 & -0.285 & -0.287 & -0.002 & 0.104 & 0.104 & 0.003 \\
WrongIndependence\_QTT & 0.250 & -0.008 & -0.005 & 0.003 & 0.076 & 0.076 & 0.002 \\
WrongIndependence\_QTT & 0.500 & 0.300 & 0.302 & 0.002 & 0.071 & 0.071 & 0.002 \\
WrongIndependence\_QTT & 0.750 & 0.608 & 0.607 & -0.001 & 0.079 & 0.079 & 0.003 \\
WrongIndependence\_QTT & 0.900 & 0.885 & 0.880 & -0.005 & 0.102 & 0.102 & 0.003 \\
\bottomrule
\end{tabular}
}

\paragraph{Design 5: conditional standardization before inversion.}
Let \(X\in\{0,1\}\), with
\[
P(X=1\mid T)=.75,\qquad P(X=1\mid C)=.25.
\]
Conditional baseline outcomes are \(N(-.5,1)\) for \(X=0\) and \(N(1,.7^2)\) for \(X=1\), identically across groups. Untreated post shifts are .10 and .50 by \(X\), while treatment adds .20 and .80. Conditional distributional identification is therefore exact, but marginal treated/comparison compositions differ. Each group-period cell has 1,200 observations, \(R=1000\), seed 20269505.

\noindent\resizebox{\linewidth}{!}{\begin{tabular}{lrrrrrrr}
\toprule
estimand & tau & truth & mean\_estimate & bias & mc\_sd & rmse & mc\_se\_mean \\
\midrule
StandardizedMarginal\_QTT & 0.100 & 0.208 & 0.223 & 0.016 & 0.112 & 0.113 & 0.004 \\
StandardizedMarginal\_QTT & 0.250 & 0.593 & 0.588 & -0.004 & 0.131 & 0.131 & 0.004 \\
StandardizedMarginal\_QTT & 0.500 & 0.776 & 0.774 & -0.002 & 0.081 & 0.081 & 0.003 \\
StandardizedMarginal\_QTT & 0.750 & 0.793 & 0.796 & 0.003 & 0.074 & 0.074 & 0.002 \\
StandardizedMarginal\_QTT & 0.900 & 0.797 & 0.797 & 0.001 & 0.082 & 0.082 & 0.003 \\
UnadjustedMarginal\_QTT & 0.100 & 0.208 & 0.076 & -0.132 & 0.185 & 0.227 & 0.006 \\
UnadjustedMarginal\_QTT & 0.250 & 0.593 & 0.787 & 0.194 & 0.126 & 0.231 & 0.004 \\
UnadjustedMarginal\_QTT & 0.500 & 0.776 & 1.074 & 0.298 & 0.062 & 0.305 & 0.002 \\
UnadjustedMarginal\_QTT & 0.750 & 0.793 & 1.089 & 0.296 & 0.055 & 0.302 & 0.002 \\
UnadjustedMarginal\_QTT & 0.900 & 0.797 & 1.070 & 0.274 & 0.061 & 0.280 & 0.002 \\
\bottomrule
\end{tabular}
}

The average conditional QTT target is .65. In the prospective run its mean estimate is 0.648, illustrating that it is a well-estimated object but generally differs from the marginal QTT curve obtained after standardizing CDFs and then inverting.

\paragraph{Design 6: nonlinear staggered aggregation.}
There are four cohorts with weights
\[
(.20,.25,.25,.30),
\]
sample sizes \((400,500,500,600)\), untreated means
\[
(-1,-.2,.6,1.3),
\]
and standard deviations
\[
(.8,1,1.2,.9).
\]
At event time \(e\), cohort \(g\)'s treated distribution is generated by
\[
Y_{g,e}(1)=a_g+.15e+b_gY_g(0),
\]
where
\[
\begin{aligned}
a&=(.07730893,.24522462,.37232482,.87808500),\\
b&=(1.43373299,.92097038,.78371029,1.33228767).
\end{aligned}
\]
These values were frozen before finite-sample simulation to create an economically transparent sign reversal in the aggregation gap. \(R=1000\), seed 20269506.

\noindent\resizebox{\linewidth}{!}{\begin{tabular}{rlrrrrrr}
\toprule
event\_time & estimand & tau & truth & mean\_estimate & bias & mc\_sd & rmse \\
\midrule
0 & AverageCohort\_QTT\_e0 & 0.100 & 0.352 & 0.353 & 0.001 & 0.057 & 0.057 \\
0 & AverageCohort\_QTT\_e0 & 0.250 & 0.397 & 0.397 & -0.001 & 0.044 & 0.044 \\
0 & AverageCohort\_QTT\_e0 & 0.500 & 0.448 & 0.449 & 0.002 & 0.040 & 0.040 \\
0 & AverageCohort\_QTT\_e0 & 0.750 & 0.498 & 0.500 & 0.003 & 0.045 & 0.045 \\
0 & AverageCohort\_QTT\_e0 & 0.900 & 0.543 & 0.545 & 0.002 & 0.056 & 0.056 \\
0 & Mixture\_QTT\_e0 & 0.100 & -0.112 & -0.108 & 0.004 & 0.070 & 0.070 \\
0 & Mixture\_QTT\_e0 & 0.250 & 0.218 & 0.220 & 0.002 & 0.050 & 0.050 \\
0 & Mixture\_QTT\_e0 & 0.500 & 0.374 & 0.376 & 0.002 & 0.049 & 0.049 \\
0 & Mixture\_QTT\_e0 & 0.750 & 0.691 & 0.688 & -0.003 & 0.057 & 0.057 \\
0 & Mixture\_QTT\_e0 & 0.900 & 1.151 & 1.154 & 0.003 & 0.075 & 0.075 \\
1 & AverageCohort\_QTT\_e1 & 0.100 & 0.502 & 0.498 & -0.004 & 0.055 & 0.055 \\
1 & AverageCohort\_QTT\_e1 & 0.250 & 0.547 & 0.547 & -0.001 & 0.044 & 0.044 \\
1 & AverageCohort\_QTT\_e1 & 0.500 & 0.598 & 0.598 & 0.000 & 0.040 & 0.040 \\
1 & AverageCohort\_QTT\_e1 & 0.750 & 0.648 & 0.648 & 0.000 & 0.044 & 0.044 \\
1 & AverageCohort\_QTT\_e1 & 0.900 & 0.693 & 0.694 & 0.001 & 0.055 & 0.055 \\
1 & Mixture\_QTT\_e1 & 0.100 & 0.038 & 0.035 & -0.003 & 0.069 & 0.069 \\
1 & Mixture\_QTT\_e1 & 0.250 & 0.368 & 0.365 & -0.003 & 0.053 & 0.053 \\
1 & Mixture\_QTT\_e1 & 0.500 & 0.524 & 0.524 & -0.000 & 0.049 & 0.049 \\
1 & Mixture\_QTT\_e1 & 0.750 & 0.841 & 0.841 & 0.000 & 0.057 & 0.057 \\
1 & Mixture\_QTT\_e1 & 0.900 & 1.301 & 1.303 & 0.002 & 0.074 & 0.074 \\
2 & AverageCohort\_QTT\_e2 & 0.100 & 0.652 & 0.650 & -0.002 & 0.057 & 0.057 \\
2 & AverageCohort\_QTT\_e2 & 0.250 & 0.697 & 0.696 & -0.002 & 0.045 & 0.045 \\
2 & AverageCohort\_QTT\_e2 & 0.500 & 0.748 & 0.746 & -0.002 & 0.042 & 0.042 \\
2 & AverageCohort\_QTT\_e2 & 0.750 & 0.798 & 0.796 & -0.002 & 0.046 & 0.046 \\
2 & AverageCohort\_QTT\_e2 & 0.900 & 0.843 & 0.841 & -0.002 & 0.056 & 0.056 \\
2 & Mixture\_QTT\_e2 & 0.100 & 0.188 & 0.188 & 0.000 & 0.070 & 0.070 \\
2 & Mixture\_QTT\_e2 & 0.250 & 0.518 & 0.517 & -0.001 & 0.051 & 0.051 \\
2 & Mixture\_QTT\_e2 & 0.500 & 0.674 & 0.673 & -0.002 & 0.050 & 0.050 \\
2 & Mixture\_QTT\_e2 & 0.750 & 0.991 & 0.988 & -0.003 & 0.057 & 0.057 \\
2 & Mixture\_QTT\_e2 & 0.900 & 1.451 & 1.452 & 0.001 & 0.074 & 0.074 \\
\bottomrule
\end{tabular}
}

At \(e=0\), average-cohort minus mixture QTT equals 0.464 at \(\tau=.10\) but \(-0.608\) at \(\tau=.90\), directly illustrating Proposition 1's no-universal-sign result.

To evaluate Proposition~2 without changing the frozen DGP, we also compute the influence-function standard error of the average-minus-mixture gap using Gaussian-kernel density estimates at the relevant cohort and mixture quantiles, with Silverman bandwidth $1.06\widehat\sigma n^{-1/5}$. Cohort weights remain fixed in this design, so the \(\xi_g\) terms are zero; the separate derivative audit described below checks the estimated-weight terms. The table reports the resulting pointwise 95\% intervals over the same \(R=1000\) replications.

\begin{center}
\small
\begin{tabular}{rrrrrrr}
\toprule
$\tau$ & Truth & Mean & Bias & MC SD & Mean IF SE & Coverage \\
\midrule
0.10 & 0.464 & 0.460 & -0.004 & 0.067 & 0.069 & 0.956 \\
0.25 & 0.179 & 0.177 & -0.002 & 0.045 & 0.048 & 0.974 \\
0.50 & 0.073 & 0.073 & 0.000 & 0.044 & 0.044 & 0.946 \\
0.75 & -0.193 & -0.188 & 0.005 & 0.053 & 0.054 & 0.946 \\
0.90 & -0.608 & -0.610 & -0.001 & 0.072 & 0.072 & 0.954 \\
\bottomrule
\end{tabular}

\paragraph{Bandwidth sensitivity for the smooth influence-function standard error.}
The baseline uses the Gaussian-kernel Silverman rule $h=1.06\widehat\sigma n^{-1/5}$. Because the baseline coverage at $\tau=.25$ is 0.974, about 3.5 Monte Carlo standard errors above 0.95, we recomputed all 1,000 frozen replications under four alternatives: the robust normal-reference rule $0.90\min\{\widehat\sigma,\mathrm{IQR}/1.349\}n^{-1/5}$, Scott's $\widehat\sigma n^{-1/5}$ rule, and 0.8 and 1.2 times the baseline bandwidth. No DGP, random seed, or replication is changed.
\begin{table}[H]
\centering
\caption{Design 6: bandwidth sensitivity of pointwise gap coverage}
\label{tab:design6bwsens}
\small
\begin{tabular}{lrrrrrr}
\toprule
Bandwidth rule & $\tau=.10$ & $.25$ & $.50$ & $.75$ & $.90$ & Mean IF SE at $.25$ \\
\midrule
Silverman 1.06 sd & 0.956 & 0.974 & 0.946 & 0.946 & 0.954 & 0.048 \\
Robust 0.90 min(sd,IQR/1.349) & 0.956 & 0.973 & 0.944 & 0.945 & 0.953 & 0.048 \\
Scott sd & 0.956 & 0.973 & 0.945 & 0.945 & 0.954 & 0.048 \\
0.8 x Silverman & 0.956 & 0.973 & 0.943 & 0.945 & 0.953 & 0.048 \\
1.2 x Silverman & 0.957 & 0.974 & 0.946 & 0.948 & 0.954 & 0.049 \\
\bottomrule
\end{tabular}

\end{table}
Coverage at $\tau=.25$ remains 0.973--0.974 and the mean influence-function standard error remains about 0.048 under every rule. The inflation therefore does not disappear under reasonable bandwidth changes. In this smooth Gaussian DGP it is better interpreted as finite-sample conservativeness of the first-order studentized approximation than as a Silverman-bandwidth artifact. This robustness exercise does not establish optimal density tuning, and it does not justify density-based inference at an atom.

\begin{table}[htbp]
\centering
\caption{Design 6: tilt--dispersion diagnostic}
\label{tab:design6tilt}
\small
\begin{tabular}{crrrr}
\toprule
$\tau$ & Exact gap & Population $\mathcal D_{\mathrm{tilt}}$ & Mean plug-in $\widehat{\mathcal D}_{\mathrm{tilt}}$ & MC SD\\
\midrule
0.10 &  0.464 & 0.359 & 0.386 & 0.090\\
0.25 &  0.179 & 0.364 & 0.375 & 0.069\\
0.50 &  0.073 & 0.370 & 0.378 & 0.061\\
0.75 & -0.193 & 0.377 & 0.385 & 0.069\\
0.90 & -0.608 & 0.382 & 0.413 & 0.094\\
\bottomrule
\end{tabular}

\begin{flushleft}\footnotesize
Notes: $\mathcal D_{\mathrm{tilt}}$ is the nonnegative first-order magnitude diagnostic in Proposition~1. The plug-in version uses sample cohort quantiles and Gaussian-kernel density estimates at those quantiles, with the same Silverman bandwidth rule used for the influence-function exercise. Design 6 is deliberately nonlocal, so the index is not expected to equal the signed exact gap and is not a confidence bound; it flags that density reweighting and cohort-quantile dispersion are jointly material. Monte Carlo statistics use the same $R=1000$ frozen replications and seed as Design 6.
\end{flushleft}
\end{table}
\end{center}

The mean influence-function standard errors track the Monte Carlo standard deviations closely. Coverage ranges from 0.946 to 0.974 over the five quantiles; the mild conservativeness at \(\tau=.25\) is finite-sample rather than built into the target. This experiment isolates the smooth aggregation layer and does not validate density-based QTT inference at mass points.

\paragraph{Design 7: assignment-level simultaneous inference.}
For state \(s\) and individual \(i\),
\[
Y_{ist}(0)=a_s+u_{ist},\qquad
a_s\sim N(0,.4^2),\quad u_{ist}\sim N(0,1).
\]
Half the states are assigned treatment; there are 60 individuals per state-period, one baseline period, and two post periods. Treatment adds .25 and .40 at event times 0 and 1. The QTT family is
\[
(e,\tau)\in\{0,1\}\times\{.25,.50,.75\}.
\]
The bootstrap resamples complete state histories within treatment status. A precision extension uses \(R=1{,}000\), \(B=199\), seed 202610007, and cluster counts 30, 40, and 50; the DGP and estimator are otherwise unchanged.

\noindent\resizebox{\linewidth}{!}{\begin{tabular}{rrrrrrr}
\toprule
clusters & R & B & family\_coverage & family\_mc\_se & mean\_pointwise\_coverage & mean\_critical \\
\midrule
30 & 1000 & 199 & 0.946 & 0.007 & 0.990 & 2.691 \\
40 & 1000 & 199 & 0.954 & 0.007 & 0.990 & 2.663 \\
50 & 1000 & 199 & 0.949 & 0.007 & 0.991 & 2.653 \\
\bottomrule
\end{tabular}
}

Family coverage is 0.946, 0.954, and 0.949 for 30, 40, and 50 clusters. The corresponding Monte Carlo standard errors are 0.0071, 0.0066, and 0.0070.

\subsection{Reproducibility}

The suite is generator-reproducible: the DGPs, sample sizes, numerical grids, seeds, and estimators are encoded in the master specification and executable script supplied with the replication files. The tables in this appendix are generated from that suite.

\section{Stop-rule size, power, and local-power calibration}

Design 8 remains the paper's direct finite-cluster experiment for the 99-coordinate gate. Independent cluster score vectors are Gaussian with
\[
\operatorname{Corr}(\psi_{e,\tau},\psi_{e',\tau'})
=
0.65^{|e-e'|}0.75^{|k(\tau)-k(\tau')|}.
\]
The fully studentized wild-cluster procedure uses 499 Rademacher draws in each of 5,000 outer replications.

\begin{table}[htbp]\centering
\caption{Finite-sample size and power of the fully studentized 99-coordinate wild-cluster rule}\label{tab:stoprulewild}
\small\begin{tabular}{rllrr}\toprule
Clusters & Scenario & Shift & Rejection rate & MC s.e.\\\midrule
20 & Null & 0.00 & 0.052 & 0.003\\
30 & Null & 0.00 & 0.051 & 0.003\\
50 & Null & 0.00 & 0.044 & 0.003\\
30 & Localized & 0.50 & 0.223 & 0.006\\
30 & Localized & 0.70 & 0.552 & 0.007\\
30 & Diffuse & 0.25 & 0.165 & 0.005\\
30 & Localized & 0.90 & 0.854 & 0.005\\
30 & Diffuse & 0.40 & 0.457 & 0.007\\
\bottomrule\end{tabular}
\begin{flushleft}\footnotesize Notes: 5,000 outer replications and 499 Rademacher draws per replication.\end{flushleft}
\end{table}

The limiting Gaussian and quadratic-omnibus comparisons reported earlier in this appendix remain unchanged because they were already fully parameterized and reproducible. Their critical values are simulated from the same correlated Gaussian null; in particular the quadratic statistic is not compared with a naive $\chi^2_{99}$ cutoff, so the comparison is degrees-of-freedom-free with respect to that approximation.

\section{Same-Object Staggered-QTT Software Audit}\label{app:qtesoftware}

This section documents the software-facing exercise in the main text and records a source-order issue discovered during replication. The substantive comparison itself does not depend on that issue: all reported paper objects are reconstructed independently, explicitly keyed by $(g,t)$, and aggregated by label.

\paragraph{Independent reconstruction of the group--time distributions.}
We use the public \texttt{mpdta} panel and the no-covariate, not-yet-treated specification documented for \texttt{qte::ddid}. For each group $g$ and evaluation date $t$, the two-period subset uses baseline $t-1$ in pre-treatment placebo cells and $g-1$ in post-treatment cells. The Callaway--Li--Oka construction evaluates each control unit's baseline outcome rank in the control baseline distribution, maps that rank into the treated baseline quantile, and adds the control unit's realized outcome change. The resulting pseudo-outcomes form the recovered untreated distribution $\widehat F_{g,t}^{0}$; the empirical treated outcomes form $\widehat F_{g,t}^{1}$.

As a check that the local distributions are reconstructed correctly, their post-treatment means reproduce the public ATT aggregation. With groups 2004, 2006, and 2007, the group-size/post-length weights are $0.026178$ for each of the four 2004 post cells, $0.104712$ for each of the two 2006 post cells, and $0.685864$ for the 2007 post cell. The resulting overall ATT is $-0.0452763$, compared with $-0.0453$ in the package documentation. The reconstructed event-time-zero ATT is $-0.0323719$, compared with $-0.0324$ in the documentation; the $e=1,2,3$ values are $-0.0637330,-0.1377314,-0.1086553$, respectively, again matching the published rounded values.

\paragraph{Operator comparison after explicit label alignment.}
For the main text's supported event-time-zero cohort set $g\in\{2006,2007\}$, we retain the observed supported-sample weights $40/171$ and $131/171$. The average-cohort QTT is formed by generalized inversion cohort by cohort and then weighted averaging. The mixture QTT first forms the two explicitly label-aligned finite mixtures and then applies the generalized inverse. No first-stage object changes between the two columns.

\begin{center}
\small
\begin{tabular}{rrrr}
\toprule
$\tau$ & Average cohort QTT & Mixture QTT & Gap \\
\midrule
0.1 & -0.050 & 0.068 & -0.118 \\
0.2 & -0.038 & -0.048 & 0.010 \\
0.3 & -0.050 & -0.016 & -0.035 \\
0.4 & -0.069 & -0.067 & -0.002 \\
0.5 & -0.096 & -0.088 & -0.007 \\
0.6 & -0.074 & 0.005 & -0.078 \\
0.7 & 0.012 & -0.009 & 0.020 \\
0.8 & -0.033 & -0.007 & -0.026 \\
0.9 & -0.103 & -0.066 & -0.037 \\
\bottomrule
\end{tabular}

\end{center}

The lower-tail sign reversal is particularly transparent: at $\tau=.10$ the average-cohort QTT is about $-0.050$, whereas the mixture QTT is about $+0.068$. Opposite signs also occur at $\tau=.60$ and $.70$. These calculations use the exact generalized inverse of the finite mixtures. A separate all-post-period aggregation, using the standard group-size/post-length weights, is supplied in \texttt{qte\_ddid\_same\_object.csv}; at the median its average-QTT and explicitly aligned mixture-QTT values are approximately $-0.125$ and $-0.088$.

\paragraph{State-delete-one sensitivity.}
To distinguish an operator pattern from dependence on one state, we delete each of the 29 states in turn and recompute the complete event-time-zero construction: comparison samples, recovered $F_0$ and $F_1$, cohort QTTs, supported-sample weights, mixture CDFs, and generalized inverses. This is a descriptive sensitivity exercise rather than a fixed-$L$ confidence procedure. The lower-tail sign reversal is highly stable: at $\tau=.10$, the two aggregands have opposite signs after 28 of the 29 state deletions. The corresponding counts are 8 of 29 at $\tau=.60$ and 18 of 29 at $\tau=.70$. The full summary is:
\begin{center}
\scriptsize
\begin{tabular}{crrrr}\toprule
$\tau$ & Full avg. & Full mix & Opposite sign after state deletion & Gap range \\
\midrule
0.1 & -0.050 & 0.068 & 28/29 & [-0.367,0.016] \\
0.2 & -0.038 & -0.048 & 0/29 & [-0.064,0.059] \\
0.3 & -0.050 & -0.016 & 3/29 & [-0.061,0.029] \\
0.4 & -0.069 & -0.067 & 0/29 & [-0.037,0.037] \\
0.5 & -0.096 & -0.088 & 0/29 & [-0.070,0.032] \\
0.6 & -0.074 & 0.005 & 8/29 & [-0.112,0.025] \\
0.7 & 0.012 & -0.009 & 18/29 & [-0.054,0.089] \\
0.8 & -0.033 & -0.007 & 9/29 & [-0.054,0.041] \\
0.9 & -0.103 & -0.066 & 4/29 & [-0.103,0.041] \\
\bottomrule\end{tabular}

\end{center}
The gap itself is not sign-stable at every quantile---indeed its delete-one range crosses zero even at $\tau=.10$ because deleting state 37 makes both aggregands positive. The robust statement is therefore narrower and more informative: the lower-tail \emph{operator-induced sign disagreement} survives almost every one-state deletion, whereas the middle/upper-quantile sign disagreements are more state-sensitive. The complete 29-by-9 recomputations are supplied in \texttt{qte\_same\_object\_state\_delete.csv}.

\paragraph{Why the replication does not use the package aggregate QTT as a benchmark.}
The current public source of \texttt{ptetools} 1.0.1 creates a nontrivial positional-ordering sensitivity for staggered QTT aggregation. In \texttt{compute.pte}, the outer loop is over time periods and the inner loop is over groups, so \texttt{extra\_gt\_returns}---which contains the group--time CDFs---is stored in time-major order. The aggregation helper first row-binds the ATT results in that order but then calls base-R \texttt{merge(..., by.x="group", by.y="groups")} to attach group sizes. Base R uses \texttt{sort=TRUE} by default for a data-frame merge, so the merged result is sorted by group. The resulting overall, dynamic, and group weight vectors are therefore indexed to the post-merge order. The QTT aggregation routine, however, extracts $F_0$ and $F_1$ directly from the original \texttt{extra\_gt\_returns} list and passes those CDF lists and the post-merge weight vectors positionally to \texttt{combine\_ecdfs}, without a $(g,t)$ rejoin.

For \texttt{mpdta}, the two orders are visibly different:
\[
\begin{split}
\text{CDF-list order: }& (2004,2004),(2006,2004),(2007,2004),(2004,2005),\ldots,\\
\text{weight-table order: }& (2004,2004),(2004,2005),(2004,2006),(2004,2007),\ldots .
\end{split}
\]
The exact 12-position mapping and weights are supplied in \nolinkurl{qte_ddid_weight_alignment.csv}. As an external numerical check, a Python reconstruction that deliberately applies the post-merge weight vector positionally to the time-major CDF list reproduces six of the nine QTT entries printed in the current \texttt{qte} documentation within $4\times10^{-5}$ and is within $0.0105$ at all nine quantiles. The remaining small differences reflect that the audit emulates, rather than executes in R, the \texttt{BMisc::make\_dist}/\texttt{quantile.ecdf} 1,000-knot reconstruction. Explicit $(g,t)$ alignment instead changes the curve materially. Because this is a source audit plus an independent emulation rather than execution of the installed R stack itself, we treat the finding as an implementation-ordering sensitivity rather than make a package-correctness claim. It is nevertheless sufficient reason not to use the package's aggregate staggered-QTT output as empirical evidence for either aggregand.

The contribution of the paper does not rely on this implementation detail. The economically relevant point survives after removing it completely: with each recovered CDF and weight explicitly keyed by $(g,t)$, the average and mixture operators still give materially different summaries and can disagree in sign. The replication code therefore treats label alignment as part of the estimand definition, not as a software convenience.

\section{Real-Data Aggregation Illustration: Construction and Robustness}\label{app:mpdta}

This section documents the post-development \texttt{mpdta} illustration in the main text. The data are the public balanced county panel distributed with the \texttt{did} package accompanying Callaway and Sant'Anna (2021): 500 counties observed annually from 2003 through 2007, with \texttt{lemp} equal to log county-level teen employment, county identifier \texttt{countyreal}, and first-treatment year \texttt{first.treat}. The distributional unit in this illustration is therefore the county-level outcome, not an individual teenager. State FIPS is recovered from the county identifier because minimum-wage treatment timing is state-level. The resulting panel contains 29 states. The replication archive includes the exact CSV used by the analysis and its SHA-256 checksum.

\paragraph{Target population and support rule.}
We use cohorts $g\in\{2006,2007\}$ at event time $e=0$. The 2004 cohort is excluded by a support rule that does not use its post-treatment outcome: with $g-1=2003$ as baseline there is no earlier observed placebo period, and the cohort lies in only one treated state. The supported target population has 40 counties in the 2006 cohort and 131 in the 2007 cohort, with fixed sample-share weights
\[
\widehat\omega_{2006}=40/171=0.233918,
\qquad
\widehat\omega_{2007}=131/171=0.766082.
\]
These are treated as fixed weights for this finite support-based sample-composition target. Thus the empirical parameter conditions on the observed supported-cohort composition; it is not a plug-in estimate of a superpopulation cohort-share target. The estimated-cohort-weight terms in Proposition~2 of the main text, and the weight-region extension in Corollary~1, are therefore not needed for this fixed-weight reference illustration.
The 2006 counties are concentrated in three treated states, whose within-cohort county shares have Herfindahl index 0.341 (an effective count of only 2.93 equally weighted states); the corresponding figures for the nine-state 2007 cohort are 0.151 and 6.63. These summaries do not enter the estimator. They quantify why a county-rich dataset can nevertheless be cluster-poor for distributional inference.

The support restriction is also transparent to an all-cohort point-estimate sensitivity that ignores the missing placebo support for the 2004 cohort and therefore is \emph{not} given a causal interpretation. Adding the 20 counties first treated in 2004 changes the median average-versus-mixture gap only from $-0.085$ to $-0.088$ log points and the $\tau=.10$ gap from $-0.113$ to $-0.135$; at $\tau=.80$ the all-cohort point gap remains positive at 0.028. The full descriptive sensitivity is supplied as \texttt{mpdta\_allcohort\_descriptive\_sensitivity.csv}. This check is reported precisely to show that the main point pattern is not mechanically created by omitting the unsupported early cohort.

\paragraph{Cluster CDF scores.}
For any fixed county set $A$, year $t$, and threshold $y$, let $N_{\ell A}$ be the number of counties in state cluster $\ell$ belonging to $A$, and let
\[
S_{\ell A,t}(y)=\sum_{i\in A:\,c(i)=\ell}\mathbf 1\{Y_{it}\le y\}.
\]
For independent (not necessarily identically distributed) state clusters, define
\[
\mu_{A,L}=L^{-1}\sum_{\ell=1}^L E[N_{\ell A}],
\qquad
F_{A,t}(y)=
\frac{L^{-1}\sum_{\ell=1}^L E[S_{\ell A,t}(y)]}
     {\mu_{A,L}}.
\]
If $\inf_L\mu_{A,L}>0$, the cluster denominator obeys a law of large numbers, and the centered numerator satisfies the required cluster CLT, the ratio expansion gives
\[
\psi_{\ell A,t,L}(y)
=\frac{S_{\ell A,t}(y)-F_{A,t}(y)N_{\ell A}}{\mu_{A,L}},
\qquad
\sqrt L\{\widehat F_{A,t}(y)-F_{A,t}(y)\}
=L^{-1/2}\sum_{\ell=1}^L\psi_{\ell A,t,L}(y)+o_p(1).
\]
This formulation allows heterogeneous county counts across states. In implementation $\mu_{A,L}$ is replaced by $N_A/L$ and the finite cluster-score array is recentered across states before multiplier resampling. For cohort $g$, baseline $b=g-1$, and a placebo year $p< b$, the CDF-PT placebo score is the corresponding linear combination
\[
\psi_{\ell,g,p}(y)-\psi_{\ell,g,b}(y)
-\psi_{\ell,C_g,p}(y)+\psi_{\ell,C_g,b}(y).
\]
The same linearity gives the counterfactual-CDF score at target year $g$,
\[
\psi^0_{\ell,g,g}(y)
=\psi_{\ell,g,g-1}(y)
+\psi_{\ell,C_g,g}(y)
-\psi_{\ell,C_g,g-1}(y).
\]
Thus the empirical multiplier construction uses the same cluster-score representation as the main theory and, in particular, does not treat 500 counties as independent when the policy assignment is state-level. Whether its many-cluster regularity is a good finite-$L$ approximation remains a separate question; the three-state 2006 cohort is the main weak point.

\paragraph{Diagnostic family and sensitivity.}
The reference family stacks $e=-3,-2$ and $y=4.0,4.5,\ldots,8.0$ separately for both supported cohorts. The table records that calculation and three transparent within-dataset sensitivity variants. Because the dataset and illustrative specification were chosen after theory development and empirical screening, these $p$-values are not claimed to have nominal model-selection size across that screening process; they are compatibility diagnostics conditional on this displayed exercise.
\begin{center}
\small
\begin{tabular}{lrrrr}
\toprule
Specification & $T_{\max}$ & 95\% crit. & $p$ & Nondeg. coords. \\
\midrule
Reference: not-yet-treated, y=4.0--8.0 & 1.953 & 2.925 & 0.744 & 36 \\
Never-treated controls, y=4.0--8.0 & 2.333 & 2.906 & 0.362 & 36 \\
Not-yet-treated, wider y=3.5--8.5 & 2.013 & 2.999 & 0.728 & 43 \\
Never-treated, wider y=3.5--8.5 & 2.333 & 2.948 & 0.420 & 43 \\
\bottomrule
\end{tabular}

\end{center}
The wider-grid specifications have one deterministic zero-variance coordinate; that coordinate is removed from studentization rather than assigned an artificial standard error. All four max-$|t|$ calculations are far from rejection within the displayed dataset. As throughout the paper, failure to reject a finite placebo family does not prove CDF-PT; here it also does not retroactively turn a post-development illustration into a prespecified PROCEED branch.

\paragraph{Point estimands and state jackknife.}
At the target date, the untreated CDF estimator is
\[
\widehat F^0_{g,g}
=\widehat F_{g,g-1}+\widehat F_{C_g,g}-\widehat F_{C_g,g-1}.
\]
We clip to $[0,1]$ and use $L_2$ isotonic projection with deterministic zero/one sentinels before generalized inversion. The sample correction is modest rather than a repair of gross incoherence: the maximum absolute projection adjustment is about 0.023 for the 2006 counterfactual CDF and 0.013 for 2007. Before clipping/projection, the 2006 raw additive CDF-PT counterfactual reaches a minimum of about $-0.018$ and never exceeds one; the 2007 raw counterfactual reaches about $-0.003$ and likewise never exceeds one. The raw grids contain small local monotonicity reversals, with the largest downward increment about $0.009$--$0.010$. These diagnostics confirm that shape correction is doing limited finite-sample regularization rather than converting a grossly incoherent object into a CDF. The treated CDF is empirical. Average-cohort and mixture QTTs are then computed from the same four cohort-state CDFs and the same fixed weights. A delete-one-state jackknife recomputes the entire CDF projection, inversion, and aggregation map after removing state $\ell$ and uses
\[
\widehat{\mathrm{Var}}_{\rm jack}(\widehat\theta)
=\frac{L-1}{L}\sum_{\ell=1}^L
\bigl(\widehat\theta_{(-\ell)}-\overline\theta_{(-\cdot)}\bigr)^2.
\]
Because quantiles may be nonsmooth in finite samples, this jackknife is reported as a descriptive smooth-quantile sensitivity, not as the paper's mass-point-robust inferential guarantee. At $\tau=.50$, the delete-one-state gap ranges from $-0.127$ to $0.065$, so its sign is not stable to deleting a single state. The replication archive reports every leave-one-state value.

\begin{figure}[t]
\centering
\includegraphics[width=.76\textwidth]{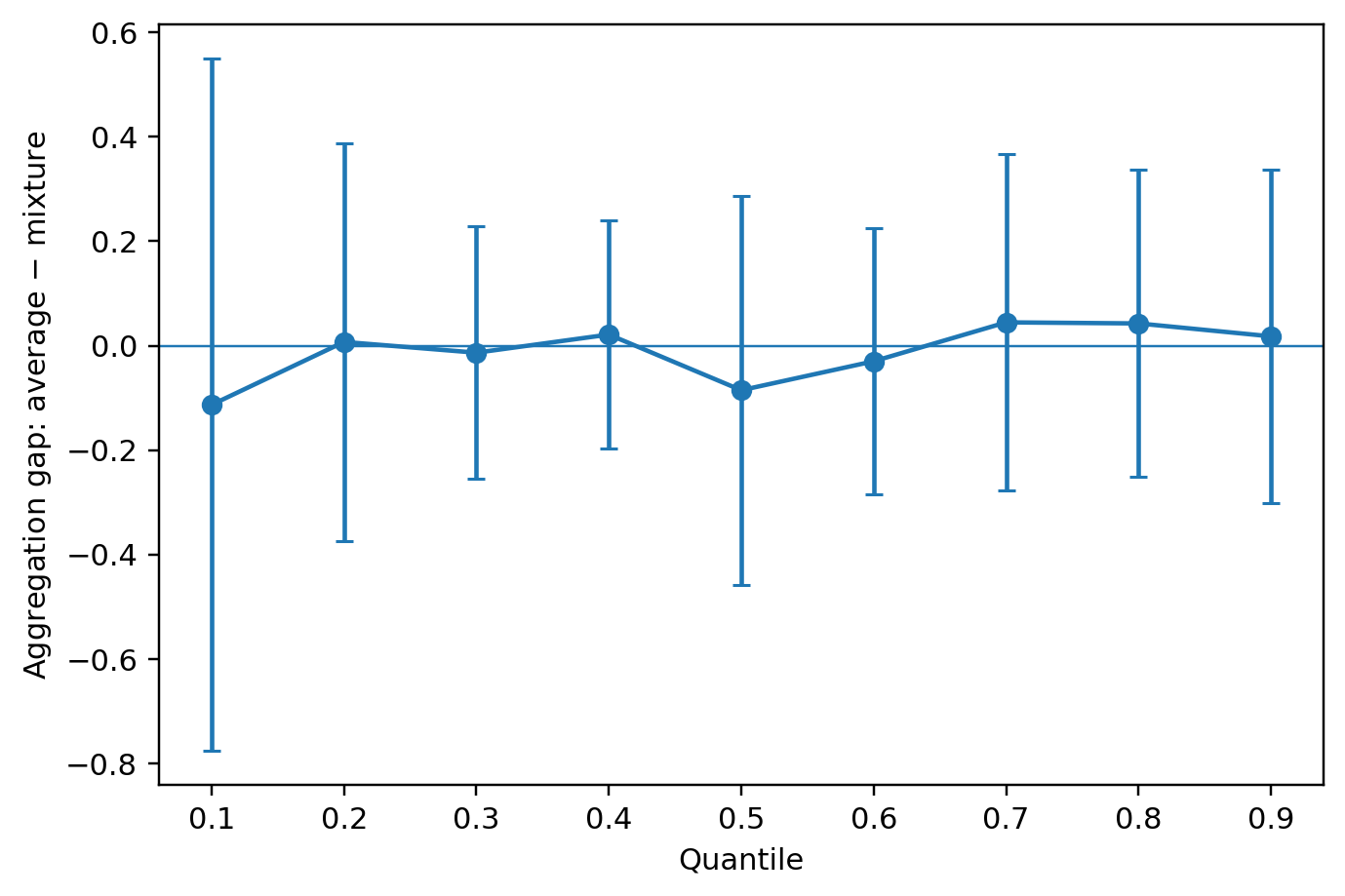}
\caption{State-delete-one smooth-quantile sensitivity for the real-data aggregation gap. Vertical bars are pointwise jackknife $\pm1.96$ standard-error ranges. They are descriptive sensitivity summaries, not the mass-point-robust simultaneous CDF-projection bands.}
\label{fig:mpdta_gap_app}
\end{figure}

\paragraph{Mass-point-safe simultaneous projection.}
For each of the four cohort-state CDF processes, we form a joint state-cluster Rademacher multiplier band over the full observed outcome support plus deterministic tail sentinels. Only nondegenerate coordinates are studentized. With 29,999 draws (seed 20260822) the common studentized critical value is 2.827. Raw lower and upper bands are tightened by monotonicity:
\[
L^{\uparrow}(y)=\sup_{x\le y}L^{\rm raw}(x),
\qquad
U^{\uparrow}(y)=\inf_{x\ge y}U^{\rm raw}(x).
\]
This tightening preserves coverage. Indeed, on the event $L^{\rm raw}(x)\le F(x)\le U^{\rm raw}(x)$ for all $x$, monotonicity of the true CDF gives $L^{\uparrow}(y)\le F(y)\le U^{\uparrow}(y)$ for every $y$. Under the many-cluster regularity conditions of the main corollary, generalized-inverse projection and its fixed-weight specialization give simultaneous outer intervals. The present application does not verify that those regularity conditions provide an accurate approximation for the 2006 component: only three treated states carry that cohort. We therefore report the following projection as an asymptotic sensitivity construction and not as a finite-$L$ confidence guarantee.
\begin{center}
\scriptsize
\begin{tabular}{rrrrrrr}
\toprule
$\tau$ & Avg. QTT & Avg. proj. band & Mix. QTT & Mix. proj. band & Gap & Gap proj. band \\
\midrule
0.1 & -0.127 & [-1.557,1.725] & -0.013 & [-1.854,1.545] & -0.113 & [-3.102,3.578] \\
0.2 & 0.030 & [-0.959,1.033] & 0.022 & [-1.190,1.103] & 0.007 & [-2.062,2.223] \\
0.3 & -0.019 & [-1.127,1.023] & -0.005 & [-1.055,1.019] & -0.013 & [-2.146,2.078] \\
0.4 & -0.047 & [-1.138,1.041] & -0.068 & [-1.091,1.057] & 0.021 & [-2.194,2.132] \\
0.5 & -0.101 & [-1.098,1.165] & -0.016 & [-1.206,1.201] & -0.085 & [-2.300,2.371] \\
0.6 & -0.045 & [-1.599,1.544] & -0.015 & [-1.475,1.289] & -0.030 & [-2.888,3.019] \\
0.7 & 0.038 & [-1.502,1.465] & -0.007 & [-1.383,1.357] & 0.045 & [-2.859,2.848] \\
0.8 & 0.023 & [-1.459,1.512] & -0.019 & [-1.523,1.559] & 0.043 & [-3.017,3.035] \\
0.9 & -0.043 & [-1.932,2.083] & -0.061 & [-1.749,1.990] & 0.018 & [-3.922,3.833] \\
\bottomrule
\end{tabular}

\end{center}
The intervals are intentionally wide. The largest pointwise CDF half-width before monotone tightening is about 0.246, reflecting state-level dependence and, especially, the fact that the 2006 cohort is represented by only three treated states. The projection intervals therefore make a useful inferential point: a visually material difference between the two aggregation orders does not imply that the gap is precisely learned from this short panel.

\paragraph{Comparison-pool robustness and discreteness.}
Recomputing all point estimands with never-treated controls only leaves the principal pattern intact; the full table is supplied as \texttt{mpdta\_control\_pool\_sensitivity.csv}. The documented outcome is log county-level teen employment. Its empirical distribution contains repeated values, although the observed atoms are small: the largest atom is about one percent in the full target-year cross sections. We therefore retain the structure-preserving CDF-projection display rather than assume smooth quantile differentiation merely because the empirical distribution looks nearly continuous.

\paragraph{Variance-equivalent precision calibration.}
At the median the point gap has magnitude $|\widehat\Delta|=0.0854$ and the delete-one-state smooth-quantile sensitivity standard error is about $0.1902$ with $L=29$ states. As a deliberately mechanical calibration, hold per-cluster information and the covariance structure fixed and scale the standard error at the canonical $L^{-1/2}$ rate. A two-sided 5\% interval would then just exclude zero at roughly
\[
29\left\{\frac{1.96(0.1902)}{0.0854}\right\}^2\approx553
\]
state-cluster equivalents. The corresponding normal-approximation requirements are about 1,130 cluster equivalents for 80\% power and 1,513 for 90\% power against a gap of the same magnitude. These are not forecasts of literal treated-state counts: cohort composition, treated shares, dependence, and information per cluster would all change in a different design. They quantify why hundreds of county observations do not imply precise aggregation-gap inference when treatment variation is state-level.

\paragraph{Interpretive scope.}
The dataset was screened and added after development of the theory and after requests for a real-data aggregation illustration. Consequently, the route, threshold grid, and dataset choice are not described as prospectively frozen. The displayed placebo $p$-values condition on this selected illustration and Diagnostic Result D4's selective coverage is not invoked for the dataset-screening step. The exercise demonstrates the estimand distinction on genuine staggered data; it is not a new definitive evaluation of U.S. minimum-wage policy. State clustering, the short panel, and the small number of treated states in the 2006 cohort remain substantive limitations.

\section{Frozen Minimum-Wage Design}
\begin{table}
\caption{Treatment cohorts in the frozen baseline minimum-wage design.}
\label{tab:policy_cohorts}
\begin{tabular}{lr}
\toprule
Cohort date & Treated states \\
\midrule
2015-01 & 8 \\
2016-01 & 1 \\
2016-07 & 1 \\
2017-01 & 5 \\
2018-01 & 1 \\
2019-01 & 1 \\
2019-07 & 1 \\
2020-01 & 2 \\
2020-07 & 1 \\
2021-05 & 1 \\
2021-09 & 1 \\
\bottomrule
\end{tabular}
\end{table}

\paragraph{Policy-blind validation.}
The full prespecified data-quality gate used before the route-specific outcome diagnostics is reported below.
\begin{table}[htbp]
\centering
\caption{Policy-blind data-quality gate}\label{tab:policyblind}
\small
\begin{tabular}{lr}\toprule
Diagnostic & Frozen-build result\\\midrule
ORG wage-and-salary observations & 1,310,406\\
Observations with measurable hourly earnings & 1,235,205\\
Outcome availability & 94.3\%\\
Monthly analytical sample range & 8,582--12,955\\
State/DC units present each month & 51\\
Unmatched state-month policy cells & 0\\
Minimum state-month hourly-wage cell & 42\\
Prespecified minimum-cell support threshold & 15\\
Maximum monthly change in direct-hourly share & 0.031\\
\bottomrule\end{tabular}
\begin{flushleft}\footnotesize Notes: All entries are computed before examining post-treatment distributional effects. DC is retained as a comparison unit and is not among the 23 treated states. Passing this table is a data-construction check, not an identifying-assumption test.\end{flushleft}
\end{table}

\begin{figure}[htbp]
\centering
\includegraphics[width=.78\textwidth]{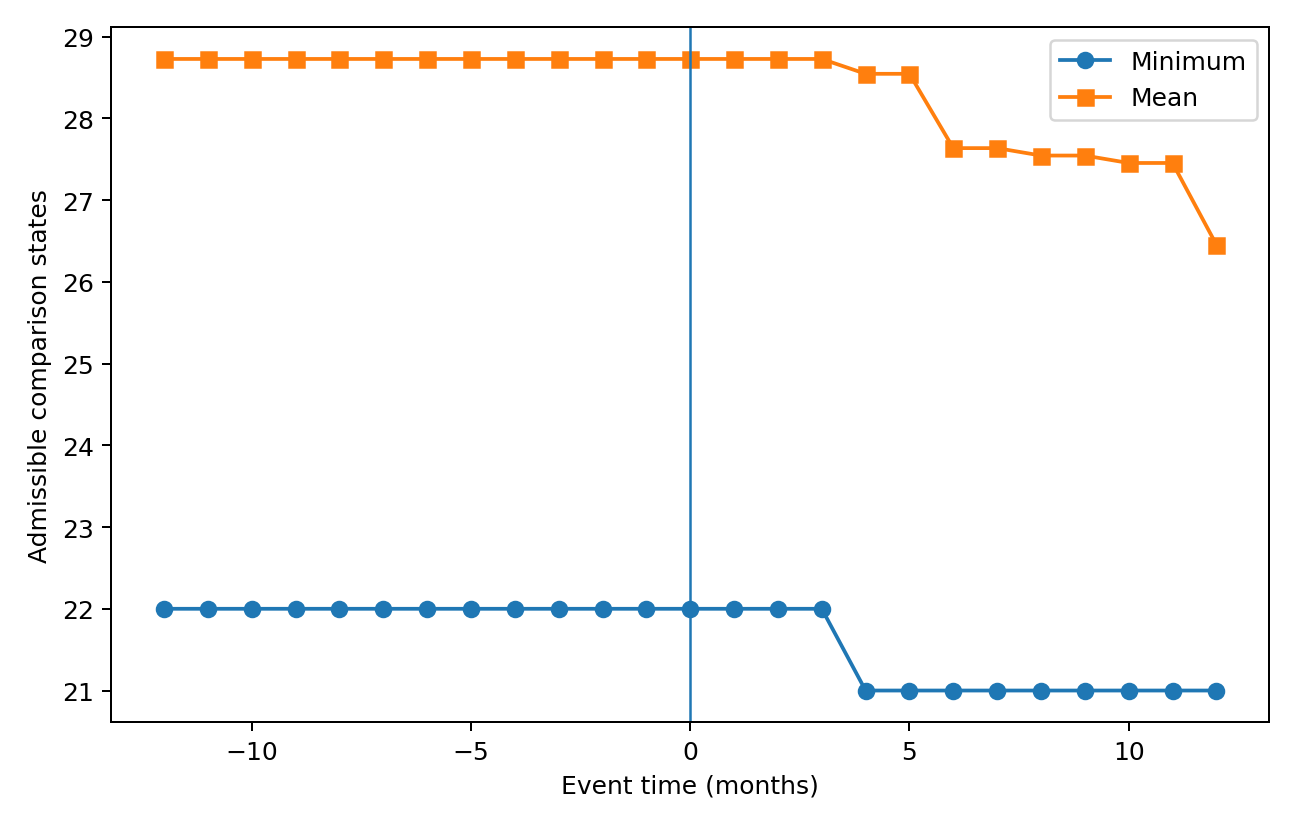}
\caption{Admissible comparison-state support in the frozen minimum-wage design.}
\label{fig:policy_support_app}
\end{figure}

\begin{figure}[htbp]
\centering
\includegraphics[width=.78\textwidth]{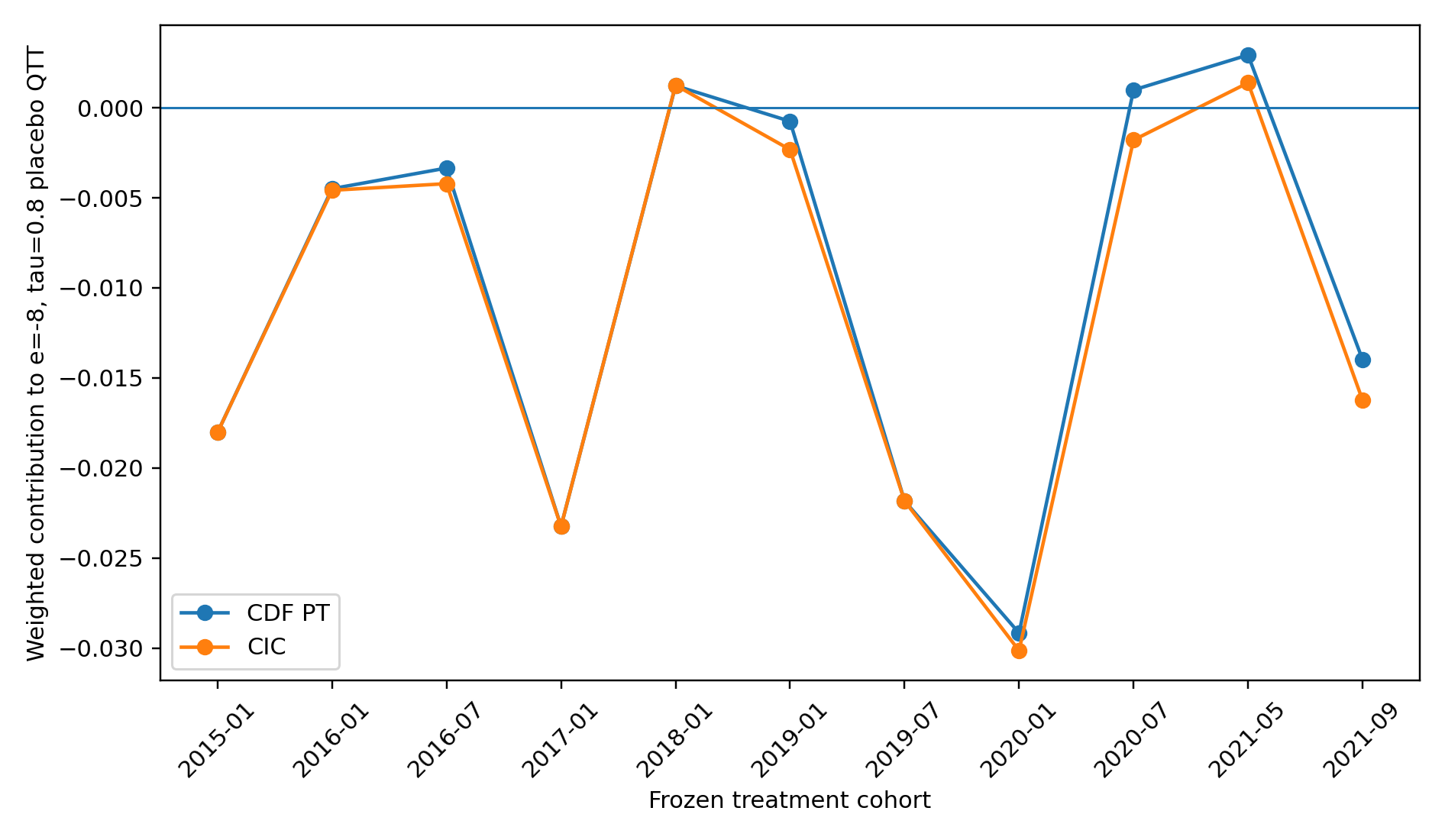}
\caption{Cohort contributions to the common $e=-8,\tau=0.8$ placebo signal. The decomposition is diagnostic only.}
\label{fig:cohort_driver_app}
\end{figure}

\section{Mass-Point-Robust Fixed-Threshold CDF Diagnostics}

The smooth QTT corollaries require local densities because they differentiate the inverse CDF. Reported wages exhibit heaping, so we also evaluate CDF-PT directly at fixed real-wage thresholds, where no quantile inversion or density estimate is required. The frozen family uses \(e=-12,\ldots,-2\) and thresholds
\[
y\in\{\$8,\$10,\$12,\$15,\$20,\$25,\$30,\$40,\$50\}
\]
in December-2022 dollars. Fully studentized leave-one-state scores are combined with 19,999 Rademacher draws and Romano--Wolf stepdown.

\begin{table}[htbp]\centering
\caption{Fixed-dollar-threshold CDF diagnostics robust to wage mass points}\label{tab:cdfthreshold}
\small\begin{tabular}{lrrrr}\toprule
Route & Coordinate & CDF gap & Jackknife s.e. & Stepdown $p$\\\midrule
Unconditional CDF PT & $e=-8,y=\$25$ & 0.042 & 0.013 & 0.0302\\
Conditional/standardized CDF PT & $e=-8,y=\$40$ & 0.037 & 0.014 & 0.3428\\
\bottomrule\end{tabular}
\begin{flushleft}\footnotesize Notes: The conditional row reports its smallest adjusted $p$-value; no conditional coordinate is rejected at 5\%.\end{flushleft}
\end{table}

The unconditional CDF family rejects only \(e=-8,y=\$25\), while the conditional/standardized family has no 5\% stepdown rejection. Accordingly, conditional CDF-PT is treated as diagnostic-family dependent rather than as a robust rejection. This result is deliberately reported next to the QTT-grid findings: the smooth-quantile and fixed-threshold views need not produce identical route-level conclusions for a heaped outcome.

\section{Additional Empirical Robustness}

The primary pairs bootstrap uses 499 state-history draws. With a \(+1\) finite-bootstrap correction, the unconditional CDF-PT familywise \(p\)-value is approximately \(0.08\), \(0.04\), and \(0.048\) using 99, 199, and 499 draws, respectively; the CIC route rejects throughout. This confirms that the unconditional route is genuinely near the decision boundary.

A fully studentized jackknife wild-cluster-\(t\) analysis with 9,999 Rademacher draws yields familywise \(p\)-values \(0.046\), \(0.006\), and \(0.005\) for unconditional CDF-PT, conditional/standardized CDF-PT, and CIC. A separate 19,999-draw Romano--Wolf calculation is used for coordinate-level adjusted \(p\)-values.

\begin{table}[htbp]\centering
\caption{Small-cluster robustness with jackknife wild-cluster studentization}\label{tab:wildcluster}
\small\begin{tabular}{lrrr}\toprule
Route & Max $|t|$ & 95\% critical & Familywise $p$\\\midrule
Unconditional CDF PT & 3.331 & 3.308 & 0.046\\
Conditional/standardized CDF PT & 3.886 & 3.355 & 0.006\\
Changes-in-changes & 3.839 & 3.305 & 0.005\\
\bottomrule\end{tabular}
\begin{flushleft}\footnotesize Notes: 9,999 Rademacher wild-cluster draws over 51 state/DC histories. Studentization uses leave-one-state jackknife standard errors.\end{flushleft}
\end{table}

\begin{table}[htbp]\centering
\caption{Romano--Wolf stepdown localization of the QTT-grid placebo failures}\label{tab:rwstepdown}
\small\begin{tabular}{lrrrr}\toprule
Route & $(e,\tau)$ & Estimate & Jackknife s.e. & Stepdown $p$\\\midrule
Unconditional CDF PT & $(-8,0.8)$ & -0.112 & 0.034 & 0.0463\\
Conditional/standardized CDF PT & $(-8,0.8)$ & -0.108 & 0.028 & 0.0065\\
CIC & $(-8,0.8)$ & -0.123 & 0.032 & 0.0058\\
\bottomrule\end{tabular}
\begin{flushleft}\footnotesize Notes: 19,999 Rademacher wild-cluster draws. Within each route, the table reports the only coordinate rejected at the 5\% stepdown level.\end{flushleft}
\end{table}

\begin{figure}[htbp]
\centering
\includegraphics[width=.82\textwidth]{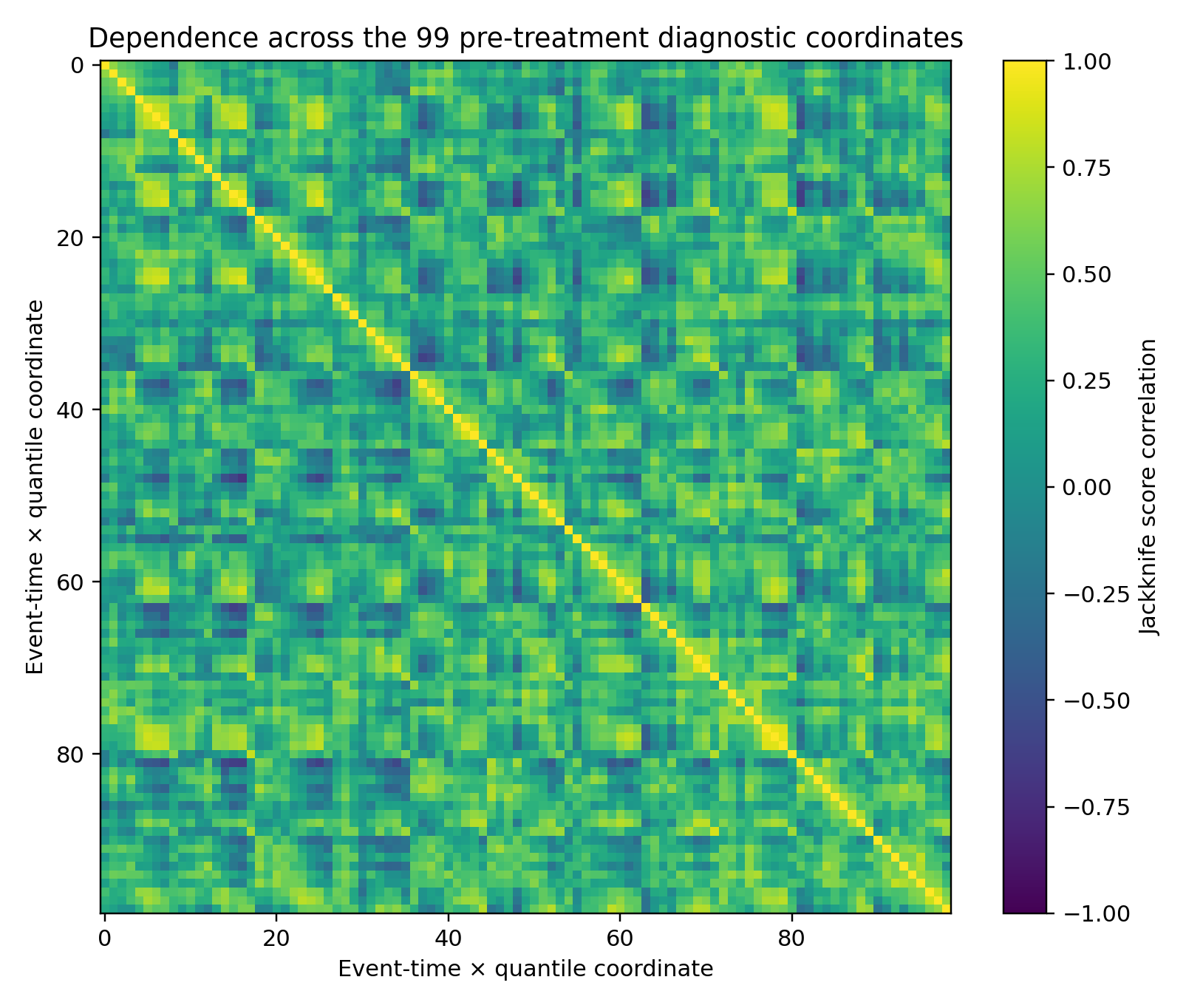}
\caption{Jackknife score correlation across the 99 unconditional CDF-PT diagnostic coordinates. Dependence is strong across nearby event-time and quantile coordinates, so the effective multiplicity burden is much smaller than under 99 independent tests.}
\end{figure}

\section{Novelty Boundary Relative to the Closest Recent Literature}

The paper's formal claims are intentionally narrower than a claim to have invented staggered-adoption distributional DiD. Li and Lin (2024) establish staggered-adoption QTT identification under distributional DiD and dependence restrictions; Ciaccio (2024) develops a multi-period distributional toolkit; Miller and Galvao (2026) develop doubly robust staggered QTT estimation and inference; Arias (2026) analyzes conditional/unconditional quantile DiD and staggered weighting; and Bhattacharjee, Li, and Xue (2026) develop an omnibus distributional treatment-heterogeneity test.

The paper-specific contribution is instead centered on the noncommutativity of two staggered-QTT aggregation operations. Proposition~1 gives a sharp one-sided finite-dispersion interval conditional on the fixed cohort quantiles and target weights, a coarser globally sharp weight--range envelope, an exact equality characterization, a density-tilted local representation, compact-uniform deterministic bounds, and a sharp first-order tilt--dispersion diagnostic; Proposition~2 and the uniform-process corollary provide pointwise and process-level sampling laws; Proposition~3 separates efficiency from target choice; and the mass-point corollary projects joint CDF--weight regions through both aggregation orders without differentiating quantiles. These results are integrated with, but conceptually distinct from, the supporting STOP/PROCEED architecture documented in this appendix.

Arias (2026) is close enough that the distinction should be explicit. His density weighting is generated by quantile-regression or recentered-influence-function TWFE decompositions; with multiple post-treatment periods and staggered adoption, those decompositions can combine cohort--time effects with non-convex weights. The density tilt in Proposition~1 is a different object: the starting cohort weights are prespecified, nonnegative policy weights, the cohort-state CDFs are already identified, and quantile inversion of their positive mixture endogenously replaces the target weights by positive state- and quantile-specific local weights. The paper studies the exact difference between that mixture quantile and the target-weighted average of cohort quantiles. Djuazon and Tsyawo (2026) independently develop uniform distribution-function/quantile-function inference under staggered distributional parallel trends, including non-continuous outcomes, and their weighted-CDF construction delivers the mixture-type QTT. Miller and Galvao (2026) develop doubly robust staggered cohort/time QTT estimation and aggregation summaries built from cohort/time QTTs. We therefore claim neither component aggregation operation, density weighting in regression decompositions, nor generic component-QTT inference as new; the distinctive object is the discrepancy between the two policy aggregation operators and the sharp finite/global/local theory, inference, and target-risk consequences attached to that discrepancy.

The pre-trend interpretation is deliberately limited. Following the warning in Roth (2022), failure to reject is not treated as validation of parallel trends. De Chaisemartin and D'Haultf\oe{}uille (2026) show in a general specification-test framework that under valid centered specification, symmetric convex pre-tests cannot make conventional post-test inference liberal; the centered-Gaussian conservativeness statement summarized in the main text is the direct specialization of that result to the route-specific max-\(|t|\) gate. Sarfati and Vilfort (2026) instead residualize target estimators against diagnostic statistics to obtain first-order pre-test independence. Mikhaeil and Harshaw (2026) obtain conditionally valid average-treatment-effect-on-the-treated (ATT) inference in a block-adoption setting under a conditional-extrapolation restriction linking acceptable pre-treatment violations to post-treatment bias, while de Chaisemartin (2026) uses observed pre-trends as a reference distribution under a distinct predictor restriction. The present paper's object is narrower and distributional: after a frozen route-specific gate has passed and the identifying route is maintained, Diagnostic Result D4(ii) specializes Gaussian selective-inference logic to the joint law of the reported QTT/CDF target and the route-specific diagnostic family. The selective pivot removes the excess conservativeness under the centered model; outside that model, including local diagnostic drift, no direction-of-distortion claim is made.

\section{Prospectively Frozen NSW Experimental Validation}
\label{app:nswvalidation}

The analyzed Dehejia--Wahba National Supported Work experimental subset has \(N=445\), with 185 randomized treated observations and 260 randomized controls. The SHA-256 cryptographic hash of the analyzed binary is
\begin{center}
\texttt{d1bd2680a1c6f799f1c6d2455bf29633fdf19be01cb19490621c20a560b4e072}.
\end{center}

\subsection{Frozen pre-treatment gate}

Let \(RE_{it}\) denote annual real earnings in the Dehejia--Wahba data. For
\[
y\in\{0,500,1000,2000,3000,5000,7500,10000,15000\},
\]
the frozen placebo coordinate is the treated-minus-control difference in the 1974--1975 change of \(1\{RE\le y\}\). The observed maximum absolute studentized statistic is 1.2968. The pre-specified reference calculation permutes the treatment labels 49,999 times while preserving 185 treated labels, using seed 20269301. It gives a 95\% critical value of 2.7192 and a \(+1\) permutation \(p\)-value of 0.7388.

The LaLonde and Dehejia--Wahba sources cited in the main text establish random assignment in the NSW experiment. The 185/260 subset is formed using pre-treatment earnings availability; Dehejia and Wahba explicitly argue that this pre-treatment restriction preserves experimental balance in the reduced sample. That supports use of the treated--control contrast as an experimental benchmark for the reduced target population. It does not, however, establish that the analyzed subset was generated by simple complete randomization over every 185-of-445 labeling. We therefore call 0.7388 a \emph{label-permutation} \(p\)-value, not an exact design-based randomization \(p\)-value. Exactness of that permutation reference would require the stronger complete-label-exchangeability condition.

As a robustness calculation, we form the two-sample influence score for each threshold coordinate and use 49,999 Rademacher multipliers (seed 20269701). The multiplier maximum is 1.2999, its 95\% critical value is 2.6674, and its \(+1\) \(p\)-value is 0.7336. This asymptotic calculation does not require complete-label exchangeability. Both reference distributions therefore place the frozen gate far from rejection.

\begin{tabular}{rrrr}
\toprule
Earnings threshold & CDF-change difference & SE & t statistic \\
\midrule
0.000 & -0.043 & 0.033 & -1.297 \\
500.000 & -0.045 & 0.035 & -1.279 \\
1000.000 & -0.041 & 0.035 & -1.184 \\
2000.000 & -0.039 & 0.036 & -1.080 \\
3000.000 & -0.017 & 0.035 & -0.475 \\
5000.000 & 0.002 & 0.032 & 0.077 \\
7500.000 & 0.019 & 0.028 & 0.664 \\
10000.000 & 0.018 & 0.021 & 0.866 \\
15000.000 & -0.020 & 0.018 & -1.115 \\
\bottomrule
\end{tabular}

\subsection{1978 distributional benchmark}

Because the pre-specified gate passes and the CDF-PT benchmark comparison remains substantively maintained as part of the frozen validation design, we execute the frozen second-stage calculation,
\[
\widehat F^0_{T,78}(y)
=
\widehat F_{T,75}(y)+\widehat F_{C,78}(y)-\widehat F_{C,75}(y),
\]
with isotonic projection before inversion. The experimental benchmark is the treated-minus-control marginal quantile difference in 1978. Under the original NSW randomization, and because the reduced Dehejia--Wahba sample is selected using pre-treatment information, this estimates the marginal experimental quantile treatment effect for the reduced target population. This causal benchmark statement is logically separate from whether the simple complete-label permutation distribution is the exact finite-sample assignment distribution for the reduced data.

\begin{tabular}{rrrrl}
\toprule
$\tau$ & DiD QTT & Experimental QTT & DiD $-$ experimental & 95\% bootstrap range \\
\midrule
0.1 & 0.0 & 0.0 & 0.0 & [0, 0] \\
0.2 & 0.0 & 0.0 & 0.0 & [-920, 0] \\
0.3 & -4.6 & 929.9 & -934.4 & [-2390, 0] \\
0.4 & -100.8 & 1177.7 & -1278.6 & [-2552, 0] \\
0.5 & 587.7 & 1148.7 & -561.1 & [-1644, 365] \\
0.6 & 1323.0 & 1466.5 & -143.5 & [-1033, 678] \\
0.7 & 2133.6 & 1795.2 & 338.4 & [-749, 1217] \\
0.8 & 2879.4 & 2278.1 & 601.4 & [-1051, 1756] \\
0.9 & 3045.3 & 3275.6 & -230.3 & [-1630, 1163] \\
\bottomrule
\end{tabular}

The QTT comparison is not uniform. At \(\tau=.3\) and \(.4\), the DiD QTT is approximately zero while the randomized experimental QTT is roughly \$930 and \$1,178; the descriptive gap ranges touch zero only at their upper endpoint. These coordinates straddle the large zero-earnings mass point.

To avoid quantile inversion at the atom, we also compare the DiD and experimental treatment effects directly on the fixed CDF-threshold family. Algebraically, before the shape projection,
\[
\widehat{\DTT}_{DiD}(y)-\widehat{\DTT}_{Exp}(y)
=
\widehat F_{C,75}(y)-\widehat F_{T,75}(y).
\]
Thus the CDF-scale validation gap is exactly the 1975 treated--control CDF imbalance. A 49,999-draw Rademacher multiplier max-\(|t|\) calculation gives a 95\% critical value of 2.540 and a global \(p\)-value of 0.220.

\begin{table}[htbp]
\centering
\caption{NSW comparison on the CDF scale}
\label{tab:nswcdfbenchmark}
\small
\begin{tabular}{rrrrr}
\toprule
Earnings threshold & DiD DTT & Experimental DTT & DiD $-$ experimental & 95\% simultaneous band\\
\midrule
0 & -0.026 & -0.111 & 0.085 & [-0.033,0.202] \\
500 & -0.055 & -0.131 & 0.076 & [-0.038,0.190] \\
1,000 & -0.007 & -0.084 & 0.077 & [-0.032,0.186] \\
2,000 & 0.019 & -0.061 & 0.079 & [-0.021,0.180] \\
3,000 & -0.032 & -0.075 & 0.042 & [-0.046,0.131] \\
5,000 & -0.065 & -0.077 & 0.012 & [-0.063,0.086] \\
7,500 & -0.120 & -0.106 & -0.014 & [-0.067,0.039] \\
10,000 & -0.064 & -0.060 & -0.004 & [-0.045,0.037] \\
15,000 & -0.048 & -0.047 & -0.001 & [-0.026,0.025] \\
\bottomrule
\end{tabular}
\begin{flushleft}\footnotesize
Notes: Earnings are nominal dollars in the NSW data. Before CDF-shape projection, the difference between the additive-CDF-DiD DTT and the randomized experimental DTT equals the 1975 control-minus-treated CDF imbalance at the same threshold. The simultaneous band uses 49,999 Rademacher multiplier draws over individual influence scores (seed 202610002). The global max-$|t|$ $p$-value for equality of the two CDF-effect curves on these nine thresholds is 0.220. This CDF-scale comparison avoids quantile inversion at the zero-earnings mass point.
\end{flushleft}
\end{table}

This threshold-scale result does not establish literal equality of the curves, but the familywise comparison does not reject on the prespecified thresholds. The lower thresholds display the same directional baseline imbalance that helps explain the lower-middle QTT discordance.

The largest absolute point gap is about \$1,279 at \(\tau=.40\). A post-gate descriptive uncertainty analysis uses 4,999 stratified nonparametric pairs-bootstrap draws within original treatment status (seed 20269601). All nine percentile ranges for the DiD-minus-experimental gap contain zero, but at \(\tau=.3\) and \(.4\) they touch zero only at the upper boundary. Because earnings have a substantial atom at zero and ordinary quantiles are nonsmooth at mass points, we do not attach smooth-quantile coverage to these ranges. The mass-point-safe CDF-inversion corollary in the main text gives a formal route, but the frozen NSW protocol did not prespecify a joint post-treatment treated/counterfactual CDF band; constructing one after viewing the benchmark would change the validation design. The reported percentile ranges therefore remain descriptive sampling-variability summaries, and Diagnostic Result D4(ii) is not invoked for them.

The validation therefore establishes neither distributional parallel trends nor equality of the population QTT curves. Its evidentiary value is procedural: the gate and analysis plan were fixed before the outcome analysis, the gate is far from rejection under two reference calculations, and the resulting distributional estimates can be compared against an independently randomized benchmark without tuning the diagnostic to that benchmark.

\section{Additional Technical Clarifications}

\paragraph{Inference after the diagnostic gate.}
Diagnostic Result D4 addresses the distributional consequence of reporting a post-treatment distributional target only after the frozen placebo gate passes. Its branch-adjusted and Gaussian-selective confidence sets account for that reporting event; they do not turn nonrejection into evidence that the maintained identifying assumption is true.

The main text distinguishes identification assumptions from their observable pre-treatment implications. A failure of a pre-treatment diagnostic is evidence against the maintained implementation; a non-rejection does not prove the missing post-treatment counterfactual restriction. CDF parallel trends and changes-in-changes are alternative, generally non-nested models. Repeated-cross-section standardization must use a fixed reference covariate distribution if composition changes over time. For a cohort contrast between baseline $s$ and evaluation date $t$, not-yet-treated controls must satisfy $G_i>\max(s,t)$ and must exclude the target cohort; this guarantees untreated status at both dates. Finally, an average of cohort-specific QTTs and the QTT obtained after mixing cohort distributions are distinct nonlinear summaries.

\end{document}